\documentclass[showpacs,preprintnumbers, twocolumn,
amsmath,amssymb,APSl,prd,nofootinbib,superscriptaddress]{revtex4-2} 
\usepackage{graphicx}
\usepackage{caption}
\usepackage{subcaption}
\usepackage{xcolor}
\usepackage{hyperref}
\usepackage{amsfonts}
\usepackage{bm}
\usepackage{mathrsfs}
\usepackage{soul}
\usepackage{makecell,multirow}
\usepackage{rotating}
\usepackage[utf8]{inputenc}
\usepackage{amsmath}
\usepackage{makecell}

\usepackage{amssymb}
\usepackage{tensor}
\usepackage{graphicx}
\usepackage{bigints}
\usepackage{bbm}
\usepackage{graphicx}
\usepackage{subcaption}
\usepackage{epsf}
\usepackage{bm}
\usepackage{amsmath}
\usepackage{amsfonts}
\usepackage{amssymb}
\usepackage{graphicx}
\usepackage{tabularx}
\usepackage{multirow}
\usepackage{color}%
\providecommand{\U}[1]{\protect\rule{.1in}{.1in}}
\newcommand{\ie}{\begin{equation}}
\newcommand{\fe}{\end{equation}}

\newcommand{\mincir}{\raise
-3.truept\hbox{\rlap{\hbox{$\sim$}}\raise4.truept\hbox{$<$}\ }}
\newcommand{\magcir}{\raise
-3.truept\hbox{\rlap{\hbox{$\sim$}}\raise4.truept\hbox{$>$}\ }}

\providecommand{\U}[1]{\protect\rule{.1in}{.1in}}

\usepackage{tikz,xcolor,hyperref}

\usepackage{hyperref}             % Enable clickable links
\hypersetup{
    colorlinks=true,              % Colored links instead of boxed
    breaklinks=true,              % Allow links to break across lines
    citecolor=green,               % Color for citation links
    linkcolor=[rgb]{0,0.5,0.9},   % Color for internal links
    urlcolor=blue,                 % Color for URLs
    filecolor=green               % Color for file links
}

\definecolor{lime}{HTML}{A6CE39}
\DeclareRobustCommand{\orcidicon}{%
	\begin{tikzpicture}
	\draw[lime, fill=lime] (0,0) 
	circle [radius=0.16] 
	node[white] {{\fontfamily{qag}\selectfont \tiny ID}};
	\draw[white, fill=white] (-0.0625,0.095) 
	circle [radius=0.007];
	\end{tikzpicture}
	\hspace{-2mm}
}

\foreach \x in {A, ..., Z}{%
	\expandafter\xdef\csname orcid\x\endcsname{\noexpand\href{https://orcid.org/\csname orcidauthor\x\endcsname}{\noexpand\orcidicon}}
}

\newcommand\orcidEdnaldo{{\href{https://orcid.org/0000-0001-7230-3666}{\orcidicon}}}
\newcommand\orcidFrancisco{{\href{https://orcid.org/0000-0002-9388-8373}{\orcidicon}}}
\newcommand\orcidManuel{{\href{https://orcid.org/0000-0001-8586-0285}{\orcidicon}}}
\newcommand\orcidTarciso{{\href{https://orcid.org/0009-0007-0450-2672}{\orcidicon}}}
\newcommand\orcidHenrique{{\href{https://orcid.org/0000-0001-7565-4277}{\orcidicon}}}
\newcommand\orcidLuis{{\href{https://orcid.org/0009-0009-4322-6484}{\orcidicon}}}
\newcommand\orcidAraujo{{\href{https://orcid.org/0000-0002-8790-3944}{\orcidicon}}}
\newcommand\orcidJorde{{\href{https://orcid.org/0009-0001-3344-2986}{\orcidicon}}}
\newcommand\orcidDiego{{\href{https://orcid.org/0000-0003-3984-9864}{\orcidicon}}}

\begin{document}

%%%%%%%%%%%%%%%%%%%%%%%%%%%%%%%%%%%%%%%%%%%%%%%%%%%%%%%%%%%%%%%% 
\title{Regular Black Holes in General Relativity from Nonlinear Electrodynamics with de Sitter Cores}
%%%%%%%%%%%%%%%%%%%%%%%%%%%%%%%%%%%%%%%%%%%%%%%%%%%%%%%%%%%%%%%% 

%%%%%%%%%%%%%%%%%%%%%%%%%%%%%%%%%%%%%%%%%%%%%%%%%%%%%%%%%%%%%%%% 
%%%%%%%%%%%%%%%%%%%%%%%%%%%%%%%%%%%%%%%%%%%%%%%%%%%%%%%%%%%%%%%% 
\author{A. A. Ara\'{u}jo Filho\orcidAraujo\!\!}
\email{dilto@fisica.ufc.br}
\affiliation{Departamento de Física, Universidade Federal da Paraíba, Caixa Postal 5008, 58051--970, João Pessoa, Paraíba,  Brazil.}
\affiliation{Departamento de Física, Universidade Federal de Campina Grande Caixa Postal 10071, 58429-900 Campina Grande, Paraíba, Brazil.}
\affiliation{Center for Theoretical Physics, Khazar University, 41 Mehseti Street, Baku, AZ-1096, Azerbaijan.}
%%%%%%%%%%%%%%%%%%%%%%%%%%%%%%%%%%%%%%%%%%%%%%%%%%%%%%%%%%%%%%%% 
\author{Ednaldo L. B. Junior\orcidEdnaldo\!\!} \email{ednaldobarrosjr@gmail.com}
\affiliation{Faculdade de F\'{i}sica, Universidade Federal do Pará, Campus Universitário de Tucuruí, CEP: 68464-000, Tucuruí, Pará, Brazil}
\affiliation{Programa de P\'{o}s-Gradua\c{c}\~{a}o em F\'{i}sica, Universidade Federal do Sul e Sudeste do Par\'{a}, 68500-000, Marab\'{a}, Par\'{a}, Brazil}
%%%%%%%%%%%%%%%%%%%%%%%%%%%%%%%%%%%%%%%%%%%%%%%%%%%%%%%%%%%%%%%% 
\author{José Tarciso S. S. Junior\orcidTarciso\!\!}
\email{tarcisojunior17@gmail.com}
\affiliation{Faculdade de F\'{i}sica, Programa de P\'{o}s-Gradua\c{c}\~{a}o em F\'{i}sica, Universidade Federal do Par\'{a}, 66075-110, Bel\'{e}m, Par\'{a}, Brazill}
%%%%%%%%%%%%%%%%%%%%%%%%%%%%%%%%%%%%%%%%%%%%%%%%%%%%%%%%%%%%%%%% 
\author{Francisco S. N. Lobo\orcidFrancisco\!\!} 
\email{fslobo@ciencias.ulisboa.pt}
\affiliation{Instituto de Astrof\'{i}sica e Ci\^{e}ncias do Espa\c{c}o, Faculdade de Ci\^{e}ncias da Universidade de Lisboa, Edifício C8, Campo Grande, P-1749-016 Lisbon, Portugal}
\affiliation{Departamento de F\'{i}sica, Faculdade de Ci\^{e}ncias da Universidade de Lisboa, Edif\'{i}cio C8, Campo Grande, P-1749-016 Lisbon, Portugal}
%%%%%%%%%%%%%%%%%%%%%%%%%%%%%%%%%%%%%%%%%%%%%%%%%%%%%%%%%%%%%%%% 
\author{Jorde A. A. Ramos\orcidJorde\!\!}
 \email{jordealves@ufpa.br}
\affiliation{Faculdade de F\'{i}sica, Programa de P\'{o}s-Gradua\c{c}\~{a}o em F\'{i}sica, Universidade Federal do Par\'{a}, 66075-110, Bel\'{e}m, Par\'{a}, Brazill}
%%%%%%%%%%%%%%%%%%%%%%%%%%%%%%%%%%%%%%%%%%%%%%%%%%%%%%%%%%%%%%%% 
\author{\\Manuel E. Rodrigues\orcidManuel\!\!} 
\email{esialg@gmail.com}
\affiliation{Faculdade de F\'{i}sica, Programa de P\'{o}s-Gradua\c{c}\~{a}o em F\'{i}sica, Universidade Federal do Par\'{a}, 66075-110, Bel\'{e}m, Par\'{a}, Brazill}
\affiliation{Faculdade de Ci\^{e}ncias Exatas e Tecnologia, Universidade Federal do Par\'{a}, Campus Universit\'{a}rio de Abaetetuba, 68440-000, Abaetetuba, Par\'{a}, Brazil}
%%%%%%%%%%%%%%%%%%%%%%%%%%%%%%%%%%%%%%%%%%%%%%%%%%%%%%%%%%%%%%%% 
\author{Diego Rubiera-Garcia\orcidDiego\!\!} 
\email{drubiera@ucm.es}
\affiliation{Departamento de Física Téorica and IPARCOS, Universidad Complutense de Madrid, E-28040 Madrid, Spain}
%%%%%%%%%%%%%%%%%%%%%%%%%%%%%%%%%%%%%%%%%%%%%%%%%%%%%%%%%%%%%%%% 
\author{Luís F. Dias da Silva\orcidLuis\!\!} 
\email{fc53497@alunos.fc.ul.pt}
\affiliation{Instituto de Astrof\'{i}sica e Ci\^{e}ncias do Espa\c{c}o, Faculdade de Ci\^{e}ncias da Universidade de Lisboa, Edifício C8, Campo Grande, P-1749-016 Lisbon, Portugal}
%%%%%%%%%%%%%%%%%%%%%%%%%%%%%%%%%%%%%%%%%%%%%%%%%%%%%%%%%%%%%%%% 
\author{Henrique A. Vieira\orcidHenrique\!\!} 
\email{henriquefisica2017@gmail.com}
\affiliation{Faculdade de F\'{i}sica, Programa de P\'{o}s-Gradua\c{c}\~{a}o em F\'{i}sica, Universidade Federal do Par\'{a}, 66075-110, Bel\'{e}m, Par\'{a}, Brazill}
%%%%%%%%%%%%%%%%%%%%%%%%%%%%%%%%%%%%%%%%%%%%%%%%%%%%%%%%%%%%%%%% 
%%%%%%%%%%%%%%%%%%%%%%%%%%%%%%%%%%%%%%%%%%%%%%%%%%%%%%%%%%%%%%%% 
%%%%%%%%%%%%%%%%%%%%%%%%%%%%%%%%%%%%%%%%%%%%%%%%%%%%%%%%%%%%%%%% 

%%%%%%%%%%%%%%%%%%%%%%%%%%%%%%%%%%%%%%%%%%%%%%%%%%%%%%%%%%%%%%%% 
\begin{abstract}
We present new regular black hole solutions in general relativity (GR) within a static, spherically symmetric framework governed by a variable equation of state, following the approach of [{\it Class. Quant. Grav. 42, 025024 (2025)}]. The matter supporting these geometries is identified as a purely magnetic monopole configuration of the Maxwell–Faraday tensor in the context of nonlinear electrodynamics (NLED). We explicitly reconstruct the corresponding NLED Lagrangian and analyze the asymptotic and central behaviors of the solutions. The geometric structure is examined through the metric functions, the regularity of the Kretschmann scalar, and the profiles of energy density and pressures, including a discussion of the resulting energy conditions. Using Event Horizon Telescope observations of Sgr A$^*$, we constrain the model parameters by comparing the predicted size of the central dark region with the inferred observational images, taking into account the effective geometry experienced by photons in the presence of NLED. Finally, we investigate the dynamical stability of these configurations under scalar perturbations by computing the quasinormal mode spectrum and performing a time-domain analysis.
\end{abstract}
%%%%%%%%%%%%%%%%%%%%%%%%%%%%%%%%%%%%%%%%%%%%%%%%%%%%%%%%%%%%%%%% 

%%%%%%%%%%%%%%%%%%%%%%%%%%%%%%%%%%%%%%%%%%%%%%%%%%%%%%%%%%%%%%%% 
\maketitle
%%%%%%%%%%%%%%%%%%%%%%%%%%%%%%%%%%%%%%%%%%%%%%%%%%%%%%%%%%%%%%%% 

%\tableofcontents

%%%%%%%%%%%%%%%%%%%%%%%%%%%%%%%%%%%%%%%%%%%%%%%%%%%%%%%%%%%%%%%%
\section{Introduction}\label{sec1}
%%%%%%%%%%%%%%%%%%%%%%%%%%%%%%%%%%%%%%%%%%%%%%%%%%%%%%%%%%%%%%%%

Einstein's General Relativity (GR), published in 1916 \cite{Einstein:1916vd}, is nowadays recognized as an observationally successful theory, since every gravitational phenomena tested so far is in agreement with its predictions. Canonical tests include the  precession of Mercury’s perihelion \cite{Will:2018mcj} and the deflection of light due to the curvature of space-time in the vicinity of massive bodies \cite{Will:2014zpa}. Furthermore, recent high-precision astrophysical tests in the strong-field regime have provided additional support for the validity of the theory. Such tests include the detection of gravitational waves from binary black hole mergers by the LIGO and Virgo collaborations~\cite{LIGOScientific:2016aoc, LIGOScientific:2017ync}, and the imaging of the plasma around supermassive objects at the center of the M87 and Milky Way galaxies by the Event Horizon Telescope (EHT) \cite{EventHorizonTelescope:2019dse,EventHorizonTelescope:2022wkp}.

Despite the remarkable success of GR, the theory still faces significant challenges. From a theoretical point of view, one of the most problematic features of the theory is that some of its physically relevant solutions, such as those describing cosmological evolution or black holes, contain space-time singularities. Such singularities are characterized by the incompleteness of geodesics (see \cite{Senovilla:2014gza} for a review) and are typically accompanied by divergences in some curvature scalars. Since GR loses its predictive power for trajectories approaching such singularities, it is widely believed that fully consistent physical solutions should be free of these pathologies everywhere.

An approach developed to resolve the singularities present in classical Maxwell electrodynamics was proposed by Born and Infeld in 1934~\cite{Born:1933pep,Born:1934gh}. In this context, they extended Maxwell’s theory in order to remove the divergences associated to the central point charge and the self-energy of point-like particles. This framework is nowadays known as nonlinear electrodynamics (NLED). Following this formulation, the work of Euler and Heisenberg, motivated by effective quantum field theory~\cite{Heisenberg:1936nmg}, suggested that NLEDs could be capable to incorporate physical features into the description of electromagnetism when strong fields are present. The work of J.~Plebański~\cite{Plebanski} gave further support to this formulation, and further generalizations were subsequently introduced, see e.g.~\cite{Kruglov:2014hpa,Kruglov:2014iwa,Kruglov:2014iqa,Bandos:2020jsw}.

A natural question in this context is whether the NLED framework can yield black hole solutions that are free of singularities. In 1968, Bardeen proposed the first such nonsingular black hole solution~\cite{Bardeen}, thereby pioneering the study of regular black holes (RBHs). In his construction, Bardeen introduced a mass function depending on the radial coordinate, rather than a constant mass as in the Schwarzschild solution, chosen such that the Kretschmann scalar remains finite everywhere. It was shown much later, in the work of Ayón-Beato and García~\cite{Ayon-Beato:2000mjt}, that the Bardeen solution can be interpreted as an exact solution of GR. In this context, the regularizing parameter is associated with a magnetic charge within a NLED framework. Subsequently, this interpretation was extended to the case of an electric source, leading to an alternative NLED realization of the solution~\cite{Rodrigues:2018bdc}.

Proposals inspired by Bardeen’s idea, as well as by the work of Ayón-Beato and García, have motivated the development of several RBH models that consider NLED as the matter source in GR. In this context, we highlight some representative works, such as those by Bronnikov~\cite{Bronnikov:2000vy}, Dymnikova~\cite{Dymnikova:2004zc}, Burinskii and Hildebrandt \cite{Burinskii:2002pz}, Balart and Vagenas~\cite{Balart:2014cga}, and Culetu~\cite{Culetu:2014lca}. Analyses of the thermodynamic properties of RBHs have also been carried out, as discussed in Refs.~\cite{Breton:2004qa,Myung:2007xd,Ma:2015gpa,Kruglov:2016ymq,Fan:2016hvf,Balart:2014jia,Kruglov:2016ezw,Kruglov:2017fck,Gullu:2020ant}. Furthermore, general studies regarding the mechanism under which such RBH arise have concluded that the Schwarzschild behavior must be replaced at the innermost region of the black hole by a de-Sitter type core \cite{Ansoldi:2008jw}. Besides finding solutions of this kind and discussing their theoretical features, one of the major goals within this framework is to explore their observational viability.

One of the main observational channels to detect putative signals of RBHs are gravitational waves. Following the merger of compact binaries, the resulting space-time does not settle instantaneously into a stationary configuration. Instead, the post-merger geometry evolves through a relaxation stage that governs the final segment of the gravitational waveform. Once the strongly nonlinear dynamics fade away, the remnant black hole transitions toward equilibrium via a characteristic pattern of exponentially damped oscillations. This behavior is captured within linear perturbation theory and is encoded in a discrete spectrum of complex eigenfrequencies known as quasinormal modes, obtained by solving the perturbation equations subject to physically motivated boundary conditions~\cite{Konoplya:2007zx,Konoplya:2013rxa,karmakar2024quasinormal,AraujoFilho:2025jcu,Konoplya:2019hlu,Kokkotas:2010zd,Konoplya:2011qq,AraujoFilho:2025hnf,Heidari:2025iiv}.
The quasinormal spectrum is fully specified by the intrinsic parameters of the black hole, such as its mass and any additional charges or geometric deformations, and does not depend on how the system was perturbed. 

In this framework, the oscillation frequency is dictated by the real component of each mode, while the imaginary component governs the decay rate of the signal. As a consequence, deviations in the underlying space-time geometry translate directly into modifications of the ringdown pattern, making quasinormal modes a sensitive diagnostic of the background metric. Because of this property, quasinormal spectra are frequently examined alongside other observational and theoretical probes of black hole structure, including the morphology of the central dark region in black hole imaging ~\cite{Jusufi:2020dhz} and wave-propagation effects characterized by greybody factors~\cite{Konoplya:2024vuj,Konoplya:2024lir,AraujoFilho:2025hkm,AraujoFilho:2024ctw}. Suggestions that current gravitational-wave data may already resolve individual quasinormal contributions remain provisional, as such interpretations rely strongly on the treatment of noise, systematics, and statistical confidence~\cite{Franchini:2023eda}. Nevertheless, continued enhancements in detector performance at LIGO, Virgo, and KAGRA, together with the expanding population of observed mergers, are expected to substantially refine ringdown studies in the near future.

The aim of this work is twofold. First, we shall study new regular black hole solutions in GR with static and spherically symmetric configurations endowed with a variable equation of state, more specifically of the form $P=\zeta(r)\rho$, as recently proposed in Ref.~\cite{Vertogradov:2024seh} with subsequent developments introduced in \cite{Heidari:2024bbd}. In this framework, the matter content is modeled as an anisotropic fluid, and under suitable restrictions on its components, namely, $p_r=-\rho$ and $p_t=P$ for its density and pressure components, the equations of motion admit only two independent components. After deriving the metric functions, we  shall specify two models for the function $\zeta(r)$. We then determine the matter content that supports such solutions, in addition to recovering the matter sector of the original work. This is achieved by coupling NLEDs to GR, where we consider only magnetic monopoles as components of the Maxwell-Faraday tensor. We perform a numerical analysis of the existence of horizons, verify the regularity of the solutions through the Kretschmann scalar, and identify the matter sources responsible for the geometric structure of these space-times. 

Second, we establish a connection between these new structures and the results of gravitational-wave astronomy. To this end, we first constrain the parameter associated with the magnetic charge of our solutions by comparing the predicted size of the central brightness depression with the observational estimates for Sgr A$^*$ reported by the EHT collaboration. In this way, we will be able to constraint this parameter through a direct relation between the geometric structure of our solutions and astrophysical observations. We then use this constraint to study the quasi-normal mode spectra of the corresponding families of black hole solutions, complemented by a time-domain analysis of the perturbations.

This manuscript is organized as follows. In Sec.~\ref{sec2}, we briefly review the field equations of GR coupled to NLED. In Secs.~\ref{Mod1}--\ref{Mod_III}, we analyze and discuss three regular, spherically symmetric black hole geometries, two of which are introduced in this work. In Sec.~\ref{sec:EC}, we present and examine the energy conditions for the two newly developed models. In Sec.~\ref{rsh}, we constrain the magnetic charge $q$ of these models using observational results from black hole imaging by the EHT collaboration for Sgr~A*. Next, in Sec.~\ref{sec:SEP}, we analyze the effective potential for scalar perturbations, which is then employed in Sec.~\ref{S:QNM} to compute the quasi-normal mode spectra of scalar fields for the three black hole solutions. These results are further corroborated through a time-domain analysis presented in Sec.~\ref{S:TDS}. Finally, in Sec.~\ref{sec:concl}, we summarize and discuss our main findings.

%%%%%%%%%%%%%%%%%%%%%%%%%%%%%%%%%%%%%%%%%%%%%%%%%%%%%%%%%%%%%%%% 
\section{GR coupled to NLED}\label{sec2}
%%%%%%%%%%%%%%%%%%%%%%%%%%%%%%%%%%%%%%%%%%%%%%%%%%%%%%%%%%%%%%%%

%%%%%%%%%%%%%%%%%%%%%%%%%%%%%%%%%%%%%%%%%%%%%%%%%%%%%%%%%%%%%%%%
\subsection{Field equations}
%%%%%%%%%%%%%%%%%%%%%%%%%%%%%%%%%%%%%%%%%%%%%%%%%%%%%%%%%%%%%%%%

We begin our analysis from the following action:
\begin{align}
	S = \int \sqrt{-g} \, \mathrm{d}^{4}x \, \big[\mathcal{R} + 2 \kappa^{2} \mathcal{L}(F) \big],
	\label{action}
\end{align}
where $g$ denotes the determinant of the space-time metric $g_{\mu\nu}$, and $\mathcal{L}(F)$ is the Lagrangian density of a NLED theory, which depends on the electromagnetic invariant
\begin{equation}
	F = \frac{1}{4} F^{\mu\nu} F_{\mu\nu}. \label{F}
\end{equation}
Here, $F_{\mu\nu}$ is the Maxwell-Faraday antisymmetric tensor, defined in terms of the vector potential $A_\mu$ as
\begin{equation}
	F_{\mu\nu} = \partial_\mu A_\nu - \partial_\nu A_\mu.
\end{equation}

The field equations arising from the action \eqref{action} are obtained by varying with respect to $A_{\mu}$ and $g_{\mu\nu}$. This yields, on the one hand, the electromagnetic equations:
\begin{align}
	\nabla_\mu \big(\mathcal{L}_F F^{\mu\nu}\big) = 0, \label{sol2}
\end{align}
where we denote $\mathcal{L}_F \equiv \partial \mathcal{L}(F)/\partial F$, and, on the other hand, the gravitational equations:
\begin{equation}
	G^\mu_{\phantom{\mu}\nu} \equiv \mathcal{R}^\mu_{\phantom{\mu}\nu} - \frac{1}{2} \delta^\mu_{\phantom{\mu}\nu} \mathcal{R} = \kappa^2 \, T^\mu_{\phantom{\mu}\nu}, \label{EqM}
\end{equation}
where the energy-momentum tensor of the NLED field takes the explicit form
\begin{align}
	T^\mu_{\phantom{\mu}\nu} = \delta^\mu_{\phantom{\mu}\nu} \mathcal{L}(F) - \mathcal{L}_F F^{\mu\alpha} F_{\nu\alpha}.
\end{align}

In this work, we consider the following static and spherically symmetric metric to construct our solutions:
\begin{equation} 
	\mathrm{d}s^2 = A(r)\, \mathrm{d}t^2 - \frac{1}{A(r)} \, \mathrm{d}r^2 - C(r) \, \mathrm{d}\Omega^2, \label{m}
\end{equation}
where $A(r)$ and $C(r)$ are functions of the radial coordinate $r$, and the line element on the two-spheres is defined as
$\mathrm{d}\Omega^2 \equiv \mathrm{d}\theta^{2} + \sin^{2}\theta \, \mathrm{d}\phi^{2}$.
The metric \eqref{m} represents the most general form compatible with staticity and spherical symmetry. The introduction of the nontrivial radial function $C(r)$ provides additional flexibility, allowing us to describe a wider class of solutions.

In this work, we focus on solutions sourced solely by a magnetic charge, $q$. In this case, the only nonvanishing components of the field strength tensor $F_{\mu\nu}$ are
\begin{align}
	F_{23} = -F_{32} = q \, \sin\theta \,,
\end{align}
which leads to the electromagnetic invariant
\begin{equation}
	F = \frac{q^2}{2\, C(r)^2}. \label{F2}
\end{equation}
Moreover, the formalism requires an important consistency condition relating the Lagrangian derivative to the radial dependence of the field:
\begin{equation}
	\mathcal{L}_F = \frac{\partial \mathcal{L}}{\partial r} \left( \frac{\partial F}{\partial r} \right)^{-1}. \label{RC}
\end{equation}

For completeness, and since we shall also examine the regularity of the solutions, we present here the explicit expression for the Kretschmann scalar:
\begin{align}
	&K=-\frac{f(r)C'(r)^{2}\left(2f(r)C''(r)+C'(r)f'(r)+2\right)}{C(r)^{3}}\nonumber
\\&+\frac{2f(r)^{2}C''(r)^{2}+C'(r)^{2}f'(r)^{2}+2f(r)C'(r)C''(r)f'(r)}{C(r)^{2}}\nonumber
\\&+\frac{4}{C(r)^{2}}+\frac{3f(r)^{2}C'(r)^{4}}{4C(r)^{4}}+f''(r)^{2}. \label{Kret}
\end{align}
We note that, in a singular space-time, this scalar typically exhibits the strongest divergence (for discussions on the hierarchy of curvature scalars, see e.g., \cite{Smolic:2026dmq,Smolic:2026dmq}). Consequently, we shall require that $K$ remains finite everywhere.

The components derived from the equations of motion, as described by Eq.~\eqref{EqM}, after substituting Eq.~\eqref{m}, are:
\begin{align}
   & \frac{A(r)C'(r)^{2}-2C(r)\big(A'(r)C'(r)+2A(r)C''(r)-2\big)}{4C(r)^{2}}
\nonumber
\\
& \qquad \qquad \qquad =\kappa^{2}{\cal L}(r),\label{EqF00}
   \\
&
\frac{C(r)\big(4-2A'(r)C'(r)\big)-A(r)C'(r)^{2}}{4C(r)^{2}}
=\kappa^{2}{\cal L}(r),
\label{EqF11}
\\
&
\frac{A(r)C'(r)^{2}-2C(r)\big(A'(r)C'(r)+A(r)C''(r)\big)}{4C(r)^{2}}
\nonumber
\\
&\qquad 
-\frac{A''(r)}{2C(r)^{2}}=\kappa^{2}\left({\cal L}(r)-\frac{q^{2}{\cal L}_{F}(r)}{C(r)^{2}}\right).\label{EqF22}
\end{align}
Solving the equations of motion \eqref{EqF00} and \eqref{EqF22}, we obtain the following expressions for the Lagrangian and its derivative:
\begin{eqnarray}
	\mathcal{L}(r) &=& -\frac{2C(r)\big(A'(r)C'(r)-2\big)+A(r)C'(r)^{2}}{4\kappa^{2}C(r)^{2}}, \label{L_RBH} \\[1mm]
	\mathcal{L}_F(r) &=& \frac{1}{2 \kappa^2 q^2} 
	\Big[C(r)\big(C(r)A''(r)+A(r)C''(r)+2\big) 
		\nonumber \\
	&& \qquad \qquad -A(r)C'(r)^{2} \Big]. \label{LF_RBH}
\end{eqnarray}

To analyze the properties of the metric functions that we will present later and to determine the presence of horizons, we will use the following condition
\begin{equation}
     A(r_{H})=0.\label{rH}
\end{equation}
where the radius $r_{H}$ denotes the presence of the horizon, while the second condition
\begin{equation}
    \frac{\mathrm{d} A(r)}{\mathrm{d}r}\bigg|_{r=r_H}=0,
    \label{der_a}
\end{equation}
allows to detect the presence of degenerate horizons.

%%%%%%%%%%%%%%%%%%%%%%%%%%%%%%%%%%%%%%%%%%%%%%%%%%%%%%%%%%%%%%%%
\subsection{Approach}
%%%%%%%%%%%%%%%%%%%%%%%%%%%%%%%%%%%%%%%%%%%%%%%%%%%%%%%%%%%%%%%%

Let us now describe our approach to find solutions and link them to the electromagnetic quantities described by  Eqs.~\eqref{L_RBH} and ~\eqref{LF_RBH}. Hereafter we consider the case in which the areal function is given by $C(r)=r$. Our approach is based in the results of \cite{Vertogradov:2024seh}.  To this end, we consider the energy-momentum tensor of our theory as given by the one of an anisotropic fluid, namely, 
\begin{equation}
	T_{\mu\nu}=(\rho+p_t)u_\mu \, u_\nu+p_t\,
	g_{\mu\nu}+(p_r-p_t)\chi_\mu \chi_\nu \,,
	\label{TME_anis}
\end{equation}
where $u^\mu$ is the four-velocity, $\chi^\mu$ is the unit
spacelike vector in the radial direction, i.e.,
$\chi^\mu=\sqrt{A(r)}\,\delta^\mu{}_r$,
$\rho(r)$ is the energy density, $p_r(r)$ is the radial pressure measured in the direction of $\chi^\mu$, and $p_t(r)$ is the transverse pressure measured in the orthogonal direction to $\chi^\mu$.
If we insert $p_t=p_r$ into Eq.~\eqref{TME_anis}, we obtain an energy-momentum tensor that represents an isotropic fluid. 

We now consider solutions of the metric \eqref{m} in which the metric function is expressed in terms of a mass function in the standard form:
\begin{equation}
	A(r) = 1 - \frac{2 \mathcal{M}(r)}{r}, \label{A}
\end{equation}
where $\mathcal{M}(r)$ is an arbitrary function of the radial coordinate $r$. Assuming an anisotropic fluid with radial and tangential pressures satisfying $p_r = -\rho$ and $p_t = P$, the symmetry of the theory allows us to express the components of the energy-momentum tensor directly in terms of $\mathcal{M}(r)$, as obtained from the equations of motion \eqref{EqF00}–\eqref{EqF22}:
\begin{align}
	\rho &= \frac{2 \mathcal{M}'(r)}{r^2}, \label{rho} \\[1mm]
	P &= -\frac{\mathcal{M}''(r)}{r}. \label{P}
\end{align}

To solve these equations, following \cite{Vertogradov:2024seh}, we introduce the relation
\begin{equation}
	P = \zeta(r) \, \rho,
\end{equation}
where $\zeta(r)$ denotes the equation of state function. Using this definition, the differential equations \eqref{rho} and \eqref{P} can be integrated to yield the mass function in the form
\begin{align}
	\mathcal{M}(r) = \int w(r) \, \mathrm{d}r + M, \label{func_M}
\end{align}
where $M$ is an integration constant, and
\begin{align}
	w(r) = w_0 \, \exp\Bigg[-2 \int \frac{\zeta(r)}{r} \, \mathrm{d}r \Bigg]. \label{w}
\end{align}

This construction allows for a complete determination of the metric functions once a specific form of $\zeta(r)$ is chosen. We now proceed to consider three such choices and discuss their properties in terms of the resulting metric, regularity, and matter Lagrangian.

%%%%%%%%%%%%%%%%%%%%%%%%%%%%%%%%%%%%%%%%%%%%%%%%%%%%%%%%%%%%%%%%
\section{Model I}\label{Mod1}
%%%%%%%%%%%%%%%%%%%%%%%%%%%%%%%%%%%%%%%%%%%%%%%%%%%%%%%%%%%%%%%%

%%%%%%%%%%%%%%%%%%%%%%%%%%%%%%%%%%%%%%%%%%%%%%%%%%%%%%%%%%%%%%%%
\subsection{Metric function}
%%%%%%%%%%%%%%%%%%%%%%%%%%%%%%%%%%%%%%%%%%%%%%%%%%%%%%%%%%%%%%%%

Our first choice is the one of \cite{Vertogradov:2024seh,Ovalle:2023ref}, where a linear function for $\zeta(r)$ is given by
\begin{equation}
    \zeta(r)=\frac{r}{R}-1,\label{zeta}
\end{equation}
where $R$ represents the surface of the compact object. Note that if we take the limit $r \to 0$, we get a de Sitter core, i.e. $P = -\rho$. Following Eq.~\eqref{func_M}, the mass function takes then the form 
\begin{equation}
       \mathcal{M}(r)=M-\frac{Rw_{0}}{4}\left(2r^{2}+2rR+R^{2}\right)\exp{\left(-\frac{2r}{R}\right)}.\label{M1}
\end{equation}

The space-time described by Eq.~\eqref{M1} is, in general, singular. For a static and spherically symmetric geometries, the presence of singularities can be assessed through the Kretschmann scalar~\cite{Lobo:2020ffi}, so we shall resort to the expression (\ref{Kret}) to perform this analysis.
A necessary condition for the Kretschmann scalar to remain finite is that, for a metric function $g_{tt}$ given by Eq.~\eqref{A}, the limit $\lim_{r \to 0} \mathcal{M}(r)/r$ must be finite. In this case, the corresponding $A(r)$ will be regular. This condition is, in fact, sufficient to ensure that the space-time is regular~\cite{Rodrigues:2023fps}.  
For all models considered here, imposing $\mathcal{M}(0) = 0$ is sufficient to ensure the regularity of the space-time.

Based on this, we can extract a relation from the analysis of $\mathcal{M}(r)$ at the center of the configuration. Specifically, in the limit $r \to 0$, imposing $\mathcal{M}(0) = 0$ yields
\begin{equation}
	\mathcal{M}(0) = M - \frac{1}{4} R w_0.
\end{equation}
With this, we can avoid the singularity at $r=0$ if we define the following relation:
\begin{equation}
    w_0=\frac{4M}{R^3} \label{w0}
\end{equation}
so that the metric function for this model reads, after using Eq.~\eqref{M1}, as
\begin{equation}
A(r)=1-\frac{2M}{r}\left[1-\frac{2}{R^{2}}\left(r^{2}+rR+\frac{R^{2}}{2}\right)\exp\left(-\frac{2r}{R}\right)\right].\label{AMod1}
\end{equation}
which also allows to find the corresponding energy density and the pressure as
\begin{eqnarray}
    \rho(r) &=& 2w_{0}\exp\left(-\frac{2r}{R}\right),\label{rho1}
    \\
    P(r) &=& 2w_{0}\left(\frac{r}{R}-1\right)\exp\left(-\frac{2r}{R}\right).\label{P1}
\end{eqnarray}

%%%%%%%%%%%%%%%%%%%%%%%%%%%%%%%%%%%%%%%%%%%%%%%%%%%%%%%%%%%%%%%%
\subsection{Kretschmann  scalar }
%%%%%%%%%%%%%%%%%%%%%%%%%%%%%%%%%%%%%%%%%%%%%%%%%%%%%%%%%%%%%%%%

We now examine the regularity of our models, starting from the Kretschmann scalar given by Eq.~\eqref{Kret}. For the present model, with the parameter restrictions specified above, we find
\begin{align}
	&K(r)= \frac{16M^{2}e^{-\frac{4r}{R}}}{r^{6}R^{8}}\Bigg[16r^{8}+32r^{6}R^{2}+32r^{5}R^{3}+
	\nonumber\\
	&
	3R^{8}\left(e^{\frac{2r}{R}}-1\right)^{2}-4r^{4}R^{4}\left(2e^{\frac{2r}{R}}-9\right)-8r^{3}R^{5}\left(e^{\frac{2r}{R}}-4\right)
	\nonumber\\
	&
	-12r^{2}R^{6}\left(e^{\frac{2r}{R}}-2\right)
	-12rR^{7}\left(e^{\frac{2r}{R}}-1\right)\Bigg]
	,\label{K_SV}
\end{align}
We observe that the Kretschmann scalar is finite at the origin:
\begin{equation}
	\lim_{r \to 0} K(r) = \frac{32 w_0^2}{3} = \frac{512 M^2}{3 R^6}. \label{K0_SV}
\end{equation}
Furthermore, for large $r$, the scalar vanishes asymptotically, indicating that the space-time is asymptotically flat and everywhere regular.

%%%%%%%%%%%%%%%%%%%%%%%%%%%%%%%%%%%%%%%%%%%%%%%%%%%%%%%%%%%%%%%%
\subsection{Matter Lagrangian}
%%%%%%%%%%%%%%%%%%%%%%%%%%%%%%%%%%%%%%%%%%%%%%%%%%%%%%%%%%%%%%%%

With the quantities presented above, we will now determine the analytical form of the Lagrangian that supports this solution. In order to reconstruct the electromagnetic quantities from Eqs.~\eqref{L_RBH} and \eqref{LF_RBH} we insert there the metric function described by Eq.~\eqref{AMod1} to find  the Lagrangian ${\cal L}(r)$ for this model as
\begin{align}
{\cal L}(r) =\frac{8 M }{\kappa ^2 R^3}	\exp{\left(-\frac{2 r}{R}\right)},
\label{L_SV} 
\end{align}
while the derivative ${\cal L}_F$ is obtained after substituting Eq.~\eqref{AMod1} into Eq.~\eqref{LF_RBH}, which yields
\begin{align}
{\cal L}_F(r) =\frac{8 M r^5 }{\kappa ^2 q^2 R^4} \exp{\left(-\frac{2 r}{R}\right)}.\label{LF_SV}
\end{align}
Inverting the relation~\eqref{F2} allows us to express $r$ as a function of $F$, so that the Lagrangian can be written explicitly in terms of $F$ as
\begin{align}
   {\cal L}(F) =	\frac{8M}{\kappa^{2}R^{3}} \exp\left(-\frac{2^{3/4}\sqrt{q}}{\sqrt[4]{F}R}\right),
 \label{L-F} 
\end{align} 
and its derivative reads
\begin{align}
   {\cal L} _F(F) =	\frac{2^{\frac{7}{4}}M\sqrt{q}}{F^{5/4}\kappa^{2}R^{4}} \exp\left(-\frac{2^{3/4}\sqrt{q}}{\sqrt[4]{F}R}\right).
 \label{LF-F} 
\end{align} 
which satisfy the consistency relation \eqref{RC}.

%%%%%%%%%%%%%%%%%%%%%%%%%%%%%%%%%%%%%%%%%%%%%%%%%%%%%%%%%%%%%%%%
\subsection{Metric function and matter in terms of $q$}
%%%%%%%%%%%%%%%%%%%%%%%%%%%%%%%%%%%%%%%%%%%%%%%%%%%%%%%%%%%%%%%%

\begin{figure}[h!]
\includegraphics[scale=0.55]{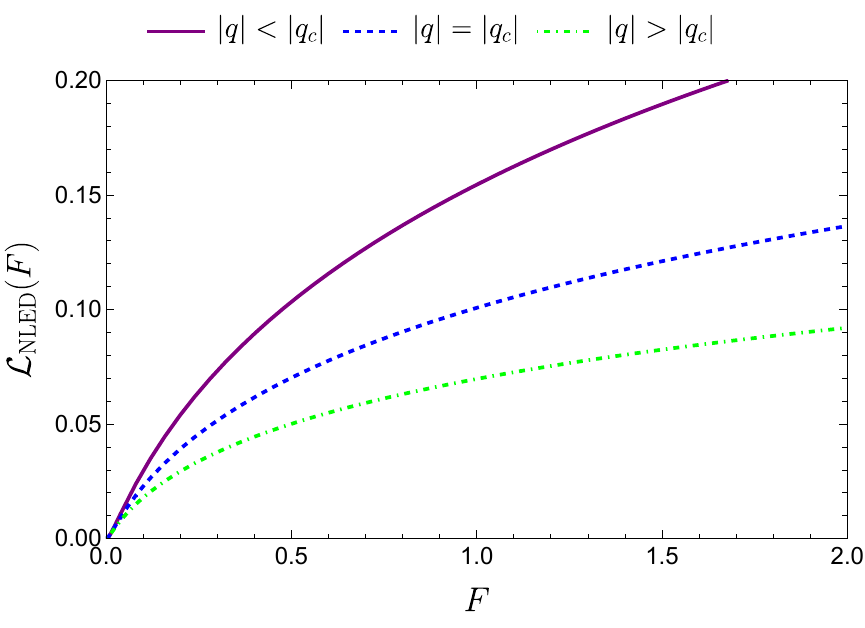}
\caption{The NLED Lagrangian ${\cal L}(F) $, given by Eq.~\eqref{L1} for different values of the electric charge    $|q|<|q_c|,\,|q|=|q_c|,\,|q|>|q_c|$.  Here we have taken $M=1$.} 
\label{fig_LxF1}
\end{figure}

In order to establish consistency with our electromagnetic solutions, we will also use the following interpretation from this model and the following ones:
\begin{equation}
    R=|q| .\label{Eq_R}
\end{equation}
This way, the most important quantities of this model, such as $A(r)$, given by Eq. \eqref{AMod1}, and ${\cal L}(F)$, given by Eq. \eqref{L-F}, have now the following form
\begin{align}
&A(r)=1-\frac{2M}{r}+2M\left(\frac{2r}{q^2}+\frac{2}{|q|}+\frac{1}{r}\right)\exp\left(-\frac{2r}{|q|}\right),\label{A1}
\\
   &{\cal L} (F) =\frac{8M}{\kappa^{2}|q|^{3}}\exp\left(-\frac{2^{3/4}}{\sqrt[4]{F}\,|q|^{1/2}}\right).\label{L1}
\end{align} 
Note that if we take the limit $|q| \rightarrow 0$ in the metric function given by Eq.~\eqref{A1}, we recover the Schwarzschild solution. In contrast, for $|q| \rightarrow \infty$, this metric function becomes asymptotically Minkowskian. We also observe that the Lagrangian given by Eq.~\eqref{L1} vanishes in the limit $F \rightarrow 0$, whereas for $F \rightarrow \infty$, it approaches $8M/\left(|q|^3 \kappa^2\right)$. The behavior of Eq.~\eqref{L1} for three different values of the charge is shown in Fig.~\ref{fig_LxF1}.

\begin{figure}[t!]
\includegraphics[scale=0.55]{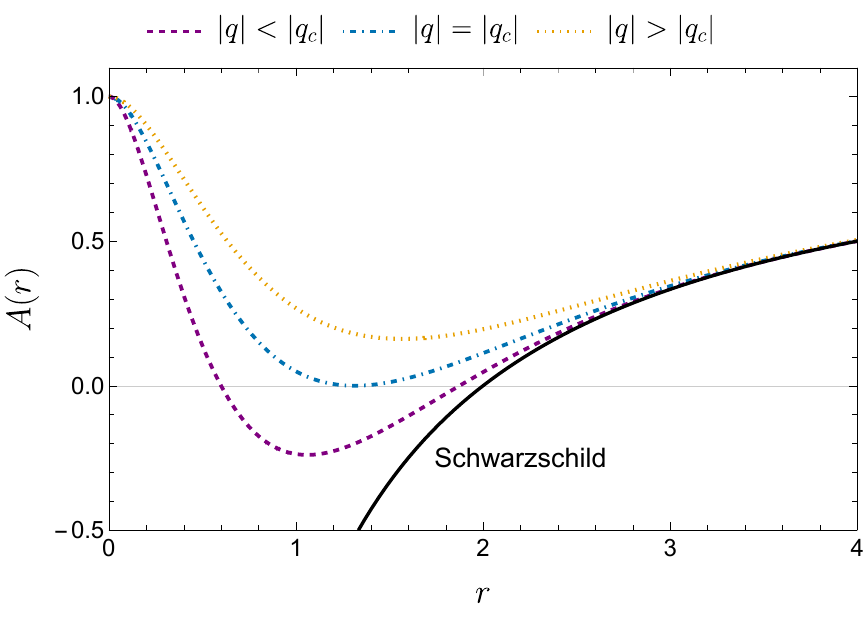}
\caption{The  metric function $A(r) $, given by Eq.~\eqref{A1} for three magnetic charge scenarios. Here we have taken $M=1$. } 
\label{fig_A}
\end{figure}

We also solve Eqs. \eqref{rH} and \eqref{der_a} simultaneously for this model, determining the critical electric charge from the metric function \eqref{A1}. Taking $M=1$, the critical charge is $|q_{c}| = 0.776$. In Fig. \ref{fig_A}, we present the behavior of the metric function~\eqref{A1} in relation to the radial coordinate $r$ for three distinct scenarios of the electric charge: $|q| < |q_{c}|$, $|q| = |q_{c}|$, and $|q| > |q_{c}|$.
When the charge is less than the critical charge, i.e., $|q| < |q_{c}|$, we observe  the formation of two horizons (purple curve), for the critical charge we depict it in blue, while the orange curve represents the case in which no horizons are present. By comparison, we also depict the Schwarzschild solution via the solid black curve, corresponding to $|q| \to 0$ in the metric function~\eqref{A1}.
%%%%%%%%%%%%%%%%%%%%%%%%%%%%%%%%%%
%%%%%%%%%%%%%%%%%%%%%%%%%%%%%%%%%%%%%%%%%%%%%%%%%

Taking into account the conditions~\eqref{w0} and~\eqref{Eq_R}, then Eqs.~\eqref{rho1} and~\eqref{P1} for the energy density and pressure, can be written explicitly as
\begin{align}
    &\rho(r)=2  w_0 e^{-\frac{2 r}{|q|}}=\frac{8 M e^{-\frac{2 r}{|q|}}}{|q|^3},\label{rho2_M1}
    \\
    &
    P(r)=2 w_0 e^{-\frac{2 r}{|q|}} \left(\frac{r}{|q|}-1\right)=\frac{8 M e^{-\frac{2 r}{|q|}} \left(r-|q|\right)}{q^4}.\label{P2_M1}
\end{align}

Expanding the fluid components~\eqref{rho2_M1} and-\eqref{P2_M1} near the center, we obtain
\begin{align}
	\rho(r) &\sim -\frac{16 M r}{q^4} + \frac{8 M}{|q|^3} + \mathcal{O}(r^2), \\
	P(r) &\sim \frac{24 M r}{q^4} - \frac{8 M}{|q|^3} + \mathcal{O}(r^2).
\end{align}

In the asymptotic regime ($r \to \infty$), the fluid components behave as
\begin{align}
	\rho(r) &= \frac{8 M}{|q|^3} \, e^{-2r/|q|} \Big[ 1 + \mathcal{O}\big(1/r\big) \Big], \\
	P(r) &= e^{-2r/|q|} \Bigg( \frac{8 M}{q^4} \, r - \frac{8 M}{|q|^3} + \mathcal{O}\big(1/r^2\big) \Bigg).
\end{align}

\begin{figure}[t!]
\includegraphics[scale=0.55]{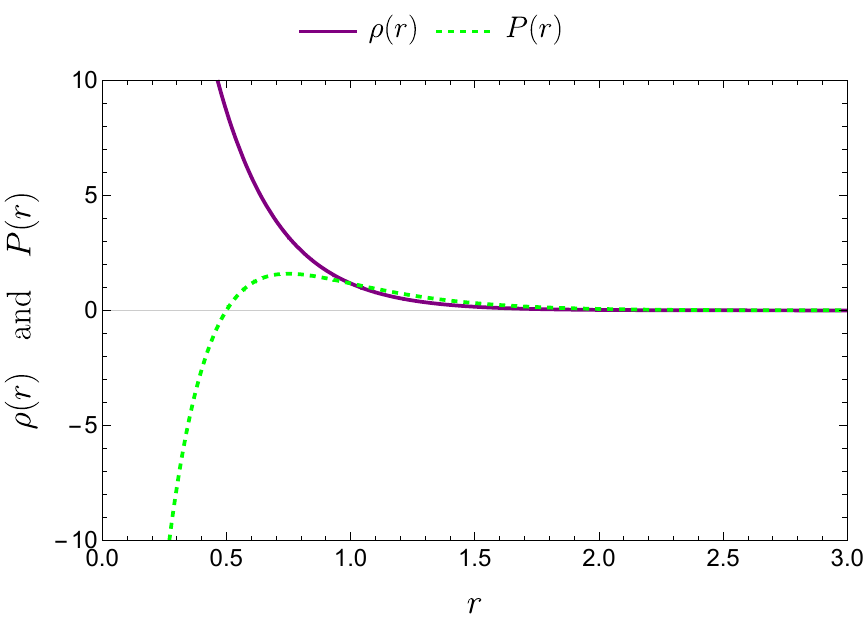}
\caption{The energy density $\rho(r)$ and pression $P(r) $, given by Eq.~\eqref{rho2_M1} and~\eqref{P2_M1}.  We have used the values of the constants  $M=1$ and $|q|=0.5$.} 
\label{fig_rhoP}
\end{figure}

The behavior of the energy density \eqref{rho2_M1} and pressure \eqref{P2_M1} is illustrated in Fig.~\ref{fig_rhoP}.

%%%%%%%%%%%%%%%%%%%%%%%%%%%%%%%%%%%%%%%%%%%%%%%%%%%%%%%%%%%%%%%%
\section{Model II}\label{Mod_II}
%%%%%%%%%%%%%%%%%%%%%%%%%%%%%%%%%%%%%%%%%%%%%%%%%%%%%%%%%%%%%%%%

%%%%%%%%%%%%%%%%%%%%%%%%%%%%%%%%%%%%%%%%%%%%%%%%%%%%%%%%%%%%%%%%
\subsection{Metric function}
%%%%%%%%%%%%%%%%%%%%%%%%%%%%%%%%%%%%%%%%%%%%%%%%%%%%%%%%%%%%%%%%

Our second model is defined by the choice $\zeta(r)$,
\begin{equation}
\zeta(r)=\frac{r-R}{r+R}.\label{zeta3}
\end{equation}
Note that at the center we have a de Sitter core, $\zeta(0)=-1$. For this model, the mass function~\eqref{func_M}, after substituting the function~\eqref{zeta3}, takes the form
\begin{align}
\mathcal{M}(r)=M-\frac{w_0 \left(3 r^2+3 r R+R^2\right)}{3 (r+R)^3}.
%=\frac{M r^3}{(r+R)^3}.\label{func_M3}
\end{align}
At the center of the configuration, i.e., in the limit $r \to 0$, $\mathcal{M}(r)$ becomes
\begin{align}
\mathcal{M}(0)=M-\frac{w_0}{3 R}.
%\label{M0}
\end{align}
From the above result, we derive the following relation:
\begin{equation}
w_{0}=3 M R.\label{w3}
\end{equation}
With this definition, our space-time becomes regular, and the mass function $\mathcal{M}(r)$ now takes the following form:
\begin{align}
\mathcal{M}(r)=
\frac{M r^3}{(r+R)^3}.\label{func_M3}
\end{align}
which is regular in both the central and the asymptotic limits. Accordingly, the metric function~\eqref{A} takes the form
\begin{eqnarray}
A(r) &=& 1-\frac{2}{r}\left[M-\frac{w_{0}\left(3r^{2}+3rR+R^{2}\right)}{3(r+R)^{3}}\right]
\nonumber \\
&=& 1-\frac{2 M r^2}{(r+R)^3}. \label{A3} 
\end{eqnarray} 

Considering the asymptotic behavior of the metric function~\eqref{A3}, we find that for large $r$ it approaches the Minkowski form. In the opposite limit, $r \to 0$, the metric function tends to unity.  
Note that upon substituting the value of $w_0$ into $A(r)$, the space-time is rendered regular. In contrast, if the limit $r \to 0$ is taken without this substitution, the metric function diverges.

Furthermore, we note that the metric function~\eqref{A3} is a particular case of the general metric function given by Eq.~(27) proposed in Ref.~\cite{Fan:2016hvf}. For the specific parameter values $\nu=1$, $\mu=3$, and $\alpha=M^{-1}q^3$ in the metric function~(27), we recover our metric function~\eqref{A3}.

%%%%%%%%%%%%%%%%%%%%%%%%%%%%%%%%%%%%%%%%%%%%%%%%%%%%%%%%%%%%%%%%
\subsection{Kretschmann  scalar }
%%%%%%%%%%%%%%%%%%%%%%%%%%%%%%%%%%%%%%%%%%%%%%%%%%%%%%%%%%%%%%%%

The Kretschmann scalar for this case takes the form
\begin{equation}
	K(r) = \frac{48 M^2 \left( r^4 - 2 r^3 R + 7 r^2 R^2 + 2 R^4 \right)}{(r + R)^{10}}, 
	\label{K_3}
\end{equation}
which is finite at the origin:
\begin{equation}
	\lim_{r \to 0} K(r) = \frac{96 M^2}{R^6}.
\end{equation}

For large $r$, i.e., in the limit $r \to \infty$, the Kretschmann scalar vanishes asymptotically, indicating that the space-time is asymptotically flat.

%%%%%%%%%%%%%%%%%%%%%%%%%%%%%%%%%%%%%%%%%%%%%%%%%%%%%%%%%%%%%%%%
\subsection{Matter Lagrangian}
%%%%%%%%%%%%%%%%%%%%%%%%%%%%%%%%%%%%%%%%%%%%%%%%%%%%%%%%%%%%%%%%

From the metric function~\eqref{A3}, we get the following ${\cal L}(r)$ for this model
\begin{align}
 {\cal L}(r) =\frac{2 w_0}{\kappa ^2 (r+R)^4}
=\frac{6 M R}{\kappa ^2 (r+R)^4},
\label{L_M2} 
\end{align}
while the derivative ${\cal L}_F$ is
\begin{align}
{\cal L}_F(r) =\frac{4 r^5 w_0}{\kappa ^2 q^2 (r+R)^5}
=
\frac{12 M r^5 R}{\kappa ^2 q^2 (r+R)^5}
.\label{LF_M2}
\end{align}
Regarding the expression of the Lagrangian in terms of $F$, we use Eq.~\eqref{F2} to express $r(F)$, which provides
\begin{align}
 {\cal L}(F)& =\frac{32 F w_0}{\kappa ^2 \left(2 \sqrt[4]{F} R+2^{3/4} \sqrt{|q|}\right)^4}
\nonumber
\\
&
=\frac{96 F M R}{\kappa ^2 \left(2 \sqrt[4]{F} R+2^{3/4} \sqrt{|q|}\right)^4},
\label{L2_M2} 
\end{align}
and
\begin{align}
  {\cal L} _F(F)
&=	\frac{32\ 2^{3/4} \sqrt{|q|} \, w_0}{\kappa ^2 \left(2 \sqrt[4]{F} R+2^{3/4} \sqrt{|q|}\right)^5}
   \nonumber
\\
&
=\frac{96\ 2^{3/4} M \sqrt{|q|} R}{\kappa ^2 \left(2 \sqrt[4]{F} R+2^{3/4} \sqrt{|q|}\right)^5}.
 \label{LF2_M2} 
\end{align} 
Here, we note that Eqs.~\eqref{L2_M2} and \eqref{LF2_M2} also satisfy the relation given by Eq.~\eqref{RC}.

%%%%%%%%%%%%%%%%%%%%%%%%%%%%%%%%%%%%%%%%%%%%%%%%%%%%%%%%%%%%%%%%
\subsection{Metric function and matter in terms of $q$}
%%%%%%%%%%%%%%%%%%%%%%%%%%%%%%%%%%%%%%%%%%%%%%%%%%%%%%%%%%%%%%%%

We can again rewrite the metric function \eqref{A3} and the Lagrangian~\eqref{L2_M2} using the relation described by Eq.~\eqref{Eq_R}, so we have that 
\begin{align}
A(r)&= %1-\frac{2}{r}\left(M_{0}-\frac{w_{0}\left(|q|^{4}+3q^{2}r+3r^{2}\right)}{3\left(|q|^{2}+r\right)^{3}}\right)\nonumber\\
1-\frac{2 M r^2}{\left(|q|+r\right)^3}\,,\label{Eq_A3}
\\
   {\cal L}(F) &=  \frac{96FM|q|}{\kappa^{2}\left(2\sqrt[4]{F}|q|+2^{3/4}\sqrt{|q|}\right)^{4}}
   \, ,\label{L3_M2}
    \\
  {\cal L} _F(F)  &=  \frac{96\, 2^{3/4} M}{\kappa ^2 \left(2 \sqrt[4]{F} |q|^{1/2}+2^{3/4}\right)^5}\,,\label{LF3_M2}
   \\
   {\cal L} _{FF}(F)  &= -\frac{240\ 2^{3/4}M}{\kappa^{2}|q|^{1/2}\left(2\sqrt[4]{F}|q|^{1/2}+2^{3/4}\right)^{6}F^{3/4}}\,.\label{LFF3_M2}
\end{align} 

It is worth noting that in the limit $|q| \to 0$, the metric function in Eq.~\eqref{Eq_A3} reduces to the Schwarzschild solution. Conversely, for $|q| \to \infty$, the metric becomes asymptotically Minkowskian. Similarly, we observe that the Lagrangian given by Eq.~\eqref{L3_M2} approaches the following form in the limit $F \to \infty$:
\begin{equation}
\lim_{F \to \infty}{\cal L}(F) =\frac{6 M}{\kappa ^2 |q|^3}.
\end{equation}

\begin{figure}[t!]
\includegraphics[scale=0.55]{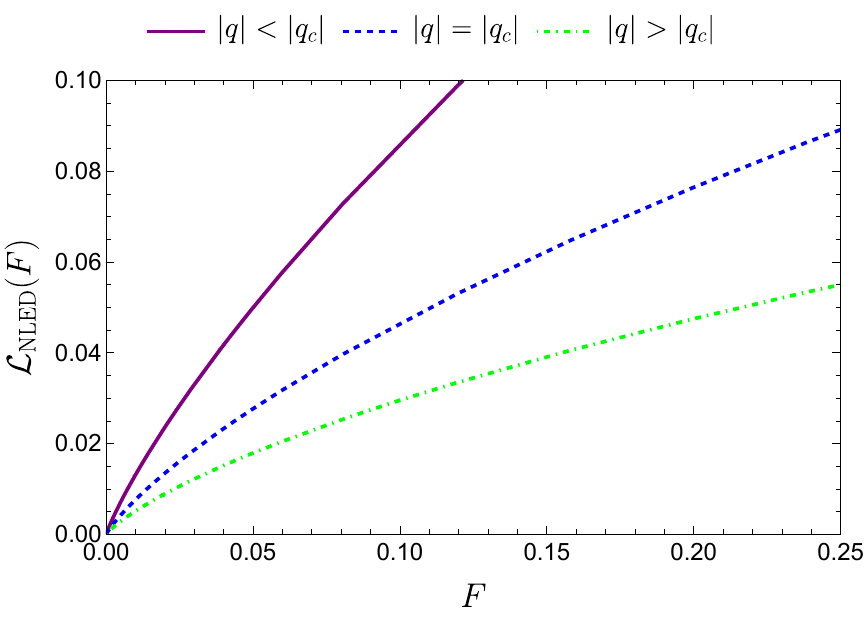}
\caption{The NLED Lagrangian ${\cal L}(F) $, given by Eq.~\eqref{L3_M2} for different values of the electric charge    $|q|<|q_c|,\,|q|=|q_c|,$ and $|q|>|q_c|$. Here we have taken $M=1$. } 
\label{fig_LxF3}
\end{figure}

In the opposite limit, $F \to 0$, we perform a series expansion of Eq.~\eqref{L3_M2} for small $F$, obtaining
\begin{equation}
	\mathcal{L}(F) \approx 
	\frac{12 M}{\kappa^2} \, F
	- \frac{48 \, 2^{1/4} M }{\kappa^2|q|^{1/2}} \, F^{5/4}
	+ \mathcal{O}\!\left(F^{3/2}\right).
\end{equation}
Notably, the leading term is linear in $F$, ensuring that the weak-field limit reproduces Maxwell’s theory.  
The behavior of Eq.~\eqref{L3_M2} for three representative values of the charge is illustrated in Fig.~\ref{fig_LxF3}.

\begin{figure}[t!]
\includegraphics[scale=0.55]{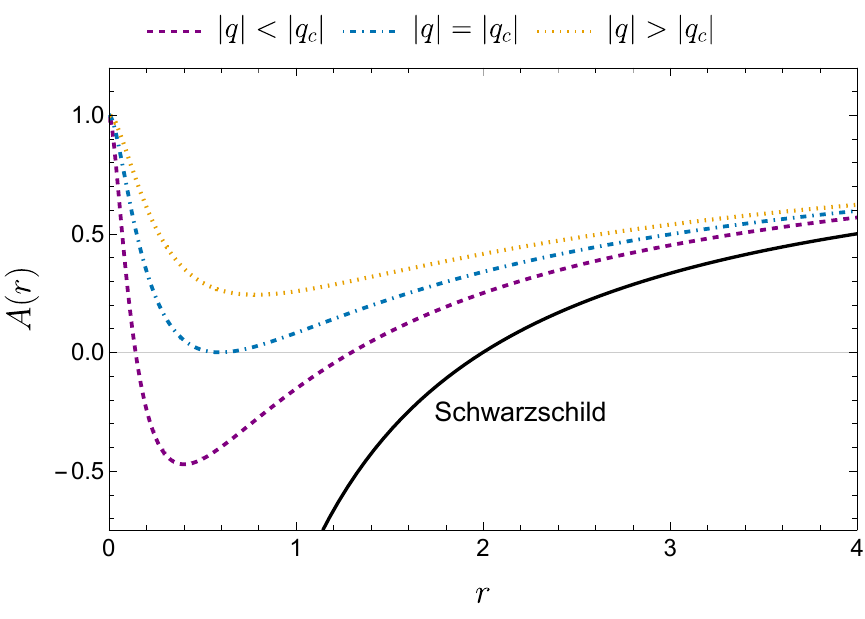}
\caption{The  metric function $A(r) $, given by Eq.~\eqref{Eq_A3} for three magnetic charge scenarios. Here we have taken $M=1$. } 
\label{fig_A2}
\end{figure}

We also solved Eqs. \eqref{rH} and \eqref{der_a} simultaneously for this model, determining the critical electric charge from the metric function \eqref{A3}. We obtained the value $|q_{c}| = 0.296$ (for $M= 1.0$). In Fig.~\ref{fig_A2}, we present the behavior of the metric function \eqref{A3} in relation to the radial coordinate $r$ for three distinct scenarios of the electric charge: $|q| < |q_{c}|$, $|q| = |q_{c}|$, and $|q| > |q_{c}|$. the behavior is similar to the one illustrated in Fig. \ref{fig_A}.

Furthermore, the quantities $\rho$ and $p$ for this model after using relation~\eqref{Eq_R} are now described as
\begin{align}
    &\rho(r)=\frac{6 M R}{(r+R)^4}=\frac{6 M |q|}{\left(|q|+r\right)^4},\label{rho2_Mod2}
    \\
    &
    P(r)=\frac{6 M R (r-R)}{(r+R)^5}=\frac{6 M |q| \left(r-|q|\right)}{\left(|q|+r\right)^5}.\label{P2_Mod2}
\end{align}
Expanding Eqs.~\eqref{rho2_Mod2} and~\eqref{P2_Mod2} near the center, we obtain
\begin{align}
 \rho(r)&\sim   \frac{6 M}{|q|^3}-\frac{24  M r}{q^4}+\mathcal{O}\left(r^2\right),
 \\
 P(r)&\sim \frac{36 M r}{q^4}-\frac{6 M}{|q|^3}+\mathcal{O}\left(r^2\right).
\end{align}
Whereas, for large values of $r$, the expansion for these components becomes
\begin{align}
 \rho(r)&\sim \frac{6 M |q|}{r^4}  -\frac{24 M q^2}{r^5}+\mathcal{O}\left(\frac{1}{r^6}\right),
 \\
 P(r)&\sim \frac{6 M |q|}{r^4}-\frac{36 M q^2}{r^5}+\mathcal{O}\left(\frac{1}{r^6}\right).
\end{align}

\begin{figure}[t!]
\includegraphics[scale=0.55]{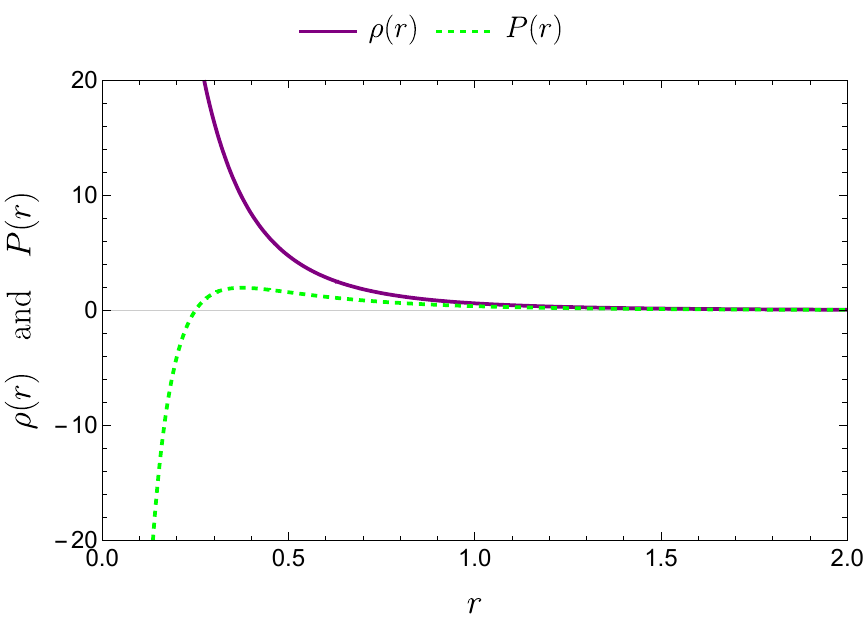}
\caption{The energy density $\rho(r)$ and pression $P(r) $, given by Eq.~\eqref{rho2_Mod2} and~\eqref{P2_Mod2}.  We have used the values of the constants as follows $M=1$ and $|q|=0.25$. } 
\label{fig_rhop2}
\end{figure}

We illustrate the behavior of these two quantities in terms of the radial coordinate $r$ in Fig.~\ref{fig_rhop2}. We use the values of the constants as: $M=1$ and $|q|=0.25$.

%%%%%%%%%%%%%%%%%%%%%%%%%%%%%%%%%%%%%%%%%%%%%%%%%%%%%%%%%%%%%%%%
\section{Model III}\label{Mod_III}
%%%%%%%%%%%%%%%%%%%%%%%%%%%%%%%%%%%%%%%%%%%%%%%%%%%%%%%%%%%%%%%%

%%%%%%%%%%%%%%%%%%%%%%%%%%%%%%%%%%%%%%%%%%%%%%%%%%%%%%%%%%%%%%%%
\subsection{Metric function}
%%%%%%%%%%%%%%%%%%%%%%%%%%%%%%%%%%%%%%%%%%%%%%%%%%%%%%%%%%%%%%%%

We finally explore a model with the following function
\begin{equation}
    \zeta(r)=\frac{\sqrt{r}-R}{\sqrt{r}+R}\, ,\label{zeta4}
\end{equation}
where for the limit of $r$ very small, we have $\zeta(0)=-1$, i.e. a de Sitter core. Therefore, for the model given by Eq.~\eqref{zeta4}, the mass function \eqref{func_M} now has the form 
\begin{eqnarray}
\mathcal{M}(r) & = & M-\frac{w_{0}}{21\left(\sqrt{r}+R\right)^{7}}\Big(35r^{3/2}R^{2}+21r^{5/2}
\nonumber \\
 && +35r^{2}R +7\sqrt{r}R^{4}+21rR^{3}+R^{5} \Big).
\end{eqnarray}
In the limit $r \to 0$, the mass function takes the form
\begin{equation}
	\mathcal{M}(0) = M - \frac{w_0}{21 R^2}.
	\label{M0}
\end{equation}
Based on this result, we impose the following relation to ensure that the space-time is everywhere regular:
\begin{equation}
	w_0 = 21 M R^2.
	\label{w4}
\end{equation}
Consequently, the mass function $\mathcal{M}(r)$ for this model can be written as
\begin{equation}
	\mathcal{M}(r) = \frac{M \left( r^{7/2} + 7 r^3 R \right)}{\left( \sqrt{r} + R \right)^7}.
	\label{func_M4}
\end{equation}
Using this mass function $\mathcal{M}(r)$, we verify that the space-time exhibits regular behavior both at the center and asymptotically. From Eq.~\eqref{func_M4}, the corresponding metric function can be obtained via Eq.~\eqref{A}, yielding
\begin{align}
A(r)=
1-\frac{2 M \left(r^{5/2}+7 r^2 R\right)}{\left(\sqrt{r}+R\right)^7}
.\label{A4}
\end{align}
In the limit of large $r$, the metric function \eqref{A4} approaches the Minkowski form, while for $r \to 0$, it tends to unity.

The energy density~\eqref{rho} and pressure~\eqref{P} are now given by
\begin{eqnarray}
    \rho(r) &=& \frac{2 w_0}{\left(\sqrt{r}+R\right)^8}=\frac{42 M R^2}{\left(\sqrt{r}+R\right)^8},\label{rho4}
    \\
    P(r) &=& \frac{2 w_0 \left(\sqrt{r}-R\right)}{\left(\sqrt{r}+R\right)^9}=\frac{42 M R^2 \left(\sqrt{r}-R\right)}{\left(\sqrt{r}+R\right)^9},\label{P4}
\end{eqnarray}
respectively.

%%%%%%%%%%%%%%%%%%%%%%%%%%%%%%%%%%%%%%%%%%%%%%%%%%%%%%%%%%%%%%%%
\subsection{Kretschmann  scalar}
%%%%%%%%%%%%%%%%%%%%%%%%%%%%%%%%%%%%%%%%%%%%%%%%%%%%%%%%%%%%%%%%

The Kretschmann scalar for this case is given by
\begin{align}
	K(r) = \frac{48 M^2}{\left( \sqrt{r} + R \right)^{18}} 
	\Big( & -108 r^{3/2} R^3 + 18 r^{5/2} R + r^3 \nonumber \\
	& + 69 r^2 R^2 + 330 r R^4 + 98 R^6 \Big),
	\label{K_4}
\end{align}
which is finite at the origin:
\begin{equation}
	\lim_{r \to 0} K(r) = \frac{4704 M^2}{R^{12}}.
\end{equation}

Moreover, in the asymptotic limit $r \to \infty$, the Kretschmann scalar vanishes, indicating that the space-time is asymptotically flat.

%%%%%%%%%%%%%%%%%%%%%%%%%%%%%%%%%%%%%%%%%%%%%%%%%%%%%%%%%%%%%%%%
\subsection{Matter Lagrangian}
%%%%%%%%%%%%%%%%%%%%%%%%%%%%%%%%%%%%%%%%%%%%%%%%%%%%%%%%%%%%%%%%

Using the metric function~\eqref{A4} in Eqs. \eqref{L_RBH} and \eqref{LF_RBH}, we can write the Lagrangian as
\begin{align}
 {\cal L}(r) &=\frac{2 w_0}{\kappa ^2 \left(\sqrt{r}+R\right)^8}
=\frac{42 M R^2}{\kappa ^2 \left(\sqrt{r}+R\right)^8}\,,
\label{L_4} 
\end{align}
and its derivative 
\begin{align}
{\cal L}_F(r) &=\frac{4 r^{9/2} w_0}{\kappa ^2 q^2 \left(\sqrt{r}+R\right)^9}= \frac{84 M r^{9/2} R^2}{\kappa ^2 q^2 \left(\sqrt{r}+R\right)^9}\,
.\label{LF_3}
\end{align}
As in previous models, we can write the analytical form of the Lagrangian~\eqref{L_4} in terms of the electromagnetic scalar
\begin{align}
 {\cal L}(F)& =\frac{512 w_0}{\kappa ^2 \left(2^{7/8} \sqrt{\frac{\sqrt{|q|}}{\sqrt[4]{F}}}+2 R\right)^8}
\nonumber
\\
&
=\frac{10752 \,M R^2}{\kappa ^2 \left(2^{7/8} \sqrt{\frac{\sqrt{|q|}}{\sqrt[4]{F}}}+2 R\right)^8}\,,
\label{L_F4} 
\end{align}
and its derivative is
\begin{align}
  {\cal L} _F(F)
&=	\frac{512\ 2^{7/8} w_0 \sqrt{\frac{\sqrt{|q|}}{\sqrt[4]{F}}}}{F \kappa ^2 \left(2^{7/8} \sqrt{\frac{\sqrt{|q|}}{\sqrt[4]{F}}}+2 R\right)^9}
   \nonumber
\\
&
=\frac{10752\ 2^{7/8} M R^2 \sqrt{\frac{\sqrt{|q|}}{\sqrt[4]{F}}}}{ \kappa ^2 F \left(2^{7/8} \sqrt{\frac{\sqrt{|q|}}{\sqrt[4]{F}}}+2 R\right)^9}\, .
 \label{LF_F4} 
\end{align} 
In this model, we also note that Eqs.~\eqref{L_F4} and \eqref{LF_F4} satisfy Eq.~\eqref{RC} identically.

%%%%%%%%%%%%%%%%%%%%%%%%%%%%%%%%%%%%%%%%%%%%%%%%%%%%%%%%%%%%%%%%
\subsection{Metric function and matter in terms of $q$ }
%%%%%%%%%%%%%%%%%%%%%%%%%%%%%%%%%%%%%%%%%%%%%%%%%%%%%%%%%%%%%%%%

Now, in order to ensure dimensional consistency with the units adopted throughout this work (and consequently with the dimension of all physical functions), we will adopt the following interpretation for this model:
\begin{equation}
    R=\sqrt{|q|} .\label{Eq_R2}
\end{equation}

\begin{figure}[t!]
\includegraphics[scale=0.55]{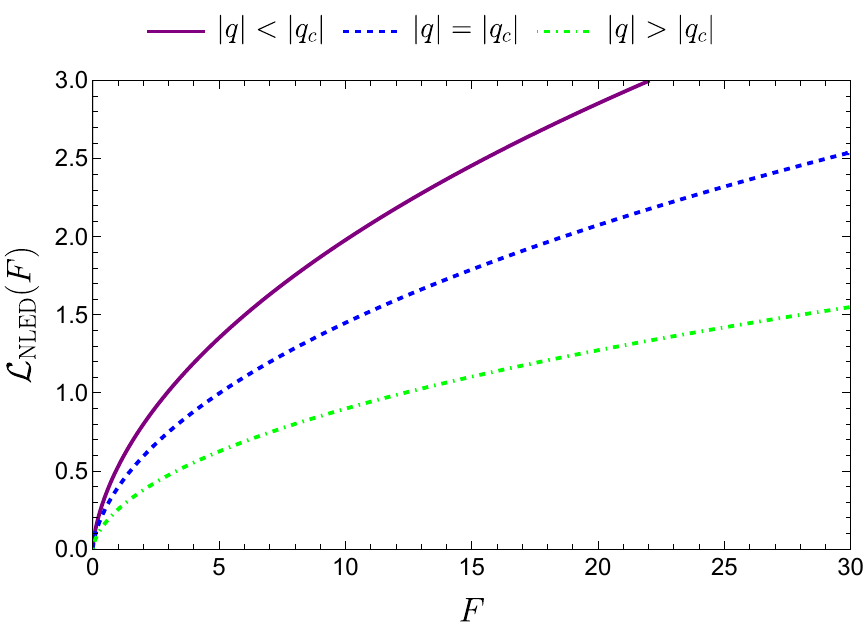}
\caption{The NLED Lagrangian ${\cal L}(F) $, given by Eq.~\eqref{L_M3} for different values of the electric charge    $\{|q|<|q_{c}|,\,|q|=|q_{c}|,\,|q|>|q_{c}|\}$.  Here we have taken $M=1$. } 
\label{fig_LxF4}
\end{figure}

Using the interpretation \eqref{Eq_R2}, the most relevant quantities of this model are described as
\begin{align}
    &A(r)=1-\frac{2 M \left(7 \sqrt{|q|} r^2+r^{5/2}\right)}{\left(\sqrt{|q|}+\sqrt{r}\right)^7}\,,\label{A41}
\\
&
\rho(r)=\frac{42 M |q|}{\left(\sqrt{|q|}+\sqrt{r}\right)^8}\label{rho2_Mod3}
,
\\
&
P(r)=\frac{42 M |q| \left(\sqrt{r}-\sqrt{|q|}\right)}{\left(\sqrt{|q|}+\sqrt{r}\right)^9}\label{P2_Mod3}
,
\\
&
{\cal L}(F) =\frac{10752 M |q|}{\kappa ^2 \left(2^{7/8} \sqrt{\frac{\sqrt{|q|}}{\sqrt[4]{F}}}+2 \sqrt{|q|}\right)^8}\,.\label{L_M3}
\end{align}
We observe that the Lagrangian~\eqref{L_M3} goes to zero when taking the limit for small values of $F$. For large values of $F$ we obtain
\begin{align}
 \lim_{F\to \infty}  {\cal L}(F) =\frac{42 M}{\kappa ^2 R^6}\,.
\end{align}
We also compute the series expansion of Eq.~\eqref{L_M3} for small values of $F$, which yields
\begin{equation}
   \mathcal{L}(F) \approx 
\frac{84M}{\kappa^{2}|q|}F+{\cal O}\left(F^{9/8}\right)\,.
\end{equation}
In this model the Maxwell limit is also recovered, since the dominant term is linear in $F$.
The behavior of Eq.~\eqref{L_M3} for three different values of the charge is shown in Fig.~\ref{fig_LxF4}.

The limit of the metric function \eqref{A41} for small values of $|q|$ recovers the uncharged solution, whereas for large values of $|q|$ it is asymptotically Minkowskian. 

Now, if we expand the fluid components~\eqref{rho2_Mod3} and~\eqref{P2_Mod3} for values of $r\ll 1$, we obtain
\begin{align}
 \rho(r)&\sim  \frac{1512 M r}{q^4}-\frac{336 M \sqrt{r}}{|q|^{7/2}}+\frac{42 M}{|q|^{3}}+\mathcal{O}\left(r^{3/2}\right),
 \\
 P(r)&\sim -\frac{2268 M r}{q^4}+\frac{420 M \sqrt{r}}{|q|^{7/2}}-\frac{42 M}{|q|^{3}}+\mathcal{O}\left(r^{11/2}\right).
\end{align}
while for values of $r\gg1$, both $\rho(r)$ and $P(r)$ tend to 
\begin{align}
 \rho(r)\sim& \frac{1512Mq^2}{r^{5}}-\frac{336M|q|^{3/2}}{r^{9/2}}
 \nonumber\\
 &
+\frac{42Mq}{r^{4}} +\mathcal{O}\left(\frac{1}{r^{3/2}}\right),
 \\
 P(r)\sim &\frac{2268Mq^2}{r^{5}}-\frac{420M|q|^{3/2}}{r^{9/2}}
 \nonumber\\
 &
 +\frac{42M|q|}{r^{4}}+\mathcal{O}\left(\frac{1}{r^{11/2}}\right).
\end{align}

\begin{figure}[t!]
\includegraphics[scale=0.55]{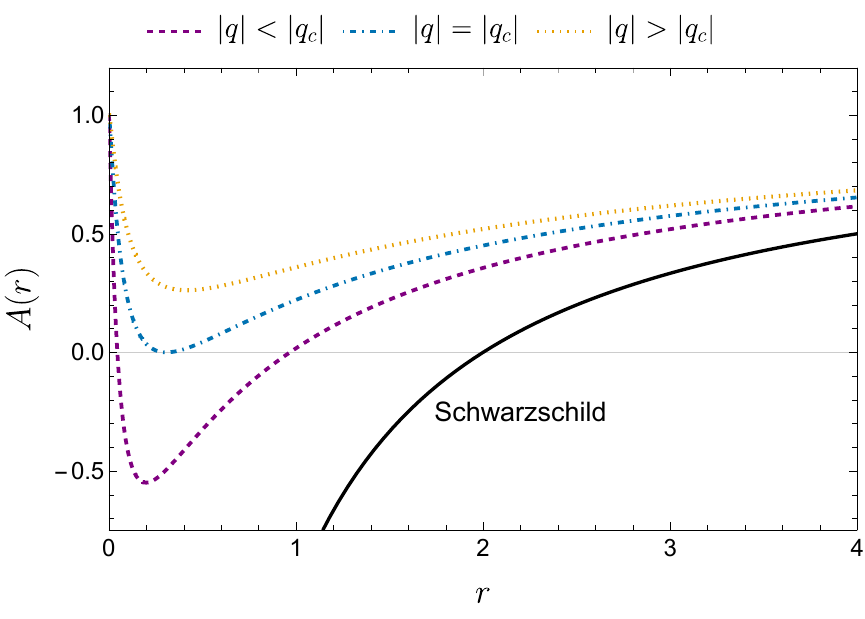}
\caption{The  metric function $A(r) $, given by Eq.~\eqref{A41} for three magnetic charge scenarios. Here we have taken $M=1$. } 
\label{fig_A3}
\end{figure}

Analogously to the previous models, we solved Eqs.~\eqref{rH} and~\eqref{der_a} simultaneously for the model described by the metric function~\eqref{A41} in order to determine $q_c$. In this case, we obtained the value $|q_{c} |= 0.141$ (for $M = 1$). The behavior of the metric function~\eqref{A41} as a function of $r$ is illustrated in Fig.~\ref{fig_A3}, for three distinct scenarios of the electric charge: $|q| < |q_{c}|$, $|q| = |q_{c}|$, and $|q| >|q_{c}|$. The behavior of the metric function in this model is also similar to that shown in Fig.~\ref{fig_A}.

\begin{figure}[t!]
\includegraphics[scale=0.55]{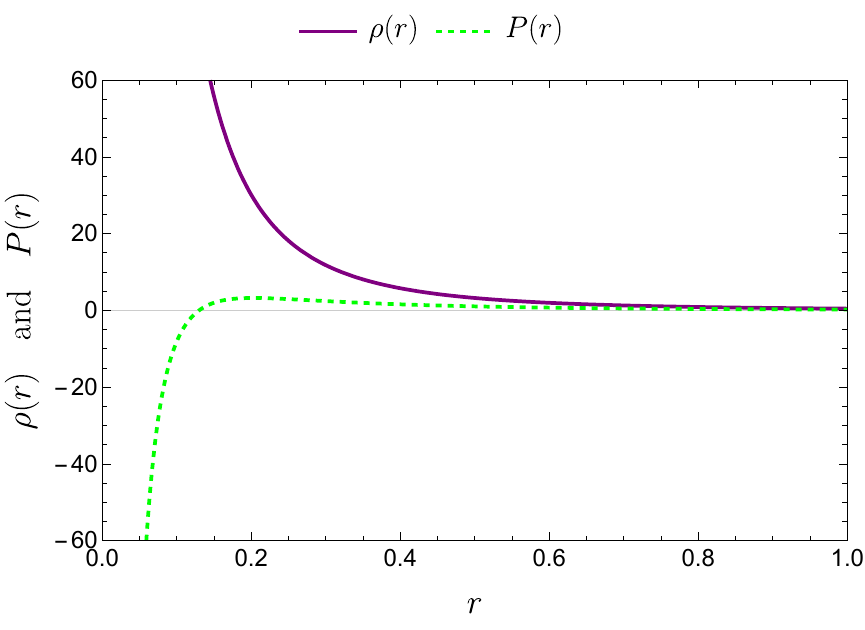}
\caption{The energy density $\rho(r)$ and pressure $P(r) $, given by Eq.~\eqref{rho2_Mod3} and~\eqref{P2_Mod3}.  We have used the values of the constants as follows $M=1$ and $|q|=0.13$. } 
\label{fig_rhop3}
\end{figure}

Finally, we illustrate in Fig.~\ref{fig_rhop3} the behavior of energy density~\eqref{rho2_Mod3} and pressure~\eqref{P2_Mod3}.

%%%%%%%%%%%%%%%%%%%%%%%%%%%%%%%%%%%%%%%%%%%%%%%%%%%%%%%%%%%%%%%%
\section{Energy conditions}\label{sec:EC}
%%%%%%%%%%%%%%%%%%%%%%%%%%%%%%%%%%%%%%%%%%%%%%%%%%%%%%%%%%%%%%%%  
To conclude the analysis of the matter sources supporting the models above, we discuss the corresponding energy conditions. For this purpose, we adopt the standard approach of modeling the matter content as an anisotropic fluid in the various black hole regions.  

Outside the horizon, where $A(r) > 0$, the energy-momentum tensor takes the form
\begin{equation}
	T_{\phantom{\mu}\nu}^{\mu} = \mathrm{diag}\left(\rho,\, -p_r,\, -p_t,\, -p_t \right),
	\label{Comp_TA}
\end{equation}
whereas inside the horizon, i.e., for $A(r) < 0$, it is instead given by
\begin{equation}
	T_{\phantom{\mu}\nu}^{\mu} = \mathrm{diag}\left(-p_r,\, \rho,\, -p_t,\, -p_t \right).
	\label{Comp_TA2}
\end{equation}

The energy conditions associated with the energy-momentum tensor~\eqref{Comp_TA} are given by:
\begin{align}
{\rm NEC}_{r,t}&=\rho+p_{r,t}\geq 0\,,\label{NEC}\\
{\rm SEC}_{(rt)}&=\rho + p_{r} + 2p_t\geq 0\,,\label{SEC}\\
{\rm DEC}_{r,t}&=\rho \, - \mid p_{r,t}\mid \geq 0\, \quad\text{or}\quad \rho\pm p_{r,t}\geq 0\,,\label{DEC}\\
{\rm WEC}_{}&=\rho\geq 0\,,\label{WEC}
\end{align}
where NEC, SEC, DEC and WEC, denote the null, strong,  dominant and weak energy conditions, respectively. Note that in these expressions the indices $r$ and $t$ refer to the radial and tangential components of the anisotropic fluid, respectively. 

Due to the symmetry imposed on the models, recalling that $p_r = -\rho$ and $p_t = P$, the energy conditions outside the horizon take the same form as those inside. Consequently, the energy conditions given by Eqs.~\eqref{NEC}–\eqref{WEC} take the simplified form of
\begin{align}
{\rm NEC}_{r} & =  0 ,\\
{\rm NEC}_{t} & = \rho + P \ge 0,\\
{\rm SEC} & = 2P\ge 0,\\
{\rm DEC}_{r} & = \rho - |p_r| \geq  0,\\
{\rm DEC}_{t} & = \rho - |P|\geq 0,\\
{\rm WEC} & = \rho \ge 0.
\end{align}
In view of this, it is sufficient to compute the energy conditions in only one of these regions. For the Model II, as defined in Sec.~\ref{Mod_II}, these conditions read as
\begin{align}
    &{\rm NEC}_r={\rm SEC}_r={\rm WEC}_r
=0; \label{NEC1F_Mod1}
    \\
    &
    {\rm NEC}_t={\rm SEC}_t={\rm WEC}_t=\frac{4 r w_0}{(r+R)^5}
    ;\label{NEC2F_Mod1}
    \\
    &
    {\rm SEC}_{(rt)}
    =
\frac{4 w_0 (r-R)}{(r+R)^5};\label{SEC3F_Mod1}
    \\
    &
    {\rm DEC}_r=    0;\label{DEC1F_Mod1}
    \\
    &
    {\rm DEC}_t=    0
    ;\label{DEC2F_Mod1}
    \\
    &
    {\rm DEC}={\rm WEC}
    =\frac{2 w_0}{(r+R)^4}
    .\label{DEC3F_Mod1}
\end{align}
while for Model III, as presented in Sec.~\ref{Mod_III}, they read as:
\begin{align}
    &{\rm NEC}_r={\rm SEC}_r={\rm WEC}_r
=0; \label{NEC1F_Mod3}
    \\
    &
    {\rm NEC}_t={\rm SEC}_t={\rm WEC}_t=\frac{4 \sqrt{r} w_0}{\left(\sqrt{r}+R\right)^9}
    ;\label{NEC2F_Mod3}
    \\
    &
    {\rm SEC}_{(rt)}
    =
\frac{4 w_0 \left(\sqrt{r}-R\right)}{\left(\sqrt{r}+R\right)^9};\label{SEC3F_Mod3}
    \\
    &
    {\rm DEC}_r=    0;\label{DEC1F_Mod3}
    \\
    &
    {\rm DEC}_t=    0
    ;\label{DEC2F_Mod3}
    \\
    &
    {\rm DEC}={\rm WEC}
    =\frac{2 w_0}{\left(\sqrt{r}+R\right)^8}
    .\label{DEC3F_Mod3}
\end{align}

Our models are consistent and exhibit qualitatively similar results to those reported in Ref.~\cite{Vertogradov:2024seh}. More specifically, for Model II, the WEC and NEC  are satisfied throughout the space-time for $w_0 \ge 0$, while the DEC is globally preserved and satisfied, identically at the center. On the other hand, the SEC is violated only within the central region $r \le R$ (or, equivalently, $r \le |q|$), and is satisfied again for $r \ge R$ ($r \ge |q|$).
Model III presents analogous behavior, the NEC, WEC, and DEC are satisfied throughout space-time, whereas the SEC is violated only in the innermost region $r \le R^2$. For $r \ge R^2$, the SEC becomes satisfied and is no longer violated.

%%%%%%%%%%%%%%%%%%%%%%%%%%%%%%%%%%%%%%%%%%%%%%%%%%%%%%%%%%%%%%%%
\section{Constraints on $q$ from shadow radius}\label{rsh}
%%%%%%%%%%%%%%%%%%%%%%%%%%%%%%%%%%%%%%%%%%%%%%%%%%%%%%%%%%%%%%%% 

%%%%%%%%%%%%%%%%%%%%%%%%%%%%%%%%%%%%%%%%%%%%%%%%%%%%%%%%%%%%%%%%
\subsection{Shadow radius}
%%%%%%%%%%%%%%%%%%%%%%%%%%%%%%%%%%%%%%%%%%%%%%%%%%%%%%%%%%%%%%%%

The images from Sgr A$^*$ obtained by the Event Horizon Telescope (EHT) have not only confirmed the predictions of GR but have also provided valuable insights about the region near black hole horizons. These observations allow us to probe extreme environments where gravitational physics enables tests of strong-field gravity, for instance, through the analysis of the central dark region in the images. In the canonical interpretation of Falcke \cite{Falcke:1999pj}, such a region is identified with the critical curve, that is, the region projected onto the image plane of a distant observer whose boundary is delineated by photons originating from a background source that, due to the intense gravitational field, are captured in the immediate vicinity of the black hole. 

The EHT Collaboration claims the existence of a correlation between the size of this dark central region (dubbed for these purposes as the black hole shadow) and the size of the observable bright region in the image, allowing to constrain the former \cite{EventHorizonTelescope:2022wkp}. Specifically, using the data reported in Table \ref{tab:my-table}, the EHT collaboration constraint the shadow's size to
\begin{equation} \label{eq:1sigma}
    4.55 \lesssim r_{sh}/M \lesssim 5.22 \ ,
\end{equation}
at $1\sigma$ deviation, while
\begin{equation} \label{eq:2sigma}
    4.21 \lesssim r_{sh}/M \lesssim 5.56 \ ,
\end{equation}
at $2\sigma$ deviation. 

Using this method (previously employed in~\cite{Vagnozzi:2022moj} for a large number of spherically symmetric black hole geometries), these observational values, which already incorporate theoretical uncertainties, can be used to assess the compatibility of our model with the current observational constraints. To this end, below we calculate and analyze the shadow radius of our models in order to constrain the charge $q$.

%%%%%%%%%%%%%%%%%%%%%%%%%%%%%%%%%%%%%%%%%%%%%%%%%%%%%%%%%%%%%%%%
\subsection{Effective metric}
%%%%%%%%%%%%%%%%%%%%%%%%%%%%%%%%%%%%%%%%%%%%%%%%%%%%%%%%%%%%%%%%

Since we are developing solutions with a matter source described by a NLED coupling, the trajectories of particles are governed by an effective metric, whose explicit form is given by~\cite{Novello:1999pg}
\begin{equation}
     g_{\rm eff}^{\mu\nu}={\cal L}_{F}g^{\mu\nu}-{\cal L}_{FF}F_{\alpha}^{\phantom{\alpha}\mu}F^{\alpha\nu}\,.\label{g_efet}
 \end{equation}
The effective metric~\eqref{g_efet} may take two distinct forms, depending on whether the NLED structure involves only an electric or a magnetic charge. In the magnetic case, which is the scenario considered in this manuscript, the effective metric takes the particular form 
\begin{align}
    ds^{2}=\bar{A}(r)dt^{2}-\bar{B}(r)dr^{2}-\bar{C}(r)d\Omega^2 \label{ds_mag}
\end{align}
where the explicit forms of the metric functions read as \cite{Toshmatov:2021fgm}
\begin{subequations}\label{met_efet_mag}
    \begin{eqnarray}
  \bar{A}(r) &=& \frac{A(r)}{{\cal L}_{F}},\label{A_bar}
\\
    \bar{B}(r)&=& \frac{1}{A(r){\cal L}_{F}},\label{B_bar}
    \\
    \bar{C}(r)&=& \frac{r^2}{{\cal L}_{F}+2F{\cal L}_{FF}}.\label{C_bar}
    \end{eqnarray}
\end{subequations}
Since the NLED acts as the matter source, it is essential to adopt the effective metric~\eqref{g_efet} to determine the shadow radius, specifically in the form given by Eq.~\eqref{met_efet_mag}, as we are dealing with magnetic charge in our solutions. Taking this into account, we incorporate the effective metric into the space-time structure in the equations governing the black hole shadow radius, in order to identify the influence of the magnetic charge on the geometry. 

To construct the shadow's radius, one first needs to find the expression of the photon sphere, namely, the surface of unstable geodesics, in the effective metric, and whose projection in the observer's plane image is the critical curve. The photon sphere radius $r_{ps}$ satisfies the equation
\begin{equation}
\bar{A}(r_{ps})\,\bar{C}'(r_{ps})=\bar{A}' (r_{ps})\,\bar{C}(r_{ps}).\label{rph}
\end{equation}
and thus we can express the shadow radius as
\begin{equation}
r_{sh}=\bar{r}_{ps}\sqrt{\frac{\bar{A}(r_{0})}{\bar{A}(\bar{r}_{ps})}}.\label{r_shadow2}
\end{equation}

We now employ the data from Table~\ref{tab:my-table} to verify whether the shadow radius predicted by the two new models introduced in this work is consistent with the shadow observed by the EHT for Sgr~A*, after constraining the parameter $q$ according to Eq.~\eqref{r_shadow2}. In this sense, below we present the results obtained by numerically computing the shadow radius for such models. In this calculation, we use the observer distance $r_0$ for the evaluation of $\tilde{A}(r_0)$, corresponding to the mean value of the measurements from the Keck and VLTI collaborations, as given in Table~\ref{tab:my-table}. Furthermore, for simplicity, we fixed the mass to $M = 1$, so that all other quantities are expressed in units of this mass.

%%%%%%%%%%%%%%%%%%%%%%%%%%%%%%%%%%%%%%%%%%%%%
\begin{table}[t!]
\centering
\resizebox{\linewidth}{!}{%
\def\arraystretch{1.4}
\begin{tabular}{cccc}
\hline \hline
\multicolumn{4}{c}{Parameter values}                                                                               \\ \hline
\hspace{1 mm}Survey\hspace{1 mm} & \hspace{1 mm}$M (\times 10^6 M_{\odot})$\hspace{1 mm} & \hspace{1 mm}$D$\hspace{1 mm} (kpc) & \hspace{1 mm}Reference\hspace{1 mm} \\ 
Keck                             & $3.951 \pm 0.047$                                     & $7.953 \pm 0.050 \pm 0.032$         & {\cite{Do:2019txf}} \\ 
VLTI                             & $4.297 \pm 0.012 \pm 0.040$                           & $8.277 \pm 0.009 \pm 0.033$         & {\cite{GRAVITY:2020gka}} \\ \hline
\end{tabular}%
}
\caption{Sgr A* mass and distance.}
\label{tab:my-table}
\end{table}

%%%%%%%%%%%%%%%%%%%%%%%%%%%%%%%%%%%%%%%%%%%%%%%%%%%%%%%%%%%%%%%%
\subsubsection{Shadow radius for model II}
%%%%%%%%%%%%%%%%%%%%%%%%%%%%%%%%%%%%%%%%%%%%%%%%%%%%%%%%%%%%%%%%

To express the shadow of the model presented in Sec.~\ref{Mod_II}, some considerations are required. In particular, we rewrite Eqs.~\eqref{LF3_M2} and~\eqref{LFF3_M2} in terms of the radial coordinate and according to the interpretation given by Eq.~\eqref{Eq_R}. In this way, we obtain
\begin{align}
  {\cal L}_{F}(r)  =-\frac{12 M r^5}{\kappa ^2 |q|\left(|q|+r\right)^5},\label{LF4_M2}
  \\
  {\cal L}_{FF}(r)  =-\frac{30 M r^9}{\kappa ^2 q^2\left(|q|+r\right)^6}.\label{LFF4_M2}
\end{align}

Thus, after substituting the metric function~\eqref{Eq_A3} together with Eq.~\eqref{LF4_M2} into the effective metric~\eqref{A_bar}, we obtain
\begin{align}
	\bar{A}(r) =& -\frac{\kappa^{2}|q|(|q|+r)}{12Mr^{6}}\Big[r^{3}(5|q|-2M)+5q^4
    \nonumber\\&
    +3|q|r^{2}(5|q|-2M)+15|q|^{3}r\Big].
	\label{A2_ef}
\end{align}
and, by considering the electromagnetic quantities~\eqref{F2}, \eqref{LF4_M2}, and~\eqref{LFF4_M2}, the metric function~\eqref{C_bar} now takes the form
\begin{align}
 \bar{C}(r)=& -\frac{\kappa ^2 |q| (|q|+r)^6}{6 M r^3 (3 |q|-2 r)}\, .\label{C2_ef}
\end{align}
With the expressions of the new metric functions given by Eqs.~\eqref{A2_ef} and~\eqref{C2_ef}, corrected for the corresponding effective metric, the shadow radius can only be determined numerically, due to the complexity of analytically solving the photon sphere from Eq.~\eqref{rph}.

\begin{figure}[t!]
   \includegraphics[scale=0.43]{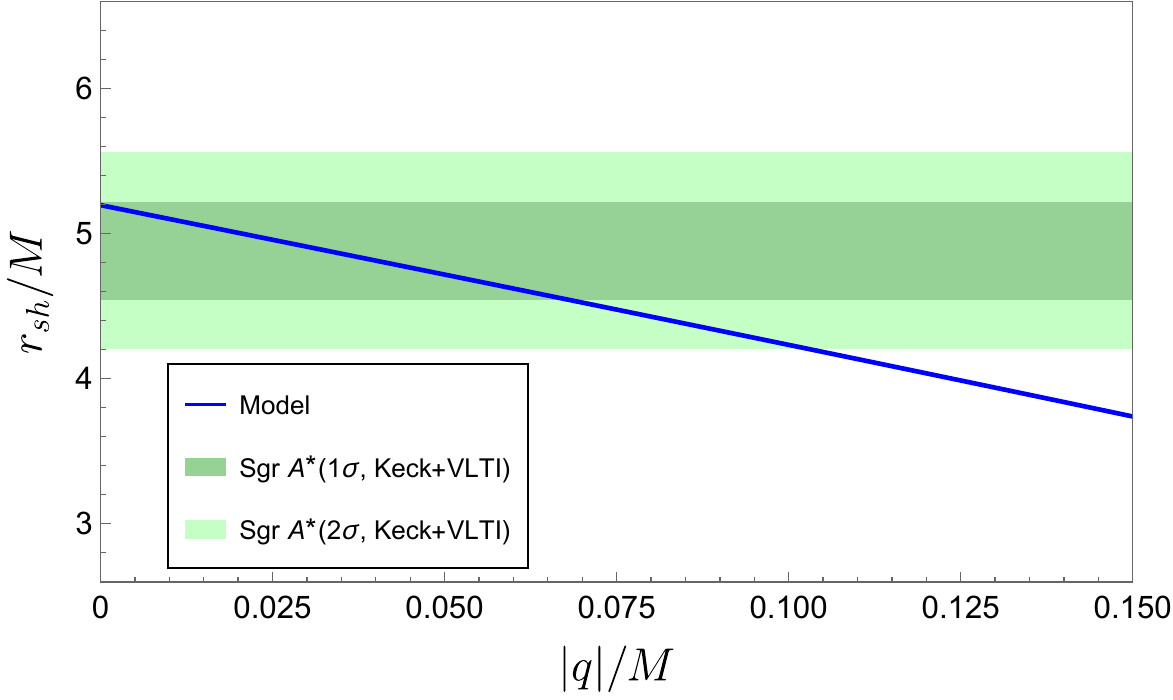}
   %\hfill{}%
   %\includegraphics[scale=0.75]{rsh2p2.pdf}
    \caption{The shadow radius $r_{sh}$ for our black hole model (solid blue curve), described by the metric function given in Eq.~\eqref{A2_ef}, calculated according to Eq.~\eqref{r_shadow2}. The shaded regions in blue and light blue represent the EHT constraints for the shadow radius of Sgr A* at confidence levels of $1\sigma$ and $2\sigma$, respectively. The white regions, beyond the $2\sigma$ values, are excluded by the EHT observations.  }
    \label{figshad}
\end{figure}

In Fig.~\ref{figshad}, we analyze the behavior of the shadow radius $r_{sh}$ of our model for the metric functions~\eqref{A2_ef} and~\eqref{C2_ef}, and we compute the shadow radius numerically from Eq.~\eqref{r_shadow2} (using $M=1$) as a function of the parameter $q/M$ and follow the uncertainties given in Eqs.~\eqref{eq:1sigma} and~\eqref{eq:2sigma}. We observe that for values of $|q|/M \approx  0.102359$, the shadow radius overcomes the bound  (\ref{eq:2sigma}). 
 %\end{figure*}

%%%%%%%%%%%%%%%%%%%%%%%%%%%%%%%%%%%%%%%%%%%%%%%%%%%%%%%%%%%%%%%%
\subsubsection{Shadow radius for model III}
%%%%%%%%%%%%%%%%%%%%%%%%%%%%%%%%%%%%%%%%%%%%%%%%%%%%%%%%%%%%%%%%

For the second model, as presented in Sec.~\ref{Mod_III}, we also compute the shadow radius following the procedure developed above. To obtain the effective geometry, we calculate the following electromagnetic quantities:
\begin{eqnarray}
  {\cal L}_{F}(r)  &=& \frac{84 M r^{9/2}}{\kappa ^2 |q| \left(\sqrt{|q|}+\sqrt{r}\right)^9},\label{LF4_M3}
  \\
  {\cal L}_{FF}(r)  &=& -\frac{189 M r^{17/2}}{\kappa ^2 |q|^{5/2} \left(\sqrt{|q|}+\sqrt{r}\right)^{10}}.\label{LFF4_M3}
\end{eqnarray}
By substituting the metric function~\eqref{Eq_A3} and the derivative of the Lagrangian~\eqref{LF4_M3} into the effective metric~\eqref{A_bar}, we obtain
\begin{align}
	\bar{A}(r) = &\left[1-\frac{2M\left(7\sqrt{|q|}r^{2}+r^{5/2}\right)}{\left(\sqrt{|q|}+\sqrt{r}\right)^{7}}\right]
    \nonumber\\&
    \times \frac{\kappa^{2}|q|\left(\sqrt{|q|}+\sqrt{r}\right)^{9}}{84Mr^{9/2}},
	\label{A3_ef}
\end{align}
whereas by considering the electromagnetic quantities~\eqref{F2}, \eqref{LF4_M3}, and~\eqref{LFF4_M3} in the metric function~\eqref{C_bar}, the metric function takes the form
\begin{align}
 \bar{C}(r)=& \frac{\kappa ^2 |q| \left(\sqrt{|q|}+\sqrt{r}\right)^{10}}{84 M r^3-105 M \sqrt{|q|} r^{5/2}}\, .\label{C3_ef}
\end{align}

\begin{figure}[t!]
   \includegraphics[scale=0.43]{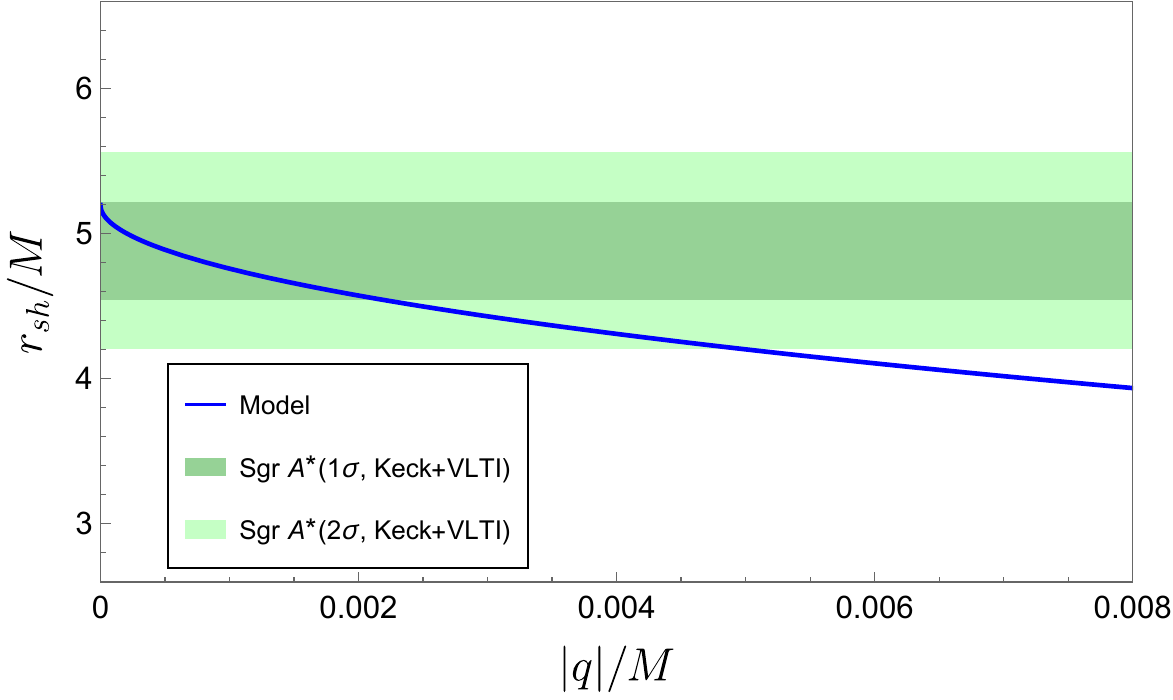}
   %\hfill{}%
   %\includegraphics[scale=0.75]{rsh2p2.pdf}
    \caption{The shadow radius $r_{sh}$ for our black hole model (solid blue curve), described by the metric function given in Eq.\eqref{A}, calculated according to Eq.~\eqref{r_shadow2}. The shaded regions in blue and light blue represent the EHT constraints for the shadow radius of Sgr A* at confidence levels of $1\sigma$ and $2\sigma$, respectively. The white regions, beyond the $2\sigma$ values, are excluded by the EHT observations.  }
    \label{figshad2}
\end{figure} %\end{figure*}

Analogously to the previous model, but now using the expressions corresponding to the effective metric given by Eqs.~\eqref{A3_ef} and~\eqref{C3_ef}, we determine the shadow radius of this model numerically and compare it with the shadow of Sgr.~A* as a function of $|q|$ for $M=1$. We illustrate the behavior of $r_{sh}$ in Fig.~\ref{figshad2}. Similarly to the previous model, the shadow radius in this case also decreases as the magnetic charge increases and for values of $|q| \gtrsim 0.00492225$, the shadow radius of our model  overcomes the bound \eqref{eq:2sigma}.

%%%%%%%%%%%%%%%%%%%%%%%%%%%%%%%%%%%%%%%%%%%%%%%%%%%%%%%%%%%%%%%%
\section{The scalar effective potential} \label{sec:SEP}
%%%%%%%%%%%%%%%%%%%%%%%%%%%%%%%%%%%%%%%%%%%%%%%%%%%%%%%%%%%%%%%%

%%%%%%%%%%%%%%%%%%%%%%%%%%%%%%%%%%%%%%%%%%%%%%%%%%%%%%%%%%%%%%%%
\subsection{General framework}
%%%%%%%%%%%%%%%%%%%%%%%%%%%%%%%%%%%%%%%%%%%%%%%%%%%%%%%%%%%%%%%%

We now turn to the dynamical response of the space-time by analyzing its characteristic resonance spectrum. This study is performed within the framework of linearized perturbations about a fixed background, which yield a discrete set of complex eigenfrequencies. These frequencies encode both the oscillatory behavior and the decay rate of perturbations induced by small excitations, and they serve as diagnostic tools to monitor how variations in the underlying parameters influence the system’s relaxation toward a stationary state.  

In the following, we focus on scalar field perturbations, which provide a clean and widely used probe of the geometry. To set the stage for the subsequent derivations, we first introduce the most general spherically symmetric line element, from which the perturbation equations and their associated boundary conditions will be formulated, gievn by 
\begin{align}
\mathrm{d}s^{2} &= -A(r)\,\mathrm{d}t^{2} + B(r)\,\mathrm{d}r^{2} + C(r)\mathrm{d}\Omega^2.
\end{align}

As is well known, the dynamics of a massless scalar field $\Phi$ are governed by the Klein-Gordon equation
\begin{align}
& \square \Phi
= \frac{1}{\sqrt{-g}}\partial_\mu\!\left(\sqrt{-g}\,g^{\mu\nu}\partial_\nu \Phi\right)=0,\\
& \sqrt{-g}=\sqrt{A(r)B(r)}\,C(r)\sin\theta .
\end{align}

In order to separate the angular, temporal, and radial sectors of the scalar dynamics and to make explicit the potential that controls the perturbative behavior, the scalar field is rewritten through 
\begin{align}
\Phi(t,r,\theta,\varphi)
= e^{-i\omega t}\,Y_{l m}(\theta,\varphi)\,\mathcal{F}_{l\omega}(r),
\end{align}
with the angular dependence encoded in spherical harmonics, which are required to obey the corresponding eigenvalue relations on the two--sphere
\begin{align}
\Delta_{S^{2}}Y_{l m}=-l(l+1)Y_{l m}.
\end{align}
As a consequence of this decomposition, the radial component is found to satisfy its own differential equation
\begin{align}
\frac{1}{\sqrt{AB}\,C}\frac{\mathrm{d}}{\mathrm{d}r}
\!\left(\sqrt{\frac{A}{B}}\,C\,\frac{\mathrm{d}\mathcal{F}}{dr}\right)
+\left(\frac{\omega^{2}}{A}-\frac{l(l+1)}{C}\right)\mathcal{F}=0 .
\end{align}
At this stage, the analysis is simplified by defining a new radial variable—the tortoise coordinate—given by
\begin{align}
\frac{\mathrm{d}r^*}{\mathrm{d}r}=\sqrt{\frac{B(r)}{A(r)}}.
\end{align}
In addition, a convenient transformation is implemented by absorbing the metric function into the radial amplitude, namely by introducing the new variable
$ \psi_{l \omega}(r)=\sqrt{C(r)}\,\mathcal{F}_{l\omega}(r)$.
With this redefinition, the original radial equation is recast into a simpler and more transparent form
\begin{align}
\frac{\mathrm{d}^{2}\psi}{\mathrm{d}r^{*2}}
+\Big(\omega^{2}-V_{s}(r)\Big)\psi=0 ,
\end{align}
so that all geometric and dynamical information is encapsulated in an effective potential, given by
\begin{align}
V_{s}(r)
= A(r)\,\frac{l(l+1)}{C(r)}
+ \frac{1}{\sqrt{C(r)}}\frac{\mathrm{d}^{2}}{\mathrm{d}r^{*2}}
\!\Big(\sqrt{C(r)}\Big).
\end{align}
When expressed explicitly in terms of derivatives with respect to the radial coordinate (denoted by primes, $' = \mathrm{d}/\mathrm{d}r$), the effective potential takes the form
\begin{eqnarray}
	V_s(r) &=& A \frac{l(l+1)}{C} 
	+ \frac{A}{B} \Bigg[
	\frac{1}{2} \frac{C''}{C} 
	- \frac{1}{4} \left( \frac{C'}{C} \right)^2 
		\nonumber \\
	&& \qquad + \frac{1}{4} \left( \frac{A'}{A} - \frac{B'}{B} \right) \frac{C'}{C}
	\Bigg],
	\label{potentialeff}
\end{eqnarray}
where $l$ is the multipole number.

In the following, we carry out an analysis of the quasinormal spectra for all configurations investigated in this work. The study is based on the explicit construction of the corresponding effective potentials for the three models considered. To compute the associated frequencies, we employ the sixth-order WKB approximation. To further corroborate our results, later we shall also perform an independent analysis in the time domain for the same set of configurations.

%%%%%%%%%%%%%%%%%%%%%%%%%%%%%%%%%%%%%%%%%%%%%%%%%%%%%%%%%%%%%%%%
\subsection{Application to Models I, II, and III}
%%%%%%%%%%%%%%%%%%%%%%%%%%%%%%%%%%%%%%%%%%%%%%%%%%%%%%%%%%%%%%%%

\begin{figure}[t!]
    \centering
    \includegraphics[scale=0.53]{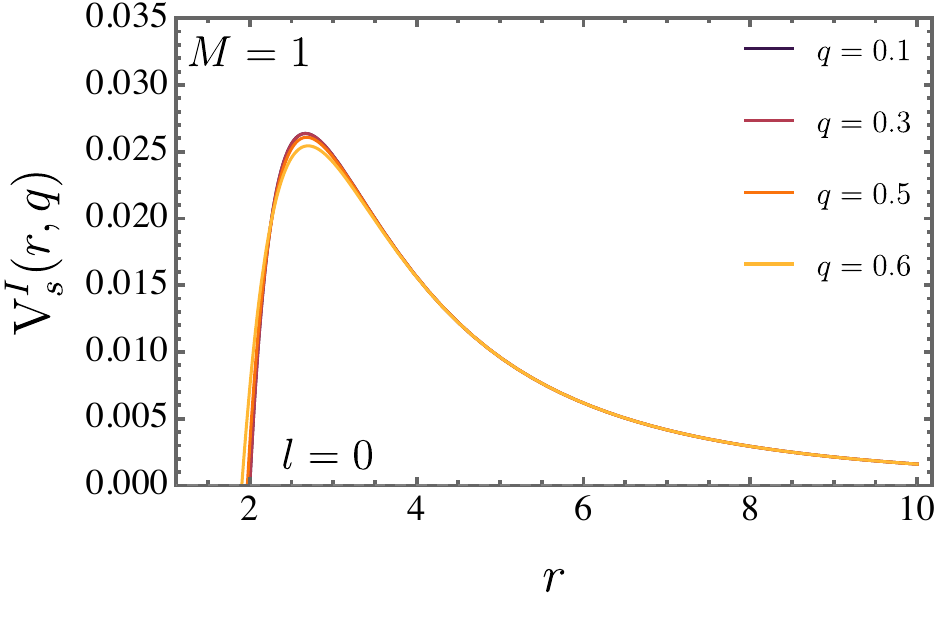}
    \includegraphics[scale=0.53]{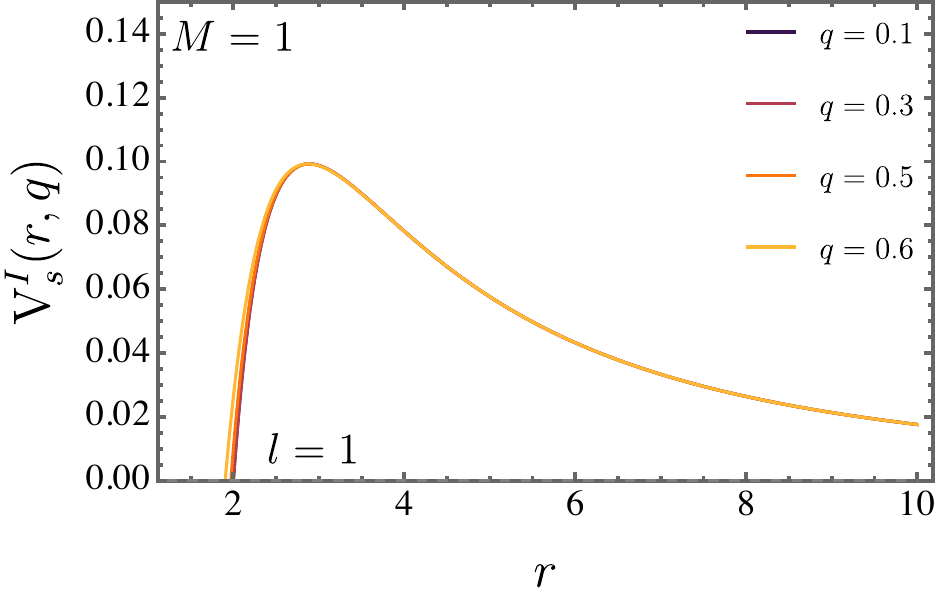}
    \includegraphics[scale=0.53]{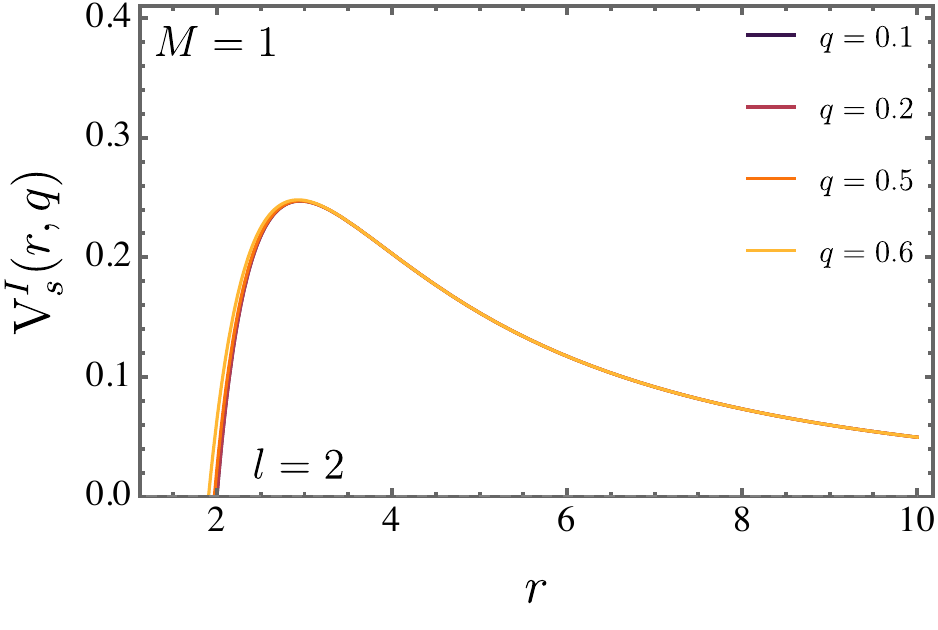}
    \caption{Radial profiles of the scalar effective potential $\mathrm{V}^{I}_{s}(r,q)$ for Model~I. From top to bottom, the panels correspond to the angular momentum modes $l=0$, $l=1$, and $l=2$.}
    \label{ploteventmode1}
\end{figure}

Using the effective potential defined in Eq.~(\ref{potentialeff}) together with the metric function corresponding to Model~I, given in Eq.~(\ref{A1}), we obtain
\begin{eqnarray}
 V^{I}_{s}(r,q) &=&  \left[2 M e^{-\frac{2 r}{|q|}} \left(\frac{2 r}{q^2}+\frac{2}{|q|}+\frac{1}{r}\right)-\frac{2 M}{r}+1\right] \nonumber  \\
	&& \quad  \times \left[\frac{l (l+1)}{r^2}-\frac{2 M e^{-\frac{2 r}{|q|}}}{r^3}-\frac{4 M e^{-\frac{2 r}{|q|}}}{|q| r^2} \right. \nonumber  \\
 && \left. -\frac{8 M e^{-\frac{2 r}{|q|}}}{q^3}-\frac{4 M e^{-\frac{2 r}{|q|}}}{q^2 r}+\frac{2 M}{r^3}\right].
\end{eqnarray}
It should be emphasized that, for fixed $r>0$, the limit $q \to 0$ yields the standard Schwarzschild scalar effective potential, because every term depending on $q$ is accompanied by the exponentially suppressed factor $e^{-2r/|q|}$. In Fig.~\ref{ploteventmode1}, the effective potential is displayed for the angular momentum values $l=0$ (top panel), $l=1$ (middle panel), and $l=2$ (bottom panel). In all cases, increasing the parameter $q$ leads to an enhancement of the potential barrier. Moreover, the potential exhibits a single--peak, bell--shaped profile, ensuring that the WKB approximation is applicable for this model, as will be demonstrated in the subsequent analysis of the quasinormal modes (QNMs).

In direct analogy with the procedure adopted for Model~I, we start from the effective potential given in Eq.~(\ref{potentialeff}) and substitute the metric function corresponding to Model~II, as defined in Eq.~(\ref{Eq_A3}), to obtain
\begin{eqnarray}
&&	\mathrm{V}^{II}_{s}(r,q) =
	\left( 1 - \frac{2 M r^2}{\left(|q| + r\right)^3} \right) \times
		\nonumber  \\
		&&
	\qquad \times\left[ \frac{l (l+1)}{r^2} 
	- \frac{4 M}{\left(|q| + r\right)^3} + \frac{6 M r}{\left(|q| + r\right)^4} \right].
\end{eqnarray}

\begin{figure}[t!]
    \centering
    \includegraphics[scale=0.53]{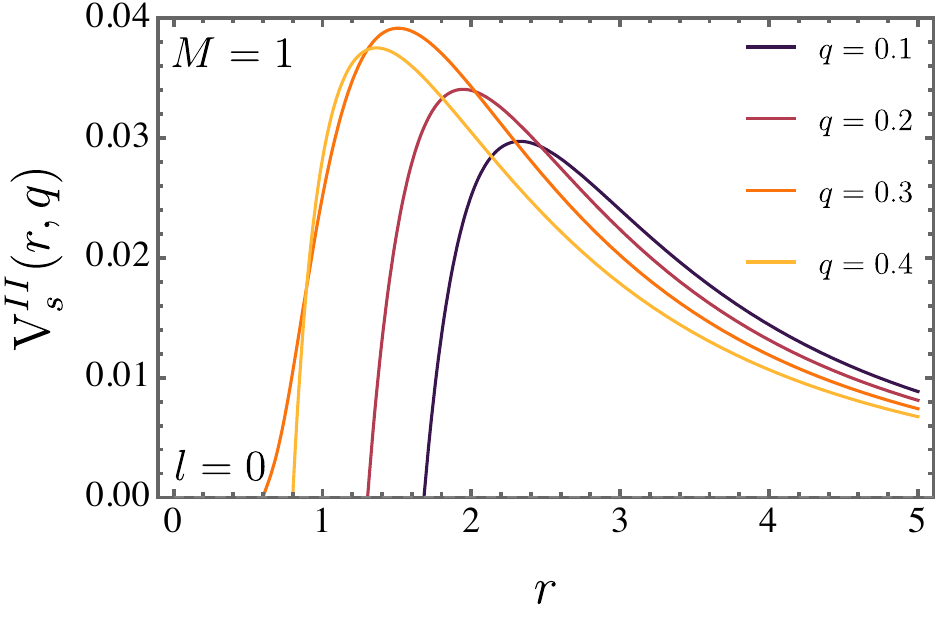}
    \includegraphics[scale=0.53]{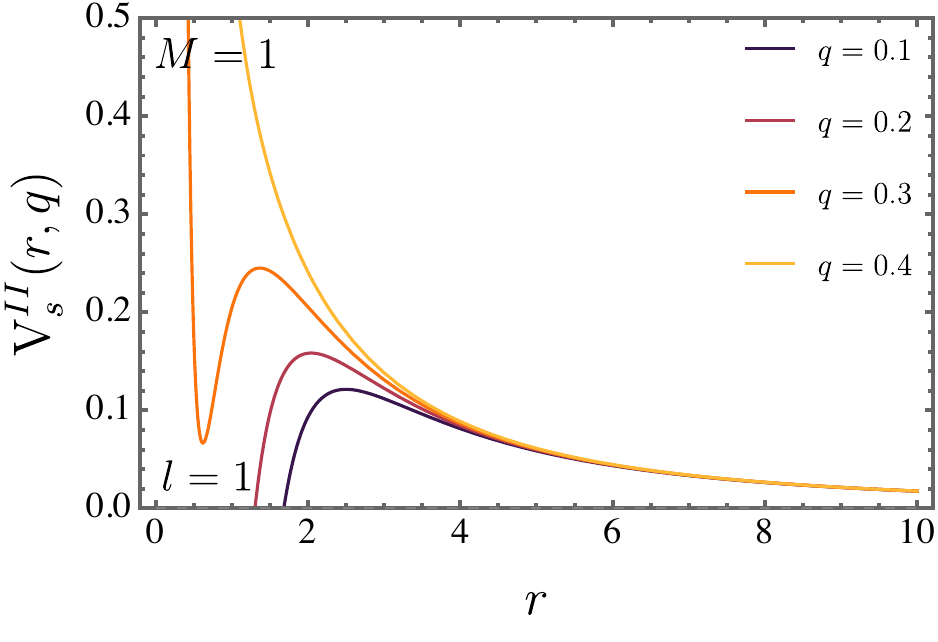}
    \includegraphics[scale=0.53]{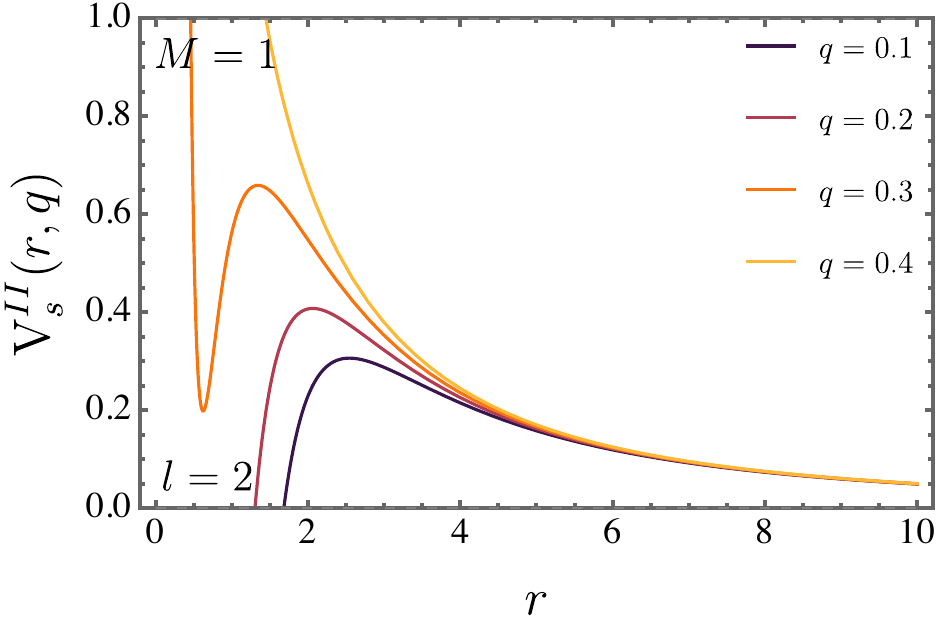}
    \caption{Scalar effective potential $\mathrm{V}^{II}_{s}(r,q)$ for Model~II as a function of the radial coordinate. The upper, middle, and lower panels correspond to $l=0$, $l=1$, and $l=2$, respectively.}
    \label{Vmodel2}
\end{figure}

For this configuration, the Schwarzschild scalar effective potential is recovered smoothly in the limit $q \to 0$. The radial behavior of the potential is displayed in Fig.~\ref{Vmodel2} for the angular momentum modes $l=0$, $l=1$, and $l=2$, shown in the top, middle, and bottom panels, respectively. As the parameter $q$ increases, the height of the potential barrier is increased for all angular modes. Similarly to the previous case, the potential retains its single--peak, bell--shaped profile.

Following the same procedure adopted for the previous configurations, we start from the effective potential given in Eq.~(\ref{potentialeff}) and substitute the metric function corresponding to Model~III, as specified in Eq.~(\ref{A41}), to obtain
\begin{eqnarray}
	&& \mathrm{V}^{III}_{s}(r,q) =
	\left( 1 - \frac{2 M \left( 7 |q| r^2 + r^{5/2} \right)}{\left( |q| + \sqrt{r} \right)^7} \right)\times
		\\
	&& \times \Bigg[
	\frac{l(l+1)}{r^2} + \frac{16 M |q| \sqrt{r}}{\left( |q| + \sqrt{r} \right)^8} 
	+ \frac{2 M r}{\left( |q| + \sqrt{r} \right)^8} 
	- \frac{28 M q^2}{\left( |q| + \sqrt{r} \right)^8}
	\Bigg]. \nonumber 
\end{eqnarray}
%\vspace{0.5cm}

This configuration exhibits the same consistency behavior observed in the preceding cases: as the deformation parameter approaches $q \to 0$, the effective potential smoothly reduces to its Schwarzschild counterpart, recovering the well known structure of the classical black hole potential.  

\begin{figure}[t!]
    \centering
    \includegraphics[scale=0.53]{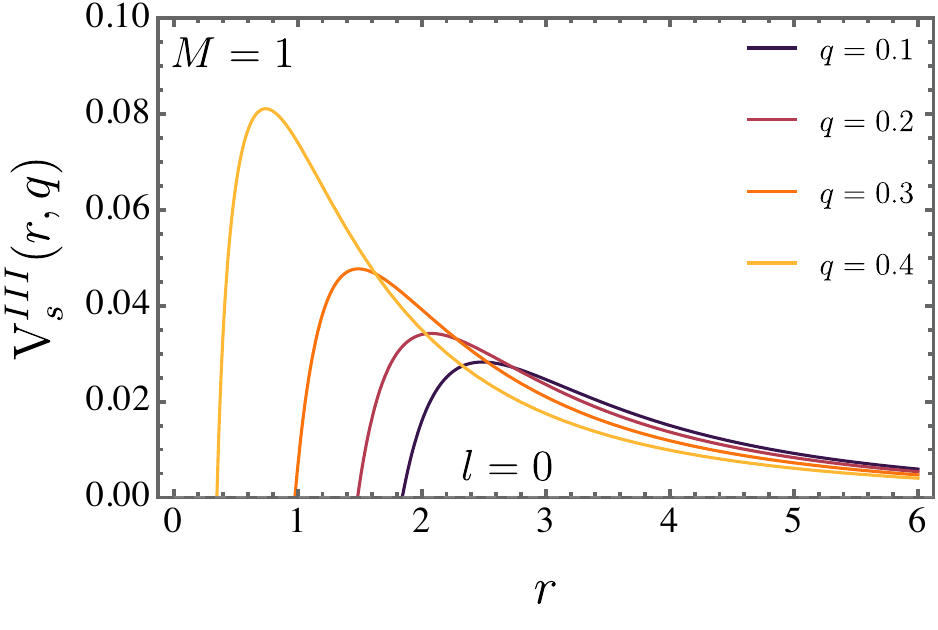}
    \includegraphics[scale=0.53]{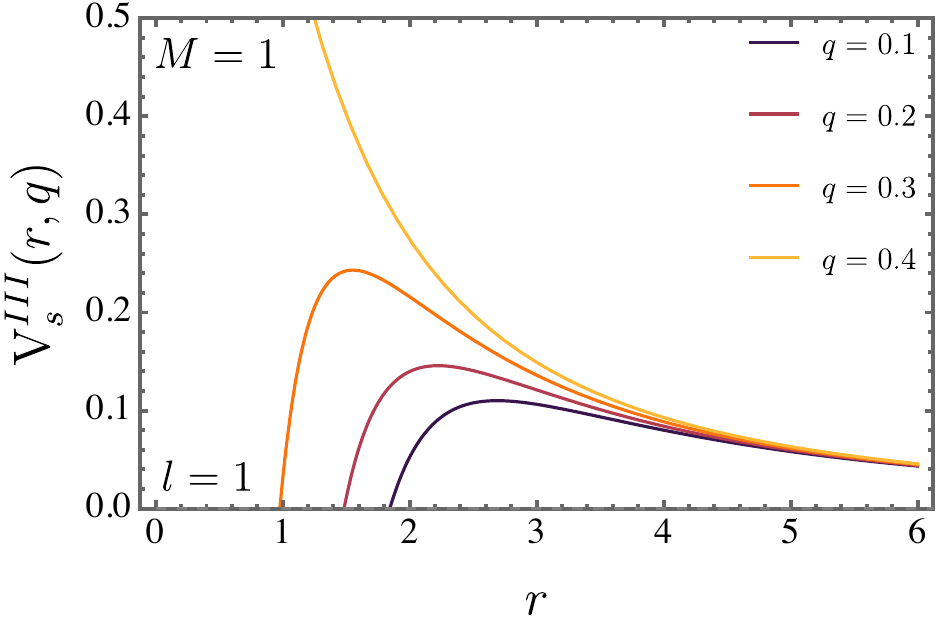}
    \includegraphics[scale=0.53]{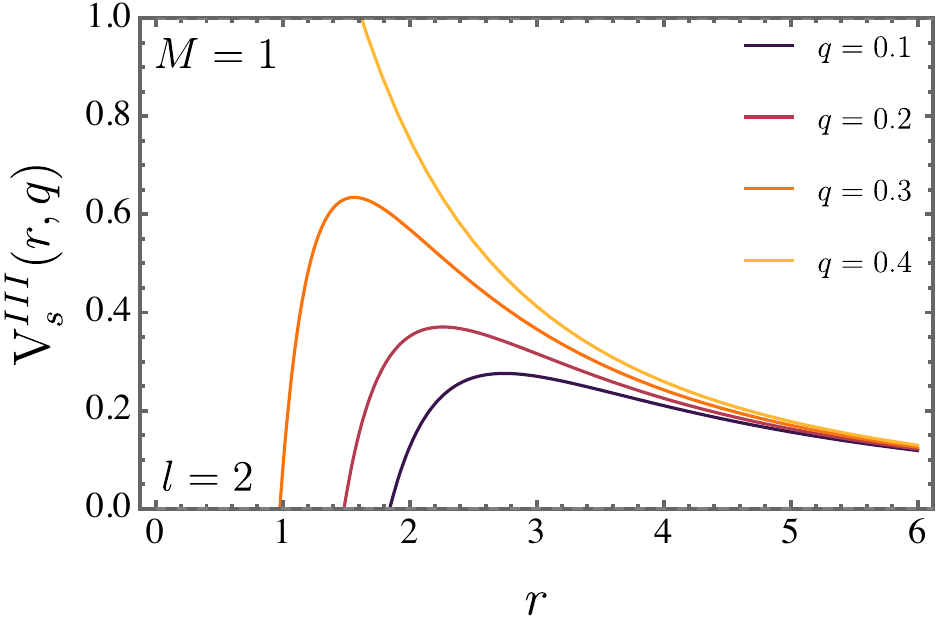}
    \caption{Radial behavior of the scalar effective potential $\mathrm{V}^{III}_{s}(r,q)$ associated with Model~III. The three panels display, from top to bottom, the angular momentum modes $l=0$, $l=1$, and $l=2$.}
    \label{ploteventmodel3}
\end{figure}

The radial profiles of the potential are displayed in Fig.~\ref{ploteventmodel3} for the angular modes $l=0$, $l=1$, and $l=2$, arranged from the top panel to the bottom. In each case, increasing the value of $q$ leads to a higher potential barrier, while the overall shape of the potential is preserved. This behavior highlights how the deformation parameter modifies the scattering properties of scalar perturbations without altering the qualitative features of the effective potential.

%%%%%%%%%%%%%%%%%%%%%%%%%%%%%%%%%%%%%%%%%%%%%%%%%%%%%%%%%%%%%%%%
\section{Quasinormal modes} \label{S:QNM}
%%%%%%%%%%%%%%%%%%%%%%%%%%%%%%%%%%%%%%%%%%%%%%%%%%%%%%%%%%%%%%%%

The determination of the QNM frequencies relies on the properties of the effective potential near its peak, where the transition between classically allowed and forbidden regions is set. By developing the wave equation in the neighborhood of this extremum and implementing a matching scheme based on the WKB approximation, one arrives at a semi--analytical formula for the complex eigenfrequencies governing the ringdown phase. In the present analysis, this procedure is carried out using the higher-order WKB formalism introduced by Konoplya~\cite{Konoplya:2011qq}, for which the resulting frequencies obey the relation
\begin{equation}
\frac{i\bigl(\omega_{n}^{2}-V_{0}\bigr)}{\sqrt{-2V_{0}^{''}}}-\sum_{j=2}^{6}\Lambda_{j}=n+\frac{1}{2}.
\end{equation}
Here, the quantity $V_{0}^{\prime\prime}$ denotes the second radial derivative of the effective potential computed at the location of its maximum, $r_{0}$. The symbols $\Lambda_{j}$ encode a hierarchy of higher--order contributions built from the value of the potential at the peak and its successive derivatives. These correction terms play a central role in refining the accuracy of the approximation. Below we describe the results of the application of this method to the three models introduced in this work.

%%%%%%%%%%%%%%%%%%%%%%%%%%%%%%%%%%%%%%%%%%%%%%%%%%%%%%%%%%%%%%%%
\subsection{Model I}
%%%%%%%%%%%%%%%%%%%%%%%%%%%%%%%%%%%%%%%%%%%%%%%%%%%%%%%%%%%%%%%%

In Tables~\ref{tabqmns0model1}, \ref{tabqmns1model1}, and \ref{tabqmns2model1}, we present the QNM spectra of scalar perturbations for $l=0$, $l=1$, and $l=2$, respectively, fixing $M=1$ and varying the deformation parameter $q$.

For all multipole numbers, the fundamental mode ($n=0$) shows a weak dependence on $q$, with both $\Re(\omega)$ (real) and $|\Im(\omega)|$ (imaginary). In contrast, higher overtones exhibit a stronger sensitivity to the deformation, particularly in the imaginary part of the frequency. This effect is most pronounced for $l=0$, where increasing $q$ leads to sizable variations in the damping rates, indicating that the deformation modifies the dissipation timescale of scalar perturbations.

As $l$ increases, the influence of $q$ becomes progressively weaker. For $l=1$ and especially for $l=2$, the spectra remain remarkably stable, with only marginal shifts in both oscillation frequencies and decay rates. This behavior reflects the reduced impact of the deformation on modes dominated by the angular momentum barrier.

These results show that the parameter $q$ mainly affects the damping of higher overtones and low–$l$ modes, while the overall structure of the QNM spectrum is preserved across the parameter space considered.

\begin{table*}[t!]
\caption{\label{tabqmns0model1} Scalar quasinormal frequencies obtained for Model~I by fixing the mass to $M=1$ and restricting the analysis to the monopole sector ($l=0$). The spectrum is displayed for several representative values of the deformation parameter $q$.}
\hspace*{-0.1cm}
\begin{tabular}{c| c | c | c} 
 \hline\hline\hline 
 $q$   & $\omega_{0}$ & $\omega_{1}$ & $\omega_{2}$  \\ [0.2ex] 
 \hline 
   0.20 & 0.110434 - 0.100829$i$ & 0.088963 - 0.344604$i$  & 0.191600 - 0.476596$i$ \\
 
  0.25  & 0.113627 - 0.096530$i$ & 0.090162 - 0.325670$i$ & 0.205306 - 0.417888$i$   \\
 
 0.30  & 0.113132 - 0.095798$i$ & 0.084925 - 0.345584$i$  & 0.167813 - 0.530735$i$  \\
 
 0.35  & 0.086200 - 0.150340$i$ & 0.088861 - 0.524490$i$ & 0.168360 - 0.971815$i$   \\
 0.40  & 0.085994 - 0.191102$i$  & 0.121042 - 0.540018$i$  &  0.274758 - 0.769677$i$  \\
   [0.2ex]  
 \hline \hline \hline 
\end{tabular}
\end{table*}

\begin{table*}[t!]
\caption{\label{tabqmns1model1} Scalar quasinormal frequencies for Model~I computed with $l=1$ and $M=1$, displayed as a function of the deformation parameter $q$.}
\hspace*{-0.1cm}
\begin{tabular}{c| c | c | c} 
 \hline\hline\hline 
 $q$   & $\omega_{0}$ & $\omega_{1}$ & $\omega_{2}$  \\ [0.2ex] 
 \hline 
   0.20 & 0.292906 - 0.0977643$i$ & 0.264455 - 0.306550$i$ & 0.230988 - 0.542293$i$  \\
 
  0.25  & 0.292872 - 0.0977683$i$ & 0.264353 - 0.306537$i$ & 0.230762 - 0.541926$i$   \\
 
 0.30  & 0.293387 - 0.0973186$i$  & 0.266134 - 0.301974$i$  & 0.232310 - 0.525666$i$  \\
 
 0.35  & 0.294248 - 0.0965185$i$ & 0.266849 - 0.298650$i$ & 0.229297 - 0.527756$i$   \\
 0.40  & 0.292469 - 0.0975965$i$  & 0.258065 - 0.317937$i$  &  0.221482 - 0.606851$i$  \\
   [0.2ex]  
 \hline \hline \hline 
\end{tabular}
\end{table*}

\begin{table*}[t!]
\caption{\label{tabqmns2model1} Quasinormal mode frequencies of scalar perturbations for Model~I evaluated at $l=2$ with $M=1$, shown for varying values of the deformation parameter $q$.}
\hspace*{-0.1cm}
\begin{tabular}{c| c | c | c} 
 \hline\hline\hline 
 $q$   & $\omega_{0}$ & $\omega_{1}$ & $\omega_{2}$  \\ [0.2ex] 
 \hline 
   0.20 & 0.483642 - 0.0967662$i$ & 0.463846 - 0.295629$i$ & 0.430383 - 0.508709$i$  \\
 
  0.25  & 0.483637 - 0.0967675$i$ & 0.463818 - 0.295641$i$ & 0.430309 - 0.508751$i$   \\
 
 0.30  & 0.48365 - 0.096737$i$ & 0.463914 - 0.295334$i$  & 0.430481 - 0.507352$i$  \\
 
 0.35  & 0.483699 - 0.0966222$i$ & 0.464046 - 0.294633$i$ & 0.430278 - 0.505548$i$   \\
 0.40  & 0.483619 - 0.0964854$i$  & 0.463244 - 0.294869$i$  &  0.427885 - 0.509773$i$  \\
   [0.2ex]  
 \hline \hline \hline 
\end{tabular}
\end{table*}

%%%%%%%%%%%%%%%%%%%%%%%%%%%%%%%%%%%%%%%%%%%%%%%%%%%%%%%%%%%%%%%%
\subsection{Model II}
%%%%%%%%%%%%%%%%%%%%%%%%%%%%%%%%%%%%%%%%%%%%%%%%%%%%%%%%%%%%%%%%

Tables~\ref{tabqmns0model2}, \ref{tabqmns1model2}, and \ref{tabqmns2model2} present the QNM spectra of scalar perturbations for Model II with $l=0,1,2$ and several values of the deformation parameter $q$. From these data, one observes that the QNM frequencies depend on $q$ through both their real and imaginary components.
For fixed angular momentum $l$ and overtone number $n$, increasing $q$ generally leads to an increase in the real part $\Re(\omega)$, indicating that scalar perturbations oscillate at higher frequencies as the deformation strengthens. This behavior reflects modifications of the effective potential, whose peak becomes higher as $q$ increases, thereby raising the characteristic oscillation scale of the modes.

The imaginary part of the frequencies exhibits a corresponding dependence on $q$. In most cases, the magnitude $|\Im(\omega)|$ increases as $q$ grows, signaling faster damping and shorter ringdown times. Physically, this indicates that the deformation facilitates energy dissipation by altering the shape of the effective potential barrier, allowing perturbations to decay more efficiently. For larger values of $q$ and higher overtones, small departures from monotonic behavior appear, suggesting that the deformation induces nontrivial changes in the potential profile that affect the decay rates.

At fixed $q$, increasing the angular momentum $l$ raises $\Re(\omega)$ while reducing the relative variation of $\Im(\omega)$. This behavior is consistent with the growing dominance of the angular momentum barrier at higher $l$, which stabilizes the spectrum and suppresses the sensitivity of the modes to deformations of the background geometry. As a result, the $l=0$ sector remains the most responsive to variations in $q$.

The overtone structure follows the expected hierarchy: higher-$n$ modes possess larger imaginary parts and therefore decay more rapidly. These modes also display an enhanced response to changes in $q$, reflecting their stronger dependence on the detailed structure of the effective potential beyond its maximum.

\begin{table*}[t!]
\caption{\label{tabqmns0model2} Quasinormal frequencies associated with scalar-field perturbations are listed for the monopole sector ($l = 0$), with the mass scale fixed to $M = 1$. The entries are organized according to the values of the deformation parameter $q$ and correspond to the configuration denoted as Model II.}
\hspace*{-0.1cm}
\begin{tabular}{c| c | c | c} 
 \hline\hline\hline 
 $q$   & $\omega_{0}$ & $\omega_{1}$ & $\omega_{2}$  \\ [0.2ex] 
 \hline 
   0.010   & 0.111625 - 0.101138$i$ & 0.090274 - 0.345118$i$ & 0.190605 - 0.480253$i$ \\
 
  0.025  & 0.113454 - 0.101577$i$ & 0.092277 - 0.345695$i$  & 0.189230 - 0.484934$i$   \\
 
 0.050  & 0.116638 - 0.102303$i$ & 0.095769 - 0.346654$i$ & 0.186741 - 0.493332$i$  \\
 
 0.075  & 0.120065 - 0.102968$i$ & 0.099583 - 0.347219$i$ & 0.184563 - 0.500979$i$    \\
 0.100  &  0.123695 - 0.103624$i$  & 0.103629 - 0.347790$i$  &  0.181969 - 0.509732$i$   \\
   [0.2ex]  
 \hline \hline \hline 
\end{tabular}
\end{table*}

\begin{table*}[t!]
\caption{\label{tabqmns1model2} Quasinormal frequencies obtained from scalar perturbations in the dipole sector ($l = 1$), evaluated by fixing the mass parameter to $M = 1$. The results are arranged according to the deformation parameter $q$ and correspond to the scenario labeled as Model II.}
\hspace*{-0.1cm}
\begin{tabular}{c| c | c | c} 
 \hline\hline\hline 
 $q$   & $\omega_{0}$ & $\omega_{1}$ & $\omega_{2}$  \\ [0.2ex] 
 \hline 
   0.010   & 0.295925 - 0.098060$i$ & 0.267809 - 0.307234$i$ & 0.234712 - 0.542901$i$ \\
 
  0.025  & 0.300621 - 0.098503$i$ & 0.273008 - 0.308286$i$ & 0.240467 - 0.54395$i$   \\
 
 0.050  & 0.308955 - 0.099225$i$ & 0.282242 - 0.309955$i$ & 0.250672 - 0.545517$i$  \\
 
 0.075  & 0.318013 - 0.099911$i$ & 0.29228 - 0.311476$i$  & 0.261746 - 0.546757$i$   \\
 0.100  &  0.327919 - 0.100542$i$  & 0.303261 - 0.312779$i$  &  0.273830 - 0.547540$i$   \\
   [0.2ex]  
 \hline \hline \hline 
\end{tabular}
\end{table*}

\begin{table*}[t!]
\caption{\label{tabqmns2model2} Quasinormal frequencies for the quadrupole configuration ($l = 2$) are compiled here with the mass parameter set to $M = 1$. The spectra are tabulated as functions of the deformation parameter $q$ and refer to the framework identified as Model II.}
\hspace*{-0.1cm}
\begin{tabular}{c| c | c | c} 
 \hline\hline\hline 
 $q$   & $\omega_{0}$ & $\omega_{1}$ & $\omega_{2}$  \\ [0.2ex] 
 \hline 
   0.010   & 0.488575 - 0.097080$i$ & 0.469011 - 0.296503$i$ & 0.43594 - 0.509943$i$  \\
 
  0.025  & 0.496265 - 0.097548$i$ & 0.477060 - 0.297800$i$  & 0.444599 - 0.511769$i$   \\
 
 0.050  & 0.509929 - 0.098312$i$ & 0.491369 - 0.299902$i$  & 0.459987 - 0.514672$i$  \\
 
 0.075  & 0.524804 - 0.099044$i$ & 0.506949 - 0.301888$i$ & 0.476741 - 0.517327$i$   \\
 0.100  &  0.541105 - 0.099724$i$  & 0.524023 - 0.303692$i$  &  0.495096 - 0.519616$i$   \\ 
 \hline \hline \hline 
\end{tabular}
\end{table*}

%%%%%%%%%%%%%%%%%%%%%%%%%%%%%%%%%%%%%%%%%%%%%%%%%%%%%%%%%%%%%%%%
\subsection{Model III}
%%%%%%%%%%%%%%%%%%%%%%%%%%%%%%%%%%%%%%%%%%%%%%%%%%%%%%%%%%%%%%%%

A consistent pattern emerges from Tables \ref{tabqmns0model3}, \ref{tabqmns1model3}, and \ref{tabqmns2model3} for Model III when the deformation parameter $q$ is varied. For all angular sectors, increasing $q$ leads to a systematic growth of the real part of the QNM frequencies. This behavior indicates that the oscillation rates become higher as the deformation strengthens, meaning that scalar perturbations probe an effectively stiffer potential barrier in the deformed geometry. At the same time, the magnitude of the imaginary part also increases for all overtones, showing that the damping becomes stronger and the perturbations decay more rapidly as $q$ grows.

For fixed $q$, the real part of the frequencies increases with the angular momentum number $l$, which is consistent with the stronger centrifugal contribution to the effective potential for higher multipoles. Correspondingly, modes with larger $l$ are less damped than those with smaller $l$, as reflected by the smaller absolute values of the imaginary parts. This hierarchy persists across all considered values of the deformation parameter, indicating that the qualitative structure of the spectrum remains stable under the deformation introduced in this model.

Regarding the overtone number, higher overtones ($n=1,2$) exhibit larger imaginary parts compared to the fundamental mode, implying faster decay rates, while their real parts follow a less uniform trend. This reflects the fact that excited modes are increasingly sensitive to the dissipative properties of the effective potential rather than to its peak structure alone.

A comparison of the three models, based on the complete set of tabulated QNM frequencies, reveals clear differences in how the deformation parameter $q$ affects the scalar spectrum. In Model I, the dependence on $q$ is generally weak, particularly for the dipole ($l=1$) and quadrupole ($l=2$) modes. In these cases, both the real and imaginary parts of the frequencies remain nearly unchanged as $q$ varies, indicating that the deformation introduces only small corrections to the effective potential. A more pronounced sensitivity arises in the $l=0$ scalar mode, where the imaginary parts display nonuniform behavior at larger values of $q$, suggesting that spherically symmetric perturbations respond more strongly to the deformation in this model.

By contrast, Models II and III exhibit a clear and regular dependence on $q$ across all angular modes. In both cases, the real parts of the QNM frequencies increase monotonically with $q$, while the magnitude of the imaginary parts generally grows, corresponding to faster damping. These effects become more pronounced for higher angular momentum and higher overtones. Between the two, Model III shows the strongest response to the deformation, with larger shifts in both the oscillation frequencies and the damping rates as $q$ increases. Overall, while Model I remains comparatively insensitive to the deformation, Models II and III encode the effects of $q$ more efficiently in their QNM spectra, with Model III displaying the largest deviations.

\begin{table*}[t!]
\caption{\label{tabqmns0model3} Quasinormal frequencies associated with scalar-field perturbations in the monopole channel ($l = 0$), computed by fixing the mass scale to $M = 1$. The entries are arranged according to the deformation parameter $q$ and correspond to the configuration denoted as Model III.}
\hspace*{-0.1cm}
\begin{tabular}{c| c | c | c} 
 \hline\hline\hline 
 $q$   & $\omega_{0}$ & $\omega_{1}$ & $\omega_{2}$  \\ [0.2ex] 
 \hline 
   0.0010 & 0.110467 - 0.100817$i$ & 0.089030 - 0.344529$i$ & 0.191831 - 0.476396$i$ \\
 
  0.0020  & 0.110466 - 0.100821$i$ & 0.089025 - 0.344558$i$ & 0.191772 - 0.476536$i$   \\
 
 0.0030  & 0.110472 - 0.100820$i$  & 0.089033 - 0.344546$i$ & 0.191796 - 0.476479$i$  \\
 
 0.0040  & 0.110483 - 0.100817$i$ & 0.089048 - 0.344513$i$ & 0.191852 - 0.476351$i$   \\
 0.0049  & 0.110486 - 0.100822$i$  & 0.089049 - 0.344540$i$  &  0.191810 - 0.476461$i$   \\
   [0.2ex]  
 \hline \hline \hline 
\end{tabular}
\end{table*}

\begin{table*}[t!]
\caption{\label{tabqmns1model3} Scalar quasinormal frequencies for the dipolar sector ($l = 1$) are collected here with the mass parameter fixed at $M = 1$. The spectrum is organized according to the values of the deformation parameter $q$ and corresponds to the framework referred to as Model III.}
\hspace*{0cm}
\begin{tabular}{c| c | c | c} 
 \hline\hline\hline 
 $q$   & $\omega_{0}$ & $\omega_{1}$ & $\omega_{2}$  \\ [0.2ex] 
 \hline 
   0.0010 & 0.292912 - 0.097761$i$ & 0.264473 - 0.306519$i$ & 0.231017 - 0.542166$i$ \\
 
  0.0020  & 0.292918 - 0.097762$i$ & 0.264480 - 0.306520$i$ & 0.231024 - 0.542160$i$   \\
 
 0.0030  & 0.292928 - 0.097763$i$  & 0.264492 - 0.306523$i$  & 0.231037 - 0.542170$i$  \\
 
 0.0040  & 0.292943 - 0.097765$i$ & 0.264507 - 0.306526$i$ & 0.231054 - 0.542174$i$   \\
 0.004900  & 0.292959 - 0.097766$i$   & 0.264525 - 0.306531$i$  &  0.231074 - 0.542178$i$   \\
   [0.2ex]  
 \hline \hline \hline 
\end{tabular}
\end{table*}

\begin{table*}[t!]
\caption{\label{tabqmns2model3} Scalar quasinormal mode frequencies associated with the quadrupolar configuration ($l = 2$), evaluated at the fixed mass scale $M = 1$. The modes are tabulated as functions of the deformation parameter $q$ within the setup identified as Model III.}
\hspace*{0cm}
\begin{tabular}{c| c | c | c} 
 \hline\hline\hline 
 $q$   & $\omega_{0}$ & $\omega_{1}$ & $\omega_{2}$  \\ [0.2ex] 
 \hline 
   0.0010 & 0.483645 - 0.096766$i$ & 0.463850 - 0.295628$i$ & 0.430390 - 0.508701$i$ \\
 
  0.0020  & 0.483655 - 0.096767$i$ & 0.463861 - 0.295629$i$ & 0.430401 - 0.508703$i$    \\
 
 0.0030  & 0.483672 - 0.096768$i$  & 0.463878 - 0.295632$i$ & 0.430420 - 0.508708$i$  \\
 
 0.0040  & 0.483696 - 0.096769$i$ & 0.463903 - 0.295637$i$ & 0.430446 - 0.508714$i$   \\
 0.0049  & 0.483722 - 0.096771$i$  & 0.463931 - 0.295642$i$  &  0.430476 - 0.508721$i$   \\
   [0.2ex]  
 \hline \hline \hline 
\end{tabular}
\end{table*}

%%%%%%%%%%%%%%%%%%%%%%%%%%%%%%%%%%%%%%%%%%%%%%%%%%%%%%%%%%%%%%%%
\section{Time-domain solution} \label{S:TDS}
%%%%%%%%%%%%%%%%%%%%%%%%%%%%%%%%%%%%%%%%%%%%%%%%%%%%%%%%%%%%%%%%

Taking into account the temporal development of scalar perturbations demands a treatment that resolves the field evolution itself, instead of extracting indirect information from frequency–space spectra alone. A time–domain analysis makes the causal propagation explicit and allows the quasinormal ringing to arise dynamically, governing both the attenuation pattern and the scattering behavior of the signal. In practice, this task is delicate: the effective potentials involved often display complex structures, so numerical instabilities can easily contaminate the results unless the integration scheme is chosen with care. A robust strategy is therefore essential to guarantee both accuracy and long–term stability. To this end, we make use of a characteristic evolution method, originally introduced by Gundlach et al.~\cite{Gundlach:1993tp}, which has proven particularly reliable for this class of dynamical problems.

Adopting the computational framework developed and refined in Refs.~\cite{Bolokhov:2024ixe,Skvortsova:2024wly,Guo:2023nkd,AraujoFilho:2024xhm,Baruah:2023rhd,Shao:2023qlt,Lutfuoglu:2025kqp,Yang:2024rms,AraujoFilho:2024lsi}, we recast the wave equation in terms of light–cone coordinates defined by $u=t-r^{*}$ and $v=t+r^{*}$. This reformulation aligns the numerical grid with the causal structure of the space-time, reducing computational complexity and improving efficiency. In the resulting variables, the perturbation equation assumes a form that is particularly amenable to stable numerical integration, which we briefly outline below
\begin{equation}
\left(4 \frac{\partial^{2}}{\partial u \, \partial v} + V(u,v)\right) \Tilde{\psi} (u,v) = 0.
\end{equation}

The computational procedure proceeds by translating the continuous evolution equation into a discrete form suitable for numerical treatment. To this end, the space-time region of interest is covered by a finite grid, and derivatives are replaced by difference operators defined on this lattice. The value of the field at a given grid point is then obtained from already known data at neighboring points, enabling the solution to be propagated iteratively across the mesh. In this way, the time evolution of the waveform is reconstructed incrementally throughout the computational domain
\begin{equation}
\begin{split}
 \Tilde{\psi}(N) =& -\Tilde{\psi}(S) + \Tilde{\psi}(W) + \Tilde{\psi}(E) \\
& - \frac{h^{2}}{8}V(S)\Big[\Tilde{\psi}(W) + \Tilde{\psi}(E)\Big] + \mathcal{O}(h^{4}).
\end{split}
\end{equation}

The time evolution is computed by constructing a uniform grid in the $(u,v)$ plane, where both null coordinates are discretized with the same step size $h$. The integration proceeds sequentially across this grid, advancing one elementary cell at a time. Within each cell, the field values at three corners are already available, while the value at the remaining corner—corresponding to the future point—is determined through the numerical update prescription. The point at $(u,v)$ defines the base of the cell, its neighboring points along the null directions supply the necessary input, and the solution at the forward vertex is then reconstructed.

To start the evolution, initial data are imposed on two intersecting null surfaces, $u=u_{0}$ and $v=v_{0}$, which fix the boundary conditions of the problem. Along the ingoing null segment $u=u_{0}$, the scalar field is initialized in the form of a localized pulse, modeled by a Gaussian profile centered at $v=v_{c}$ with width $k_2$. From these initial surfaces, the algorithm advances the solution throughout the grid, yielding the complete time–domain history of the perturbation
\begin{equation}
\Tilde{\psi}(\Tilde{u} = u_{0},v) = k_1 e^{-\frac{(v-v_{0})^{2}}{2k_2^{2}}}, \,\,\,\,\,\, \Tilde{\psi}(u,v_{0}) = \Tilde{\psi}_{0}.
\end{equation}

The numerical evolution is initiated by imposing boundary data on the null line $v=v_{0}$, where the scalar field is taken to vanish, $\tilde{\psi}(u,v_{0})=0$. This prescription establishes a simple and stable reference configuration from which the dynamics can unfold. Starting from this boundary, the solution is propagated incrementally toward larger values of $v$ at fixed $u$, in full accordance with the causal structure encoded in the double–null scheme. To keep the setup minimal and to suppress spurious numerical effects, only massless scalar perturbations are considered, and the black hole mass is fixed to $M=1$ for all simulations. The perturbation is injected through a localized Gaussian profile centered at $v=0$ with width $k_1=k_2=1$, serving as the initial trigger of the evolution. The numerical grid covers the square region $0\leq u,v\leq1000$ in the $(u,v)$ plane and is discretized with a uniform spacing $h=0.1$, which adequately captures both the quasinormal oscillations and the subsequent decay stages. In what follows, the three models are examined independently, following the same organizational structure adopted earlier in the QNM analysis.
%%%%%%%%%%%%%%%%%%%%%%%%%%%%%%%%%%%%%%%%%%%%%%%%%%%%%%%%%%%%%%%%%%%%%%%%%%%%%%%%%%%%%%%%%%%%%%%%%%%%%%%%%%%%%%%%%%%%%%%%%%%%%%%%%%%%%%%%%%%%%%%%%%%%%%%%%%%%%%%%%%%%%%%%%%%%%%%%%%%%%%%%%%%%%%%%%%%%%%%%%%%%%%%%%%%%%%%%%%%%%%%%%%%%%%%%%%%%%%%%%%%%%%%%%%%%%%%%%%%%%%%%%%%%%%%%%%%%%%%%%%%%%%%%%%%%%%%%%%%%%%%%%%%%%%%%%%%%%%%%%%%%%%%%%%%%%%%%%%%%%%%%%%%%%%%%%%%%%%%%%%%%%%%%%%%%%%%%%%%%%%%%%%%%%%%%%%%%%%%%%%%%%%%%%%%%%%%%%%%%%%%%%%%%%%%%%%%%%%%%%%%%%%%%
\subsection{Model I}

\begin{figure}[t!]
    \centering
    \includegraphics[scale=0.53]{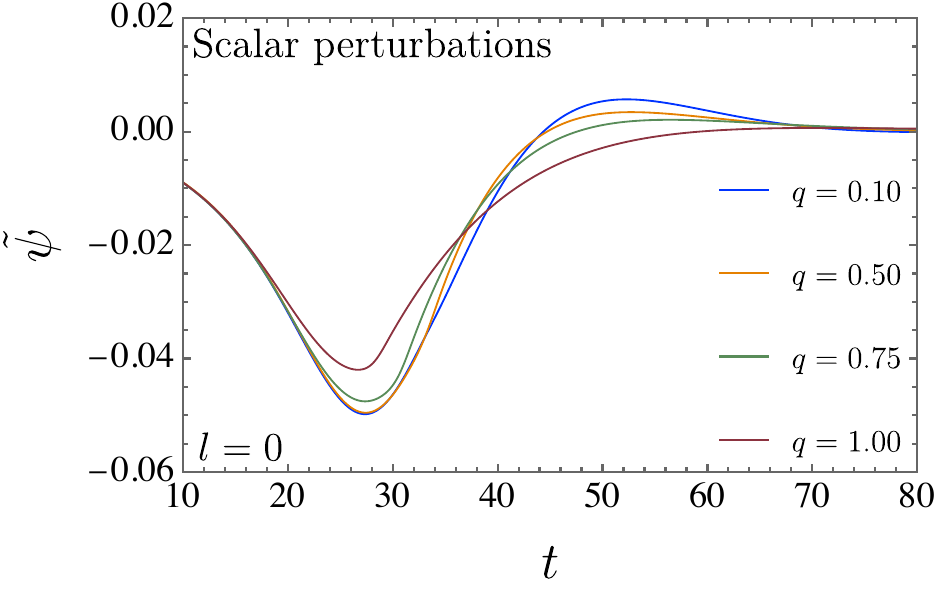}
    \includegraphics[scale=0.53]{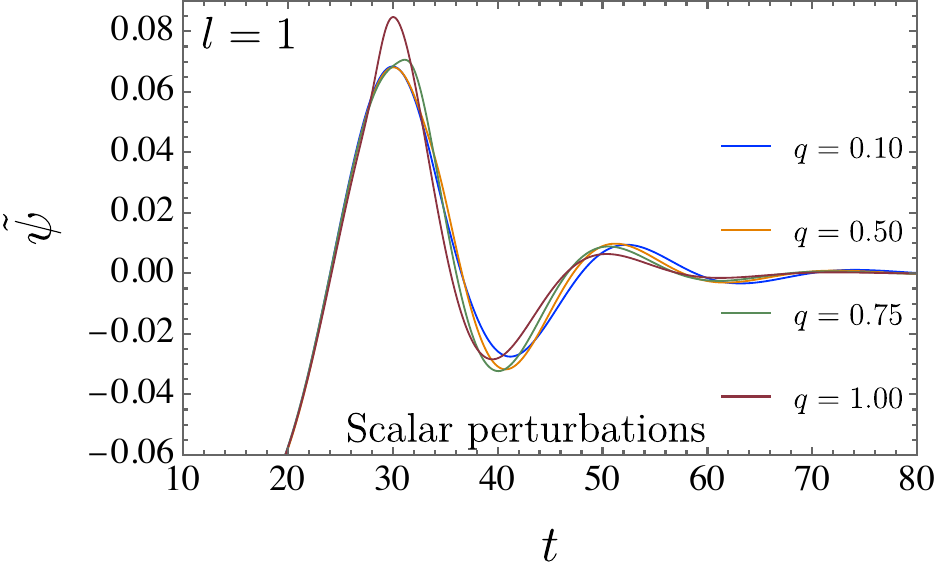}
    \includegraphics[scale=0.53]{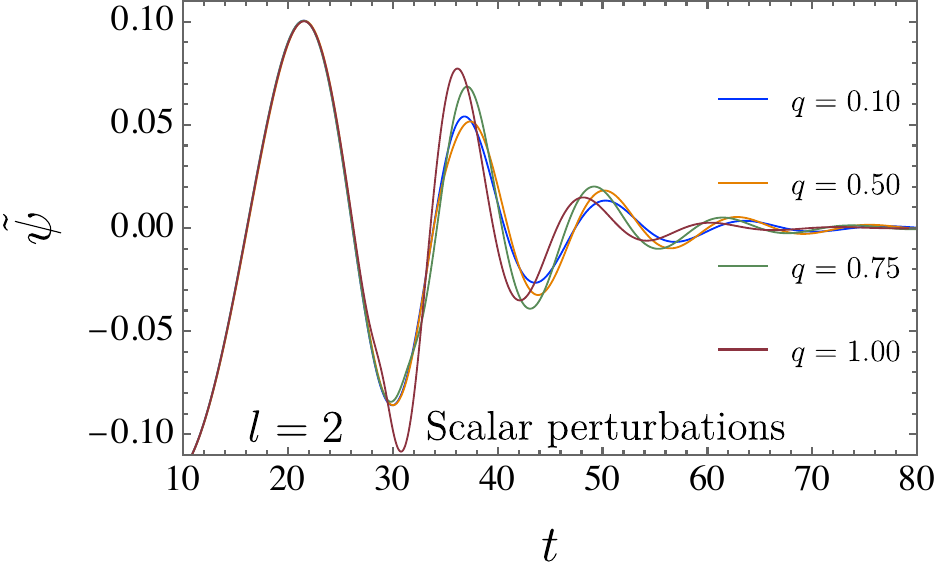}
     \caption{Time--domain profiles of the scalar field $\tilde{\psi}$ are presented for the black hole configuration of Model I, keeping the mass fixed at $M=1$ while varying the deformation parameter over $q=0.10$, $0.50$, $0.75$, and $1.00$. The signals are grouped by angular momentum, with the upper–left panel showing the $l=0$ mode, the upper–right panel the $l=1$ mode, and the lower panel the $l=2$ mode.}
    \label{psitimedomainmodel1}
\end{figure}

The time–domain behavior of scalar perturbations in the black hole geometry  of Model I is examined in this section. Fig.~\ref{psitimedomainmodel1} shows the numerical evolution of the field $\tilde{\psi}$ for a fixed mass $M=1$, while the deformation parameter is selected as $q=0.10$, $0.50$, $0.75$, and $1.00$. The results are organized by diverse multipole configurations, with $l=0$ shown in the upper left panel, $l=1$ in the upper right panel, and $l=2$ in the lower panel. In all cases, the signal is dominated at intermediate times by oscillations whose amplitude decreases exponentially, identifying the QNM ringing stage associated with scalar perturbations of the space-time defined by Eq.~(\ref{A1}).

\begin{figure}[t!]
    \centering
    \includegraphics[scale=0.53]{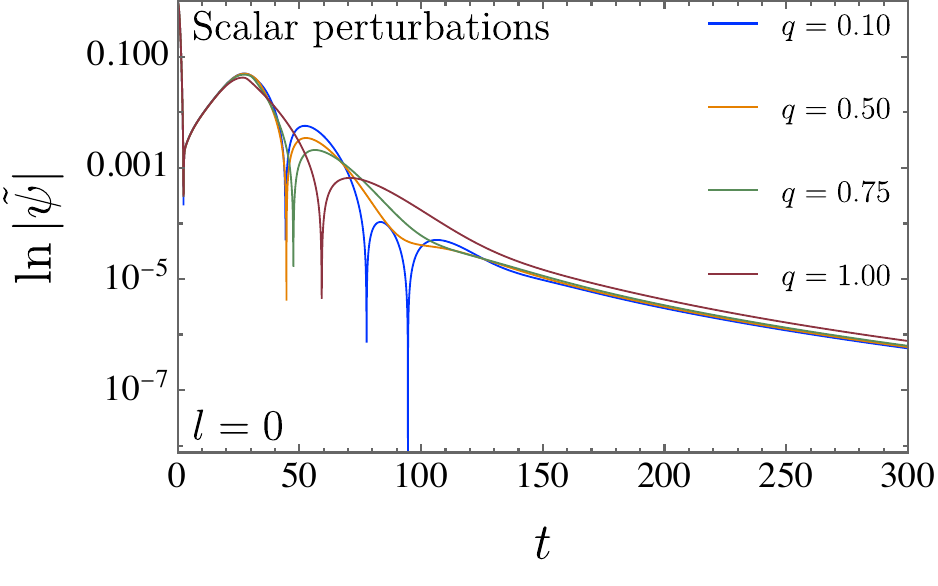}
    \includegraphics[scale=0.53]{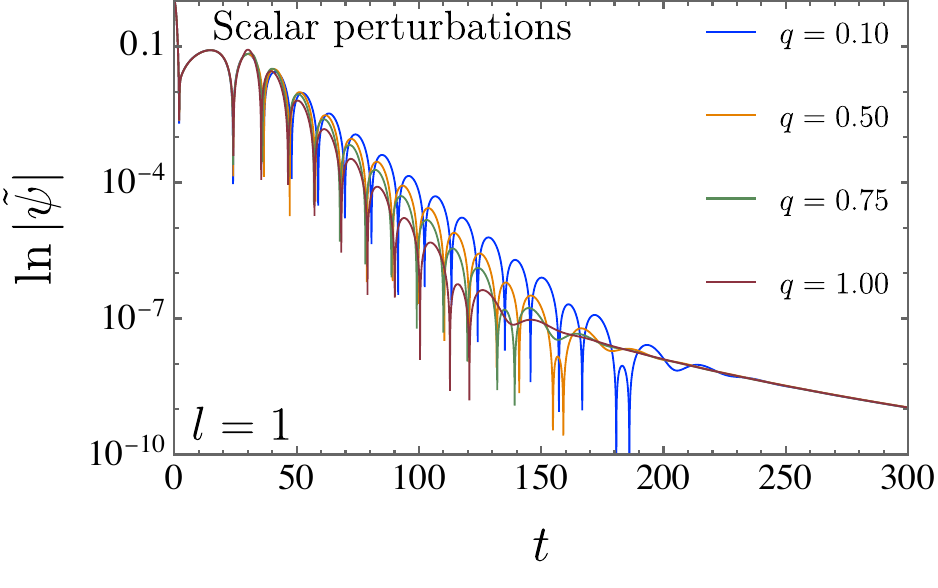}
    \includegraphics[scale=0.53]{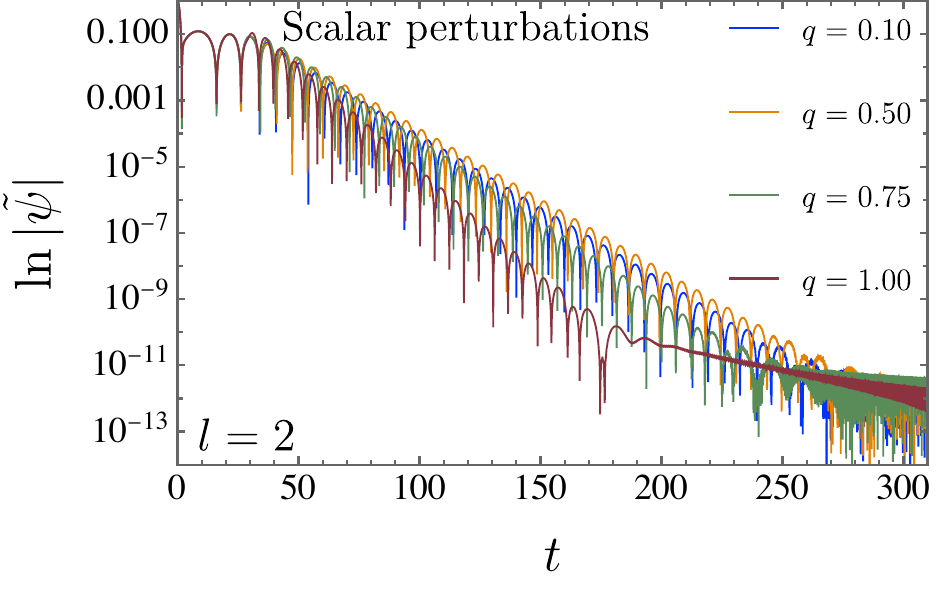}
    \caption{The time dependence of the scalar signal is illustrated through the quantity $\ln|\tilde{\psi}|$ for the Model I black hole, considering a fixed mass $M=1$ and the parameter values $q=0.10$, $0.50$, $0.75$, and $1.00$. The panels are arranged according to the angular momentum number, with the $l=0$ contribution displayed in the upper–left panel, the $l=1$ contribution in the upper–right panel, and the $l=2$ contribution in the lower panel.}
    \label{lnpsitimedomainmodel1}
\end{figure}

A more transparent characterization of the decay rate is provided in Fig.~\ref{lnpsitimedomainmodel1}, where the quantity $\ln|\tilde{\psi}|$ is displayed for the same combinations of $q$ and $l$. In this representation, the QNM regime manifests itself as an approximately linear segment, reflecting exponential damping. The departure from this linear behavior marks the transition from the oscillatory phase to the late–time regime governed by a slower decay.

\begin{figure}[t!]
    \centering
    \includegraphics[scale=0.53]{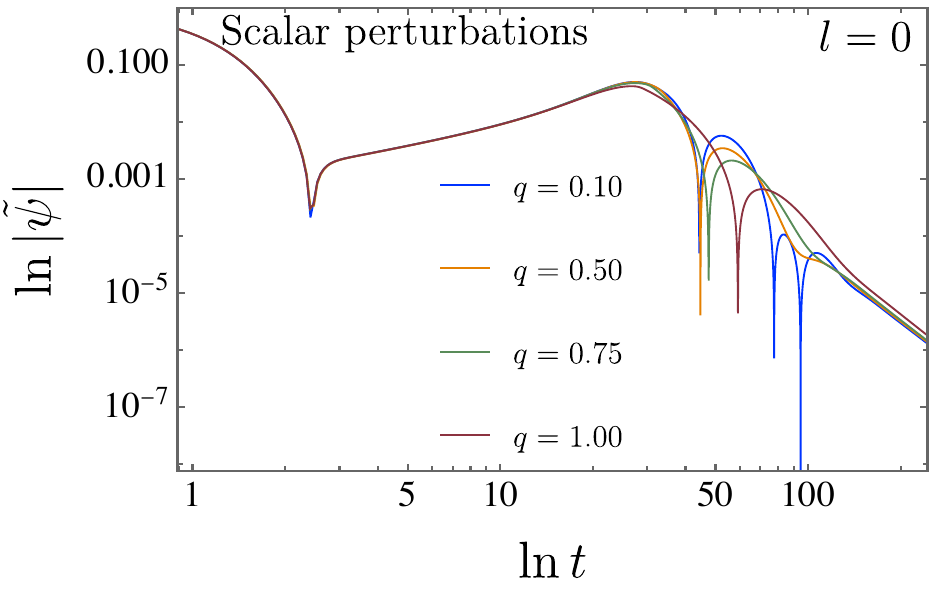}
    \includegraphics[scale=0.53]{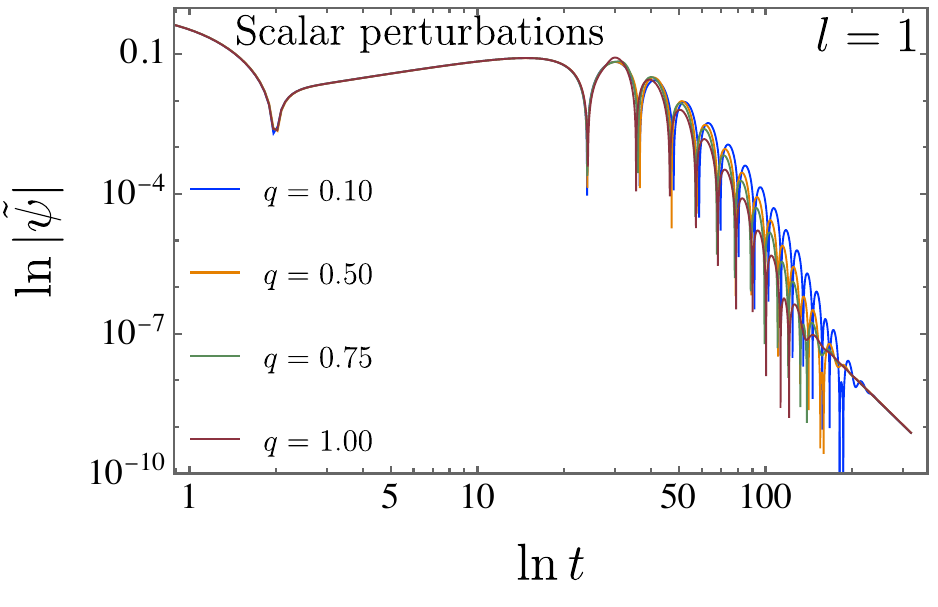}
    \includegraphics[scale=0.53]{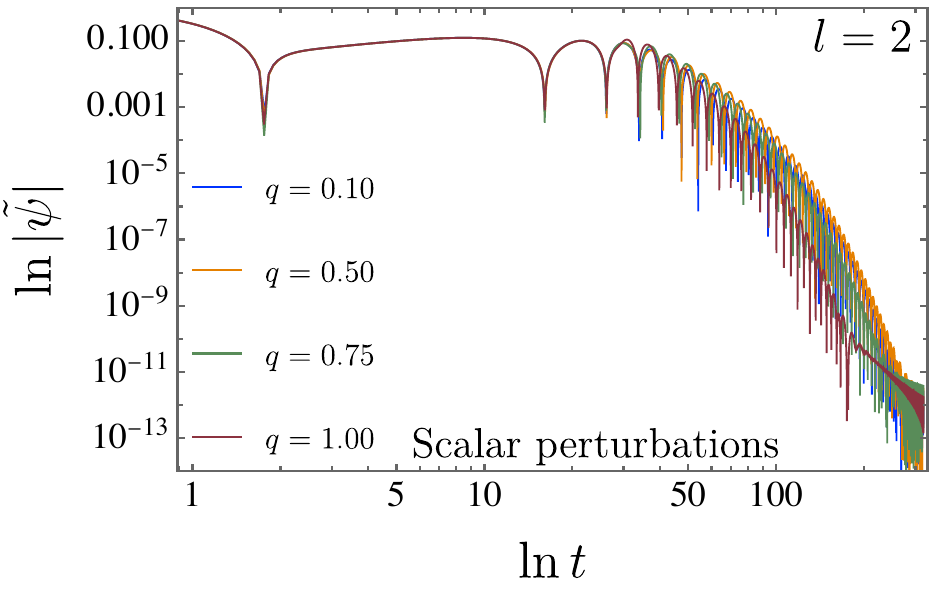}
  \caption{The asymptotic evolution of the scalar perturbation for Model I, represented on logarithmic axes by relating $\ln|\tilde{\psi}|$ to $\ln t$. The mass is fixed at $M=1$, while the deformation parameter takes the values $q=0.10$, $0.50$, $0.75$, and $1.00$. The figure is organized according to the angular index, with the $l=0$ mode presented in the upper–left panel, the $l=1$ mode in the upper–right panel, and the $l=2$ mode in the lower panel, where the late–time power–law decay becomes manifest.}
    \label{lnlnpsitimedomainmodel1}
\end{figure}

The late--time dynamics is further emphasized in Fig.~\ref{lnlnpsitimedomainmodel1}, where the evolution of $\tilde{\psi}$ is plotted on logarithmic axes, preserving the same panel arrangement. This scaling makes the asymptotic behavior explicit, revealing a clear power–law decay at large times. The emergence of these tails signals the final stage of the perturbation evolution, following the QNM ringing phase.

%%%%%%%%%%%%%%%%%%%%%%%%%%%%%%%%%%%%%%%%%%%%%%%%%%%%%%%%%%%%%%%%%%%%%%%%%%%%%%%%%%%%%%%%%%%%%%%%%%%%%%%%%%%%%%%%%%%%%%%%%%%%%%%%%%%%%%%%%%%%%%%%%%%%%%%%%%%%%%%%%%%%%%%%%%%%%%%%%%%%%%%%%%%%%%%%%%%%%%%%%%%%%%%%%%%%%%%%%%%%%%%%%%%%%%%%%%%%%%%%%%%%%%%%%%%%%%%%%%%%%%%%%%%%%%%%%%%%%%%%%%%%%%%%%%%%%%%%%%%%%%%%%%%%%%%%%%%%%%%%%%%%%%%%%%%%%%%%%%%%%%%%%%%%%%%%%%%%%%%%%%%%%%%%%%%%%%%%%%%%%%%%%%%%%%%%%%%%%%%%%%%%%%%%%%%%%%%%%%%%%%%%%%%%%%%%%%%%%%%%%%%%%%%%
\subsection{Model II}

\begin{figure}[t!]
    \centering
    \includegraphics[scale=0.53]{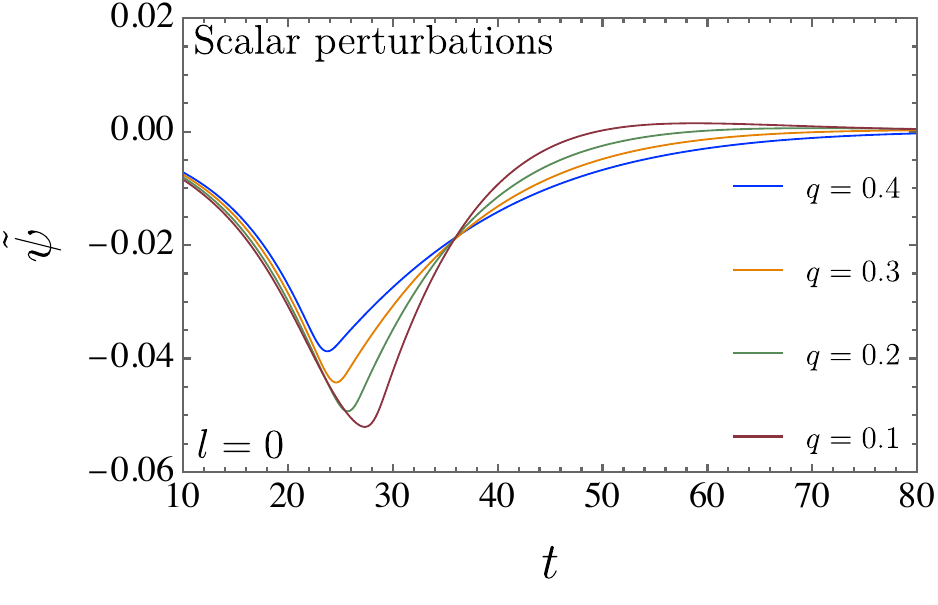}
    \includegraphics[scale=0.53]{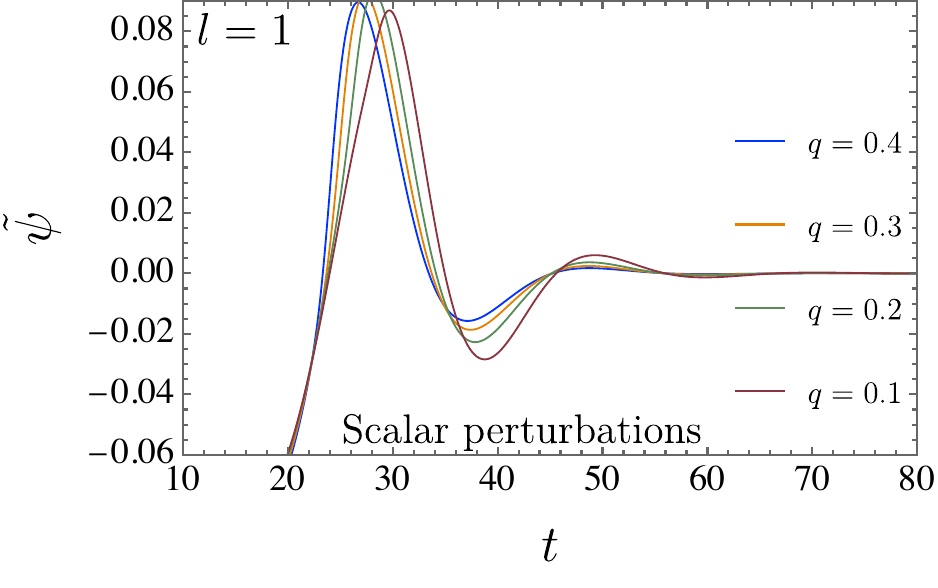}
    \includegraphics[scale=0.53]{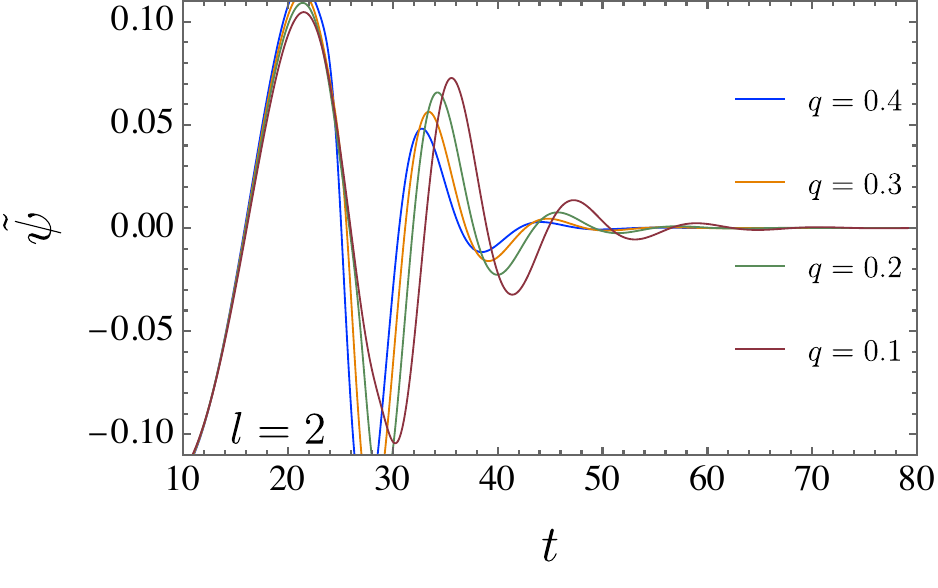}
     \caption{The temporal evolution of the scalar waveform $\tilde{\psi}$ is depicted for the Model II black hole by fixing the mass parameter to $M=1$ and assigning the deformation parameter the values $q=0.1$, $0.2$, $0.3$, and $0.4$. Each panel corresponds to a different angular contribution, with the $l=0$ component displayed in the upper–left frame, the $l=1$ component in the upper–right frame, and the $l=2$ component in the lower frame.}
    \label{psitimedomain}
\end{figure}

Here we discuss the temporal response of scalar disturbances propagating on the black hole background of Model II. The numerical signals displayed in Fig.~\ref{psitimedomain} are obtained by evolving the field $\tilde{\psi}$ with the mass fixed at $M=1$ and by assigning the deformation parameter the values $q=0.10$, $0.50$, $0.75$, and $1.00$. The panels correspond to different angular contributions, with the $l=0$ mode shown in the upper–left frame, the $l=1$ mode in the upper–right, and the $l=2$ mode in the lower frame. At intermediate times, all configurations exhibit damped oscillations, which characterize the QNM ringing produced by scalar excitations of the geometry defined in Eq.~(\ref{Eq_A3}).

\begin{figure}[t!]
    \centering
    \includegraphics[scale=0.53]{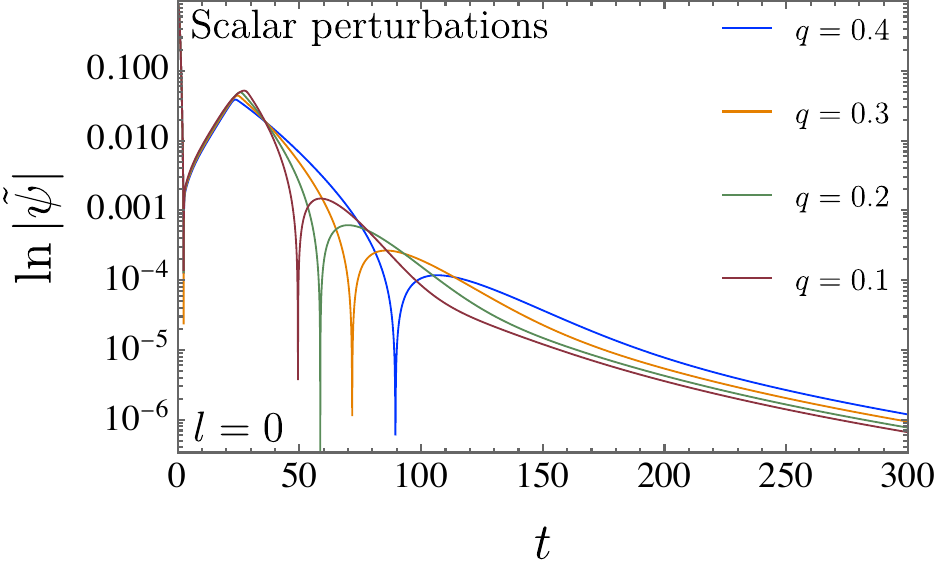}
    \includegraphics[scale=0.53]{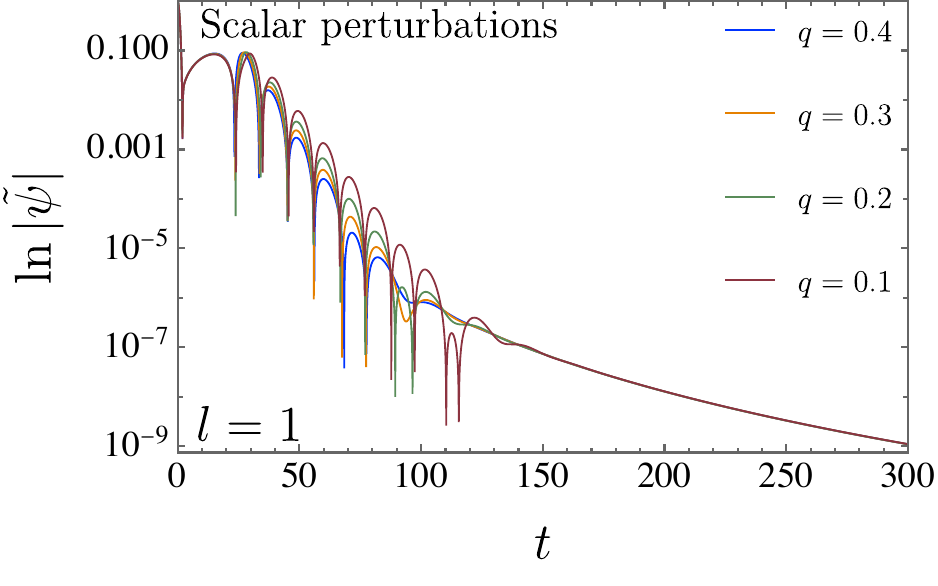}
    \includegraphics[scale=0.53]{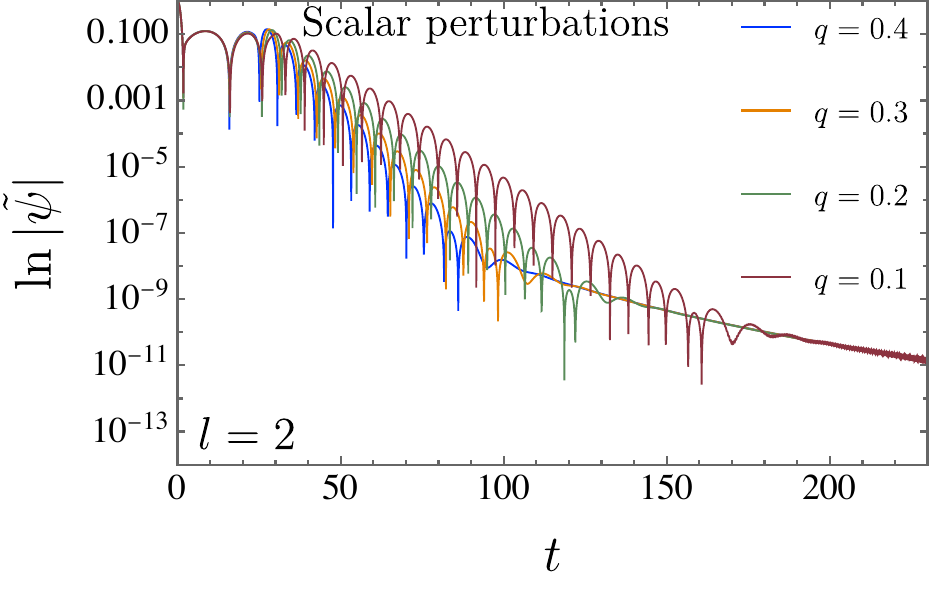}
    \caption{The logarithmic evolution of the scalar amplitude $\ln|\tilde{\psi}|$ for the Model II black hole geometry. The mass is kept fixed at $M=1$, while the deformation parameter is set to $q=0.1$, $0.2$, $0.3$, and $0.4$. The figure is organized by angular mode, with the $l=0$ sector presented in the upper–left frame, the $l=1$ sector in the upper–right frame, and the $l=2$ sector in the lower frame.}
    \label{lnpsitimedomain}
\end{figure}

The attenuation of these oscillations is more clearly exposed in Fig.~\ref{lnpsitimedomain}, where the evolution is recast in terms of $\ln|\tilde{\psi}|$. For each choice of $q$ and $l$, the QNM phase appears as a nearly straight segment, indicating exponential decay. The subsequent deviation from this behavior signals the breakdown of the oscillatory regime and the onset of a slower, late–time decay.

\begin{figure}[t!]
    \centering
    \includegraphics[scale=0.53]{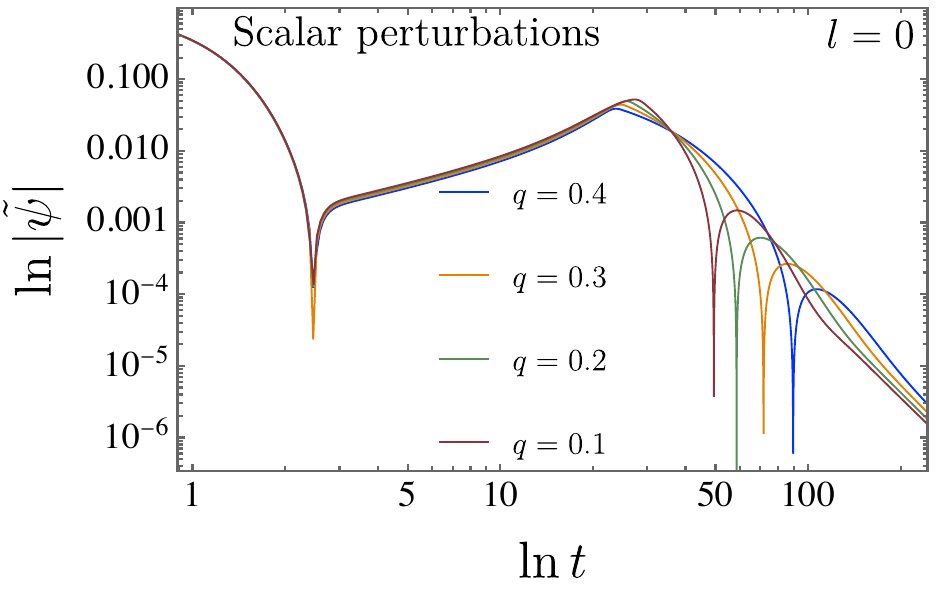}
    \includegraphics[scale=0.53]{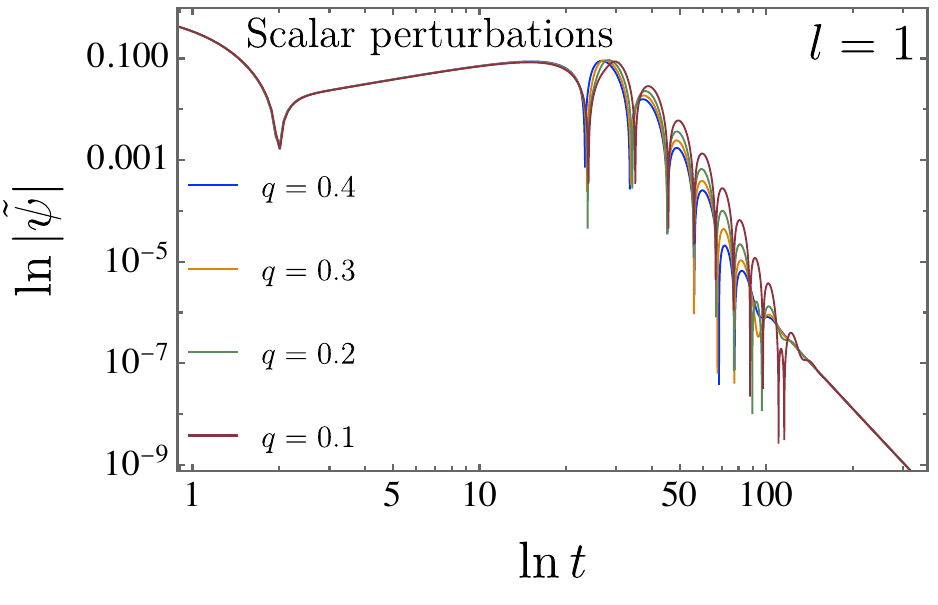}  
    \includegraphics[scale=0.53]{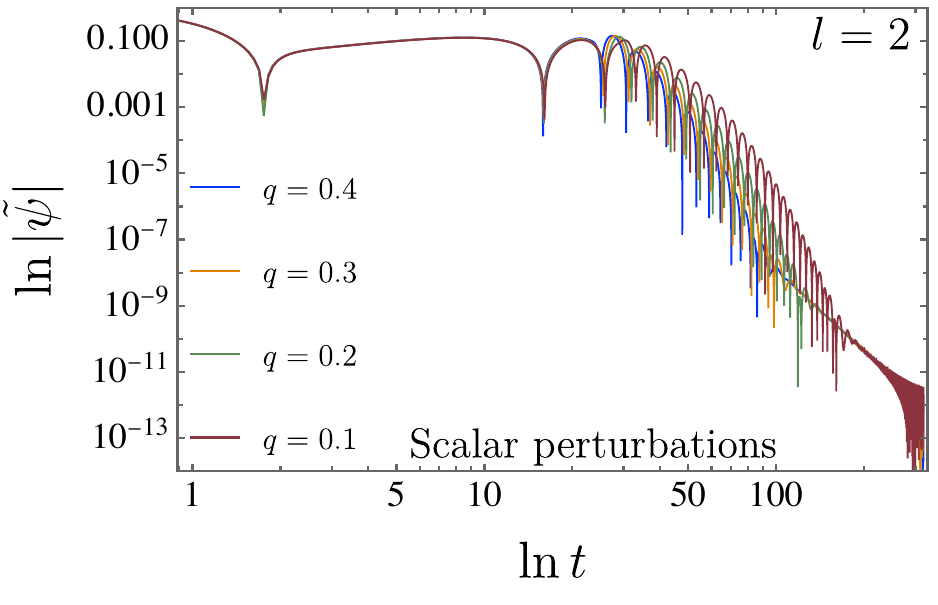}
  \caption{The late–time behavior of the scalar field in Model II is displayed using a double–logarithmic representation, where $\ln|\tilde{\psi}|$ is plotted against $\ln t$. All curves correspond to a fixed mass $M=1$, while the deformation parameter is chosen as $q=0.1$, $0.2$, $0.3$, and $0.4$. The panels are arranged by multipole order, with the $l=0$ contribution shown in the upper–left panel, the $l=1$ contribution in the upper–right panel, and the $l=2$ contribution in the lower panel, emphasizing the emergence of the late–time power–law decay.}
    \label{lnlnpsitimedomain}
\end{figure}

To isolate the asymptotic behavior, Fig.~\ref{lnlnpsitimedomain} presents the same data on double–logarithmic axes, keeping the panel layout unchanged. In this representation, the late–time evolution is clearly resolved as a power–law decrease, confirming the appearance of tail behavior that follows the QNM ringing stage.

%%%%%%%%%%%%%%%%%%%%%%%%%%%%%%%%%%%%%%%%%%%%%%%%%%%%%%%%%%%%%%%%%%%%%%%%%%%%%%%%%%%%%%%%%%%%%%%%%%%%%%%%%%%%%%%%%%%%%%%%%%%%%%%%%%%%%%%%%%%%%%%%%%%%%%%%%%%%%%%%%%%%%%%%%%%%%%%%%%%%%%%%%%%%%%%%%%%%%%%%%%%%%%%%%%%%%%%%%%%%%%%%%%%%%%%%%%%%%%%%%%%%%%%%%%%%%%%%%%%%%%%%%%%%%%%%%%%%%%%%%%%%%%%%%%%%%%%%%%%%%%%%%%%%%%%%%%%%%%%%%%%%%%%%%%%%%%%%%%%%%%%%%%%%%%%%%%%%%%%%%%%%%%%%%%%%%%%%%%%%%%%%%%%%%%%%%%%%%%%%%%%%%%%%%%%%%%%%%%%%%%%%%%%%%%%%%%%%%%%%%%%%%%%%
\subsection{Model III}

\begin{figure}[t!]
    \centering
    \includegraphics[scale=0.53]{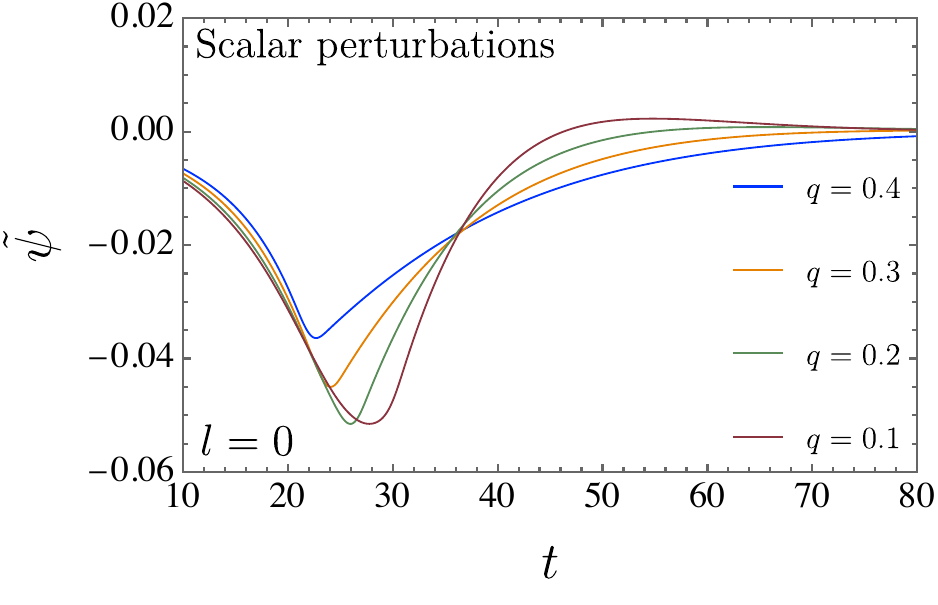}
    \includegraphics[scale=0.53]{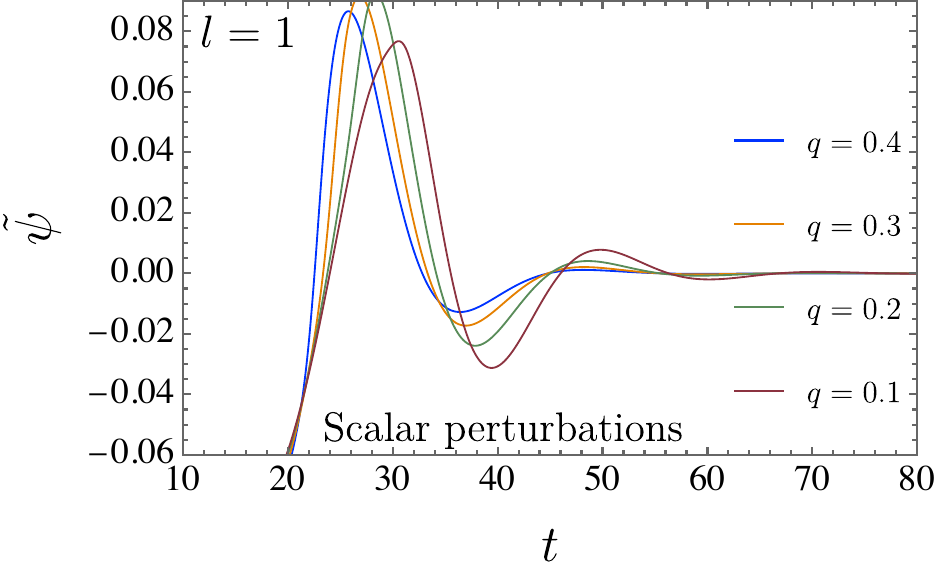}
    \includegraphics[scale=0.53]{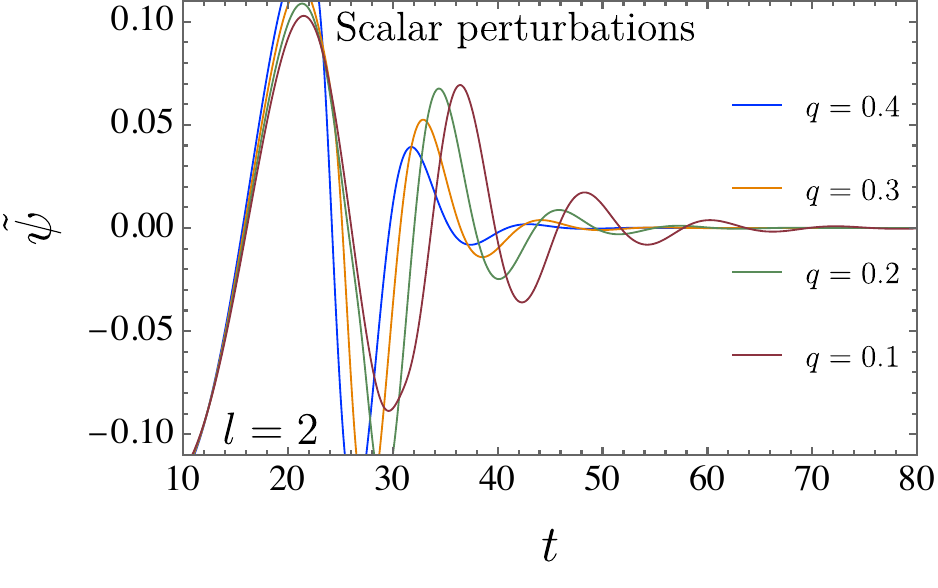}
     \caption{For Model III, the scalar field $\tilde{\psi}$ is evolved in time while the black hole mass is kept at $M=1$ and the deformation parameter is set to $q=0.1$, $0.2$, $0.3$, and $0.4$. The resulting signals are arranged according to the angular index, with the $l=0$ mode shown in the upper–left panel, the $l=1$ mode in the upper–right panel, and the $l=2$ mode in the lower panel.}
    \label{psitimedomainmodel3}
\end{figure}

The evolution in time of scalar perturbations supported by the black hole space-time of Model III is analyzed in this section. Fig.~\ref{psitimedomainmodel3} displays the numerical waveforms of the field $\tilde{\psi}$ obtained by fixing the mass parameter to $M=1$ and selecting the deformation parameter as $q=0.10$, $0.50$, $0.75$, and $1.00$. The results are arranged according to the angular momentum index, with the $l=0$ contribution shown in the upper–left panel, the $l=1$ contribution in the upper–right panel, and the $l=2$ contribution in the lower panel. For all configurations, the intermediate–time signal is dominated by oscillations whose amplitude decreases exponentially, identifying the QNM ringing associated with scalar fluctuations of the background defined by Eq.~(\ref{A41}).

\begin{figure}[t!]
    \centering
    \includegraphics[scale=0.53]{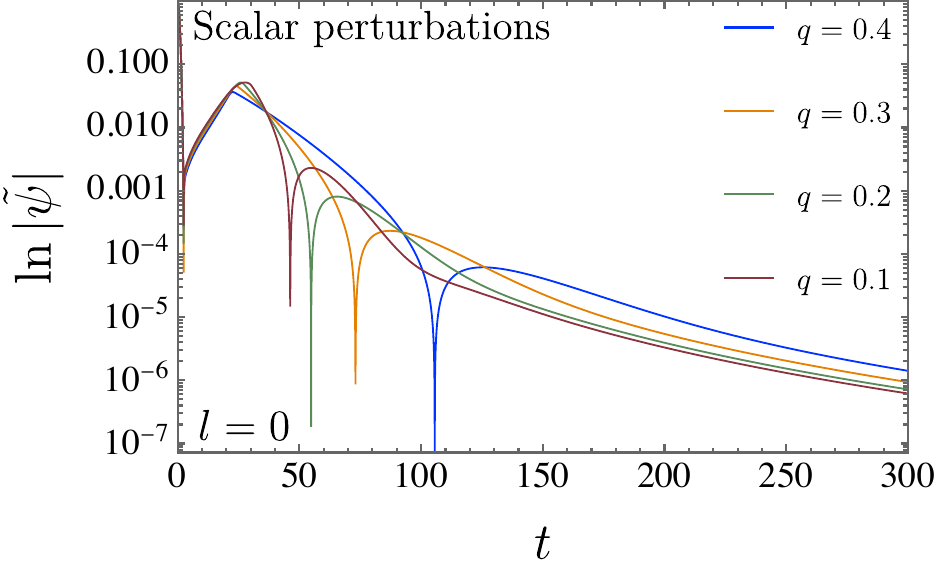}
    \includegraphics[scale=0.53]{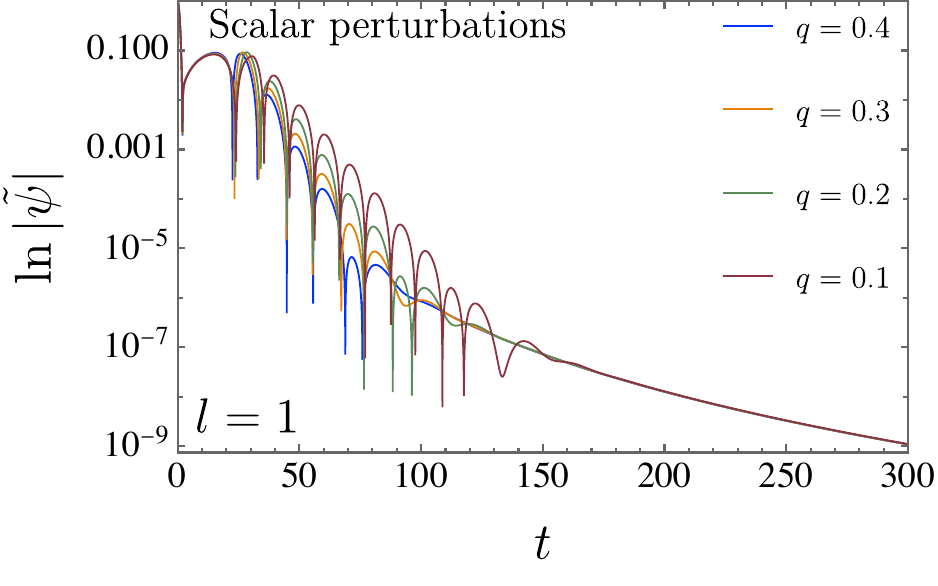}
    \includegraphics[scale=0.53]{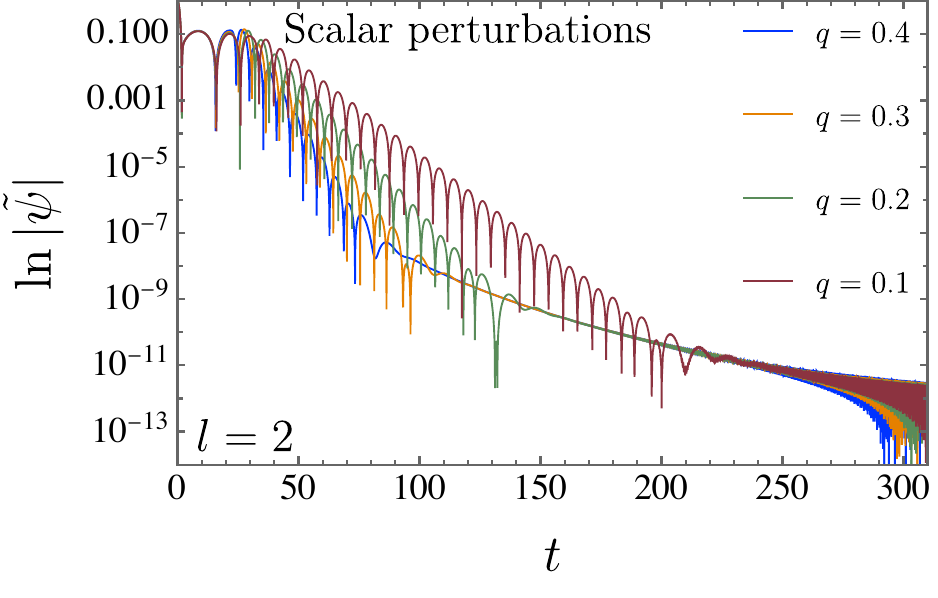}
    \caption{The time evolution of the scalar field amplitude in logarithmic form, $\ln|\tilde{\psi}|$, for the Model III black hole configuration. All curves correspond to a fixed mass $M=1$, while the deformation parameter is chosen as $q=0.1$, $0.2$, $0.3$, and $0.4$. The panels are arranged according to the multipole index, with the $l=0$ contribution shown in the upper–left panel, the $l=1$ contribution in the upper–right panel, and the $l=2$ contribution in the lower panel.}
    \label{lnpsitimedomainmodel3}
\end{figure}

A clearer view of the decay properties is obtained in Fig.~\ref{lnpsitimedomainmodel3}, where the quantity $\ln|\tilde{\psi}|$ is plotted for the same sets of $q$ and $l$. In this form, the QNM regime is characterized by an approximately linear behavior, corresponding to exponential damping. The departure from linearity marks the transition from the oscillatory phase to a regime governed by a slower decay.

\begin{figure}[t!]
    \centering
    \includegraphics[scale=0.53]{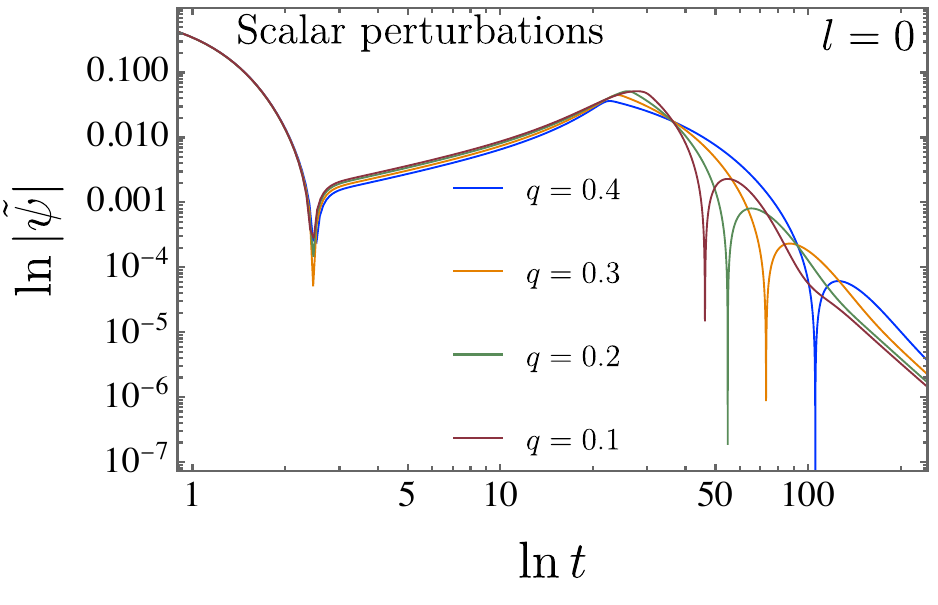}
    \includegraphics[scale=0.53]{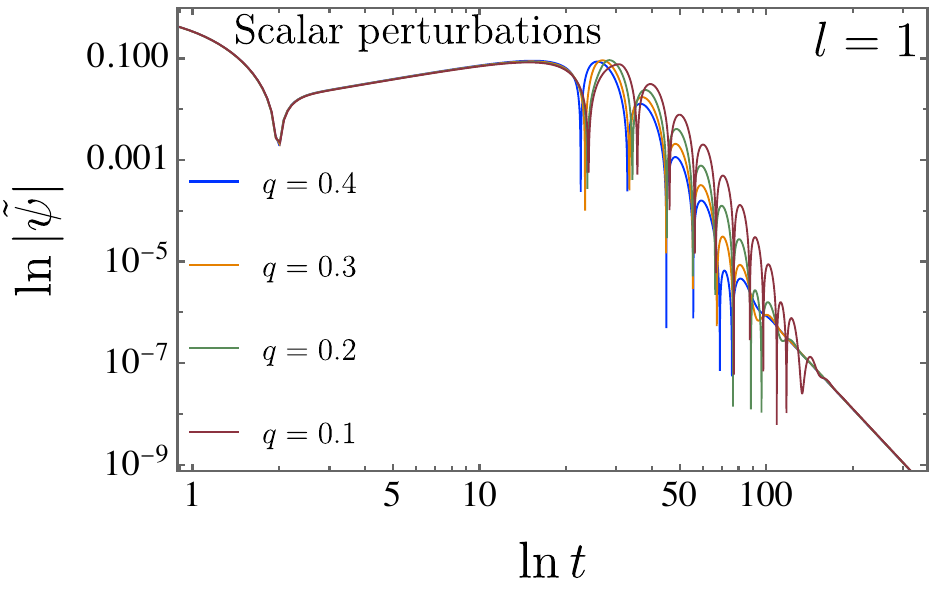}
    \includegraphics[scale=0.53]{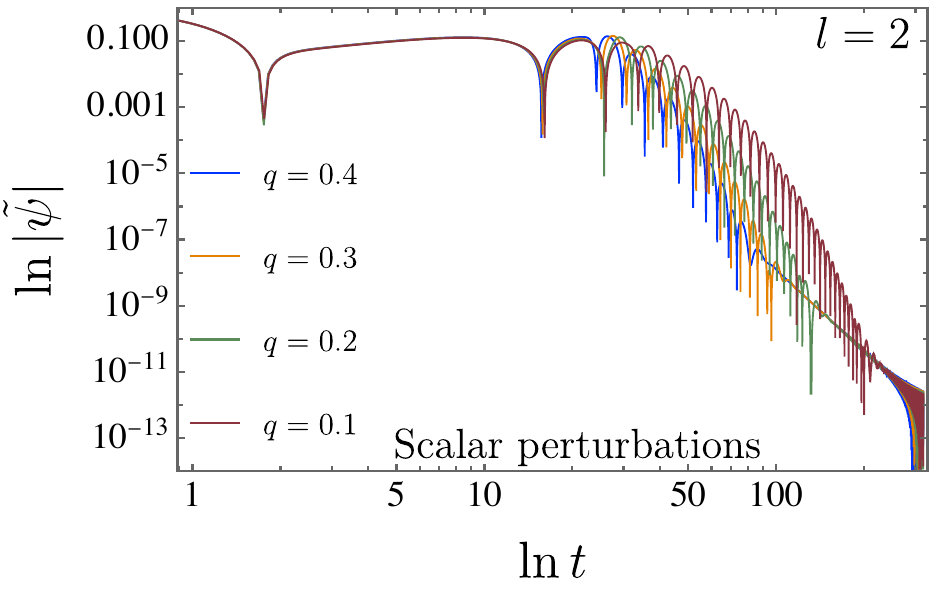}
  \caption{The asymptotic evolution of the scalar perturbation for Model III is illustrated on logarithmic scales by plotting $\ln|\tilde{\psi}|$ as a function of $\ln t$. The analysis is performed for a fixed mass $M=1$ and deformation parameter values $q=0.1$, $0.2$, $0.3$, and $0.4$. The figure is structured according to the angular momentum index, with the $l=0$ mode displayed in the upper–left panel, the $l=1$ mode in the upper–right panel, and the $l=2$ mode in the lower panel, where the characteristic power–law decay at late times becomes evident.}
    \label{lnlnpsitimedomainmodel3}
\end{figure}

The late–time response is further highlighted in Fig.~\ref{lnlnpsitimedomainmodel3}, which presents the evolution on logarithmic axes while preserving the same panel organization. This representation makes the asymptotic behavior explicit, revealing a power–law decay at large times and confirming the emergence of the tail stage that follows the QNM ringing.

%%%%%%%%%%%%%%%%%%%%%%%%%%%%%%%%%%%%%%%%%%%%%%%%%%%%%%%%%%%%%%%%
\section{Summary and Conclusion}\label{sec:concl}
%%%%%%%%%%%%%%%%%%%%%%%%%%%%%%%%%%%%%%%%%%%%%%%%%%%%%%%%%%%%%%%%

In this article we have investigated spherically symmetric black hole solutions within GR characterized by the absence of curvature singularities. This is done following an approach that was developed recently in Ref.~\cite{Vertogradov:2024seh}, based on a variable equation of state, namely, $P=\zeta(r)\rho$. Within this prescription we obtained two new regular black hole configurations, corresponding to two additional specific choices for the function $\zeta(r)$. The matter content supporting both the original solution and the two new ones is determined via the coupling the GR field equations, as given by Eq.~\eqref{EqM}, to a matter sector described by nonlinear electrodynamics (NLED), governed exclusively by a magnetic charge. Furthermore, although black hole solutions in GR are generally singular, the imposition of an appropriate condition on a parameter $w_0$ related to the mass function appearing in the metric components renders the Kretchsmann scalar finite everywhere. 

For the three models discussed in this work, described by the metric functions~\eqref{A1}, \eqref{Eq_A3}, and \eqref{A41}, we investigated the existence and structure of horizons. In all cases, the Schwarzschild solution was recovered in the limit $|q| \to 0$, while for $|q| \to \infty$ the space-times become asymptotically Minkowskian. The qualitative behavior of the metric functions turns out to be similar across the three models. In particular, the analysis of the critical charge showed that, for $|q| < |q_c|$, up to two horizons were allowed in all cases, corresponding to an outer event horizon and an inner Cauchy horizon. Each configuration admitted a degenerate horizon at $|q| = |q_c|$, whereas no horizons were present for $|q| > |q_c|$ for each of the models. Using the corresponding metric functions, given by Eqs.~\eqref{A3} and \eqref{A4}, we reconstructed the Lagrangian density associated with each model. Furthermore, the series expansions of the Lagrangians~\eqref{L3_M2} and \eqref{L_M3} for small values of the electromagnetic invariant $F$ exhibited a leading term linear in $F$, thereby recovering the Maxwell limit. Finally, we also determined the regions of space-time in which each model violates (any of) the energy conditions.

Next, we constrained the parameter of the two new models introduced in this work using the EHT constraints on the dark central region (the ``shadow") of Sgr A$^*$. To this end we followed the procedure described and systematically exploited in \cite{Vagnozzi:2022moj} for alternative black hole geometries. Since the formalism in which the metric functions were interpreted corresponds to that of NLED, we then took into account the fact that photon propagation is governed by an effective geometry rather than by the background spacetime itself. After constructing the corresponding effective metric, we computed the shadow radius for each model, identified here as the central dark region close tracked by highly-lensed light trajectories (i.e. the critical curve), and used the EHT results to constrain the magnetic charge. For Model~II, presented in Sec.~\ref{Mod_II}, the shadow radius remained within the EHT bounds for $0 \leq |q|/M \lesssim 0.102359$, whereas for Model~III, discussed in Sec.~\ref{Mod_III}, consistency with the EHT constraints was maintained for $0 \leq |q|/M \lesssim 0.00492225$. Moreover, we observed in both models that, as the value of $|q|$ increase, the shadow radius decreases, exhibiting a behavior similar to that of the Reissner–Nordström solution. 
Furthermore, regarding Model I (see Sec.~\ref{Mod1}), the constraint on the shadow radius has already been derived in Ref.~\cite{Vertogradov:2024seh}, where the parameter to be constrained is $R$. Approximately for $R \approx 0.9$, the shadow radius of Model I begins to deviate from the bounds established by the EHT. However, by considering the relation given by Eq.~\eqref{Eq_R}, we observe that the interval presents no novelty, since: $0 \leq |q|/M \lesssim 0.9$, for which the shadow radius of the model starts to deviate from the values established by the EHT.

The final analysis of our work was to examine the dynamical response of the three regular black hole configurations through the combined analysis of quasinormal modes and time-domain evolution of scalar perturbations. In all models, the QNM spectra displayed well-defined complex frequencies whose real and imaginary parts varied smoothly with the deformation parameter $q$, indicating changes in both the oscillation scale and damping rate with respect to the Schwarzschild limit. For fixed multipole number and overtone index, increasing $q$ leads to higher oscillation frequencies accompanied by stronger damping, a behavior that was directly associated with the deformation of the effective potential barrier induced by the regular core structure. These features were consistently supported by the time-domain profiles, which exhibited stable ringdown signals followed by power-law tails, with no indication of growing modes or late-time instabilities. Although the three models differed in their mass distributions and near-core behavior, their dynamical signatures remained qualitatively similar: the time evolution confirmed linear stability, while the frequency-domain analysis showed that the regularization mechanism produced quantitative shifts in the ringdown spectrum without modifying its overall structure.

As a future direction, it would be worthwhile to investigate the remaining perturbative sectors not addressed in this work, such as vector, tensor, and spinor fields. These sectors would allow a complementary analysis of both QNMs and time-domain evolution beyond the scalar case considered here. Once the full set of perturbations is available, related quantities such as greybody factors, absorption rates, and scattering cross sections could also be examined. Finally, an extension of the present analysis to gravitational lensing for the three models studied here, in particular within the strong-deflection regime accompanied by images from accretion disks, would provide further characterization of their observational properties. \\

Finally, it is important to recognize that regular black hole geometries often face challenges concerning stability and theoretical consistency. The presence of a Cauchy horizon, for instance, is well known to be associated with mass inflation instability~\cite{Poisson:1989zz,Carballo-Rubio:2024dca}. Moreover, models supported solely by pure NLED may exhibit angular Laplacian instabilities~\cite{DeFelice:2024seu} and violate causality and unitarity conditions~\cite{Schellstede:2016zue,Bronnikov:2022ofk}. On the other hand, recent developments have shown that NED theories which are causal and regularize the charge self-energy can eliminate Cauchy horizons~\cite{Hale:2025ezt}, thereby suggesting a possible route to circumvent the mass inflation instability.

Although these are fundamental issues for certain matter sectors, they do not automatically invalidate a regular geometry, since the perturbation spectrum and the theoretical consistency depend on the complete model. Indeed, the same regular geometry, such as that of Bardeen, can be supported by different sources, including linear electrodynamics with a nonminimal scalar coupling~\cite{Cordeiro:2025ivw}, which may modify the stability properties. Black-bounce solutions also emerge as an alternative avenue for investigating the stability of regular compact objects~\cite{Cordeiro:2025ydg}. The present work, by focusing on the construction and analysis of exact static solutions, provides a foundation for future investigations, in which complete perturbative and dynamical analyses may assess the global stability of these geometries.

This line of reasoning is corroborated by the study of De Felice and Tsujikawa~\cite{DeFelice:2024ops}, who demonstrate that the Laplacian instability, although unavoidable in pure NED, can be avoided in more general theories with a scalar field, under certain conditions. Although that condition eliminates horizons in that particular case, it illustrates the fundamental principle: stability is not a property of the geometry itself, but rather of the complete matter model that supports it. Therefore, the possibility of constructing the Bardeen geometry with a linear electrodynamics source nonminimally coupled to a scalar field~\cite{Cordeiro:2025ivw} represents a promising avenue to circumvent the instabilities associated with pure NLED, since the matter sector, and consequently the perturbation equations, are entirely different.

%%%%%%%%%%%%%%%%%%%%%%%%%%%%%%%%%%%%%%%%%%%%%%%%%%%%%%%%%%%%%%%%
\section*{Acknowledgments}
%%%%%%%%%%%%%%%%%%%%%%%%%%%%%%%%%%%%%%%%%%%%%%%%%%%%%%%%%%%%%%%%
%\hspace{0.5cm} 

AAAF is supported by Conselho Nacional de Desenvolvimento Cient\'{\i}fico e Tecnol\'{o}gico (CNPq) and Fundação de Apoio à Pesquisa do Estado da Paraíba (FAPESQ), project numbers 150223/2025-0 and 1951/2025. MER thanks Conselho Nacional de Desenvolvimento Cient\'ifico e Tecnol\'ogico - CNPq, Brazil, for partial financial support. This study was financed in part by the Coordena\c{c}\~{a}o de Aperfei\c{c}oamento de Pessoal de N\'{i}vel Superior - Brasil (CAPES) - Finance Code 001.
FSNL acknowledges support from the Funda\c{c}\~{a}o para a Ci\^{e}ncia e a Tecnologia (FCT) Scientific Employment Stimulus contract with reference CEECINST/00032/2018, and funding through the research grant UID/04434/2025. DRG is supported by the Spanish National Grants PID2022-138607NBI00 and CNS2024-154444, funded by MICIU/AEI/10.13039/501100011033 (“PGC Generación de Conocimiento") and FEDER, UE.

%%%%%%%%%%%%%%%%%%%%%%%%%%%%%%%%%%%%%%%%%%%%%%%%%%%%%%%%%%%%%%%%
    %\bibliographystyle{apsrev4-2}
	%\bibliography{main}
	%%\bibliographystyle{unsrt}
%%%%%%%%%%%%%%%%%%%%%%%%%%%%%%%%%%%%%%%%%%%%%%%%%%%%%%%%%%%%%%%%

%%%%%%%%%%%%%%%%%%%%%%%%%%%%%%%%%%%%%%%%%%%%%%%%%%%%%%%%%%%%%%%%
\end{document}